\documentclass[structabstract]{aa}  
\usepackage{graphicx}
\usepackage{txfonts}
\usepackage[draft=False,colorlinks=true,linkcolor=black,citecolor=blue,urlcolor=blue]{hyperref}

\usepackage{longtable,pdflscape}

\usepackage{subcaption}

\newcommand{\tempMedGdc}{17.8~K}
\newcommand{\massTotDust}{6.7$\pm$0.2$\cdot$10$^{5}$~M$_{\odot}$}
\newcommand{\massMin}{1.2$\cdot$10$^{3}$}
\newcommand{\massMax}{1.4$\cdot$10$^{4}$~M$_{\odot}$}
\newcommand{\massLimit}{3.8$\cdot$10$^{3}$~M$_{\odot}$}
\newcommand{\selAllHii}{90}

\newcommand{\selSrcAll}{146}

\newcommand{\massMed}{4.6$\cdot$10$^{3}$~M$_{\odot}$}
\newcommand{\massD}{86}
\newcommand{\massAll}{134}
\newcommand{\massBin}{0.17~dex}
\newcommand{\cumMassNr}{75}
\newcommand{\tempD}{90}
\newcommand{\tempAll}{140}
\newcommand{\sizeMed}{190~pc}
\newcommand{\sizeD}{90}

\makeatletter
\def\LongtableFooter{%
  \multicolumn{\LT@cols}{r}{\framebox[1.1\width]{\textbf{continued on the following page}}}\\}
\makeatother

\usepackage{natbib}
\bibpunct{(}{)}{;}{a}{}{,} 

\newcommand{\msun}{M$_{\odot}$}

\defcitealias{Deharveng1988}{D88}
\defcitealias{Faesi2014}{F14}

\begin{document}

   \title{Gathering dust: A galaxy-wide study of dust emission from cloud complexes in NGC~300\thanks{{\it Herschel} is an ESA space observatory with science instruments provided by European-led Principal Investigator consortia and with important participation from NASA.}}
   
   \author{M. Riener
          \inst{1,2}
          \and
          C. M. Faesi
          \inst{3}
          \and
          J. Forbrich
          \inst{4,3,2}
          \and
          C. J. Lada
          \inst{3}
          }

   \institute{Max Planck Institute for Astronomy, K\"onigstuhl 17, 69117 Heidelberg, Germany
         \and
             Institute for Astronomy (IfA), University of Vienna, T\"urkenschanzstrasse 17, 1180 Vienna, Austria
         \and
             Harvard-Smithsonian Center for Astrophysics, 60 Garden Street, Cambridge, MA 02138, USA
         \and
             University of Hertfordshire, Centre for Astrophysics Research, Hatfield AL10 9AB, UK
             }

   \date{Received ..., 2017; accepted ..., 2017}

 
  \abstract
   {}
   {We use multi-band observations by the \textit{Herschel Space Observatory} to study the dust emission properties of the nearby spiral galaxy NGC~300. 
   We compile a first catalogue of the population of giant dust clouds (GDCs) in NGC~300, including temperature and mass estimates, and give an estimate of the total dust mass of the galaxy.}
   {We carried out source detection with the multiwavelength source extraction algorithm \textit{getsources}. 
   We calculated physical properties, including mass and temperature, of the GDCs from five-band \textit{Herschel} PACS and SPIRE observations from 100-500~$\mu$m; the final size and mass estimates are based on the observations at 250~$\mu$m that have an effective spatial resolution of $\sim$170~pc.
   We correlated our final catalogue of GDCs to pre-existing catalogues of HII~regions to infer the number of GDCs associated with high-mass star formation and determined the H$\alpha$ emission of the GDCs.}
   {Our final catalogue of GDCs includes 146 sources, 90 of which are associated with known HII regions. 
    We find that the dust masses of the GDCs are completely dominated by the cold dust component and range from $\sim$~1.1$\cdot$10$^{3}$ to 1.4$\cdot$10$^{4}$~\msun.
   The GDCs have effective temperatures of $\sim$~$13-23$~K and show a distinct cold dust effective temperature gradient from the centre towards the outer parts of the stellar disk.
   We find that the population of GDCs in our catalogue constitutes $\sim$~16\% of the total dust mass of NGC~300, which we estimate to be about 5.4$\cdot$10$^{6}$~\msun .
   At least about 87\% of our GDCs have a high enough average dust mass surface density to provide sufficient shielding to harbour molecular clouds.
   We compare our results to previous pointed molecular gas observations in NGC~300 and results from other nearby galaxies and also conclude that it is very likely that most of our GDCs are associated with complexes of giant molecular clouds. }
   {}


   \maketitle
%

\section{Introduction}

Interstellar dust plays a pivotal role in the formation of molecular clouds and accordingly also in the formation of stars (e.g. \citealp{Li2002}, \citealp{Bergin2007}, \citealp{Dobbs2014}). 
Molecules predominantly form on the surfaces of dust grains, especially molecular hydrogen (H$_{2}$), the principal component of molecular clouds (e.g. \citealp{Gould1963}, \citealp{Shull1982}, \citealp{Cazaux2002}). 
The dust grains further act as an additional shielding component against the energetic interstellar radiation field that would otherwise rapidly destroy the newly formed molecules. 
Furthermore, since a significant fraction of ultraviolet and optical photons emitted by stars are re-radiated by dust in the infrared, dust is an often-used tracer of star formation in galaxies \citep[e.g.][]{Kennicutt2012}.
 
Dust is morphologically complex on galactic scales, but we can approximate its distribution into two main components: a smooth, more diffuse component of dust that is associated with the young and old stellar populations in the galactic disk and fills much of the volume of the galaxy; and a clumpy, more compact dust component that is likely associated with star-forming regions, such as giant molecular clouds (GMCs) or HII~regions.
Studies that are predominantly interested in identifying the  dust component that is associated with complexes of molecular clouds thus often use source detection algorithms that automatically filter out the diffuse dust component while focussing on identifying the more clumpy dust structures. 
This approach has proved fruitful in characterizing the properties of the interstellar medium (ISM) in star-forming regions within nearby galaxies (e.g. \citealp{Foyle2013}, \citealp{Natale2014}, \citealp{Kirk2015}).

High-resolution studies of tracers of both star formation and the cold, clumpy component of the ISM are required to obtain additional insight into the process of star formation within galaxies.
Performing such studies, however, requires observations across a range of regions within galaxies so that the role of environment on star formation can be assessed. 
Our location inside the Milky Way makes it very difficult to assemble such a sample within the Galaxy because of line of sight confusion and large inaccuracies in GMC properties due to systematic distance uncertainties. 
One obvious way to circumvent this problem is to observe other nearby galaxies, in which these limitations are minimized, while also providing crucial information about the galactic context of star formation.
Given the difficulty of surveying an entire galaxy in molecular lines to find GMCs, more easily available dust emission observations can be used as a proxy for GMCs.

In the past few years the \textit{Herschel Space Observatory} has enabled high-resolution studies of the far-infrared (FIR) dust emission in nearby galaxies including M33 \citep{Natale2014} and M31 \citep{Kirk2015}.
These dust emission observations, in addition to previous molecular gas observations, allowed a consistent and comprehensive study of the clumpy ISM within these galaxies and showed that the properties of GMCs and GMC complexes in the most nearby galaxies are consistent with those of clouds found in the Milky Way \citep[e.g.][]{Kirk2015}.

Beyond the Local Group, the spiral galaxy NGC~300 is particularly well-suited for studies connecting Galactic and extragalactic star formation.
It is located in the Sculptor Group in close proximity to the Milky Way (about $1.98$ Mpc; \citealp{Tully2013}) and has a relatively low inclination angle (48.5$^{\circ}$; see also Table \ref{tbl:ngc300} for a list of more galaxy parameters). 

NGC~300 shows many signs of recent (high-mass) star-forming activity, as is evident from its large number of HII~regions (e.g. \citealp{Deharveng1988}), supernova remnants \citep[SNRs; e.g.][]{Blair1997,Payne2004}, and OB associations \citep{Pietrzynski2001}. 
The star formation rate (SFR) in the disk of NGC~300 was estimated by \citet{Helou2004} to be $\sim 0.11$~\msun~yr$^{-1}$. 
NGC~300 shows a metallicity gradient, whose values range from $\sim 0.3$~Z$_{\odot}$ in the outskirts to $\sim 0.8$~Z$_{\odot}$ in the centre (\citealp{Kudritzki2008}, \citealp{Bresolin2009}, \citealp{Gogarten2010}).

\begin{table}
\caption{Galaxy parameters of NGC 300}
\centering
\small
\renewcommand{\arraystretch}{1.2}
\begin{tabular}{lr}
\hline\hline
 &  \\
\hline
Morphological type                              &Scd\\
R.A. (J2000)                                    &00$^{h}$ 54$^{m}$ 53.54$^{s}$\\
Dec (J2000)                                             &-37$^{\circ}$ 41$'$ 04.3$''$\\
Distance\tablefootmark{a}               &1.98 Mpc\\
Major axis position angle %
(north eastwards)                               &114.3$^{\circ}$\\
Inclination between line %
of sight and polar axis                         &48.5$^{\circ}$\\
R$_{25}$\tablefootmark{b}               &9.75$'$ ($\sim$~5.62~kpc)\\
Estimated total mass (dark and luminous)\tablefootmark{c}       &2.9$\cdot$10$^{10}$~\msun\\
Stellar mass\tablefootmark{d}   &2.1$\cdot$10$^{9}$~\msun\\
HI mass\tablefootmark{c}            &1.5$\cdot$10$^{9}$~\msun\\
\hline
\end{tabular}
\tablefoot{ 
\footnotesize 
Data from the HyperLeda database \citep{Makarov2014} except where noted otherwise.\\
\tablefoottext{a}{\citet{Tully2013}}\\
\tablefoottext{b}{length of the projected semi-major axis of the galaxy at the isophotal level 25 mag/arcsec$^{2}$}\\
\tablefoottext{c}{\citet{Westmeier2011}}\\
\tablefoottext{d}{Data from the S$^{4}$G catalogue (\citealp{Sheth2010}; \citealp{Munoz-Mateos2013}; \citealp{Querejeta2015})}
}
\label{tbl:ngc300}
\end{table}

More recently, \citet[][hereafter F14]{Faesi2014} undertook a study of molecular gas and star formation towards a sample of 76 HII~regions of NGC~300 at 250~pc scales with the Atacama Pathfinder Experiment (APEX). 
These authors concluded that the scaling relation between the SFR and total molecular gas mass is consistent with the relation found by \citet{Lada2012} for local clouds.
\citet{Faesi2016} followed this up with high angular resolution observations with the Submillimeter Array (SMA) of 11 regions already targeted by APEX and found that the GMCs they observed share similar physical properties and GMC scaling relations with GMCs in other nearby galaxies and our Milky Way.

Although a complete census of GMCs in NGC~300 would be extremely valuable, a complete mapping survey for CO emission would require prohibitive amounts of telescope time with either ALMA, the SMA, or a single dish, single mode, antenna such as APEX. 
However, we used observations by the \textit{Herschel Space Observatory} to obtain a submillimetre/FIR survey of the entire galaxy to measure its dust content. 
In this paper we present the first results from our analysis of \textit{Herschel} observations of NGC~300 and produce the first comprehensive study of the total population of giant dust clouds (GDCs) throughout this galaxy. 
The catalogue of GDCs and the analysis of their properties is intended to serve as the foundation for establishing a galaxy-wide census of GMCs in NGC~300 and will be useful in future comparisons between the dust and molecular gas characteristics in this galaxy.

This paper is structured as follows: In section~\ref{cha:Observations} we present the observations and data. 
Section~\ref{cha:sourceextraction} discusses the source extraction algorithm and our chosen settings. 
Section~\ref{cha:sourcecat} deals with the selection criteria we used to obtain the final catalogue of GDCs and describes how we determined GDC physical properties, such as temperature and mass. 
In section~\ref{cha:discuss} we discuss our results and place these results in the context of the whole galaxy. 


\section{Observations and data}
\label{cha:Observations}

\subsection{Herschel Space Observatory}
\label{sec:obs-herschel}

\begin{figure*}
\centering
\includegraphics[width=\textwidth]{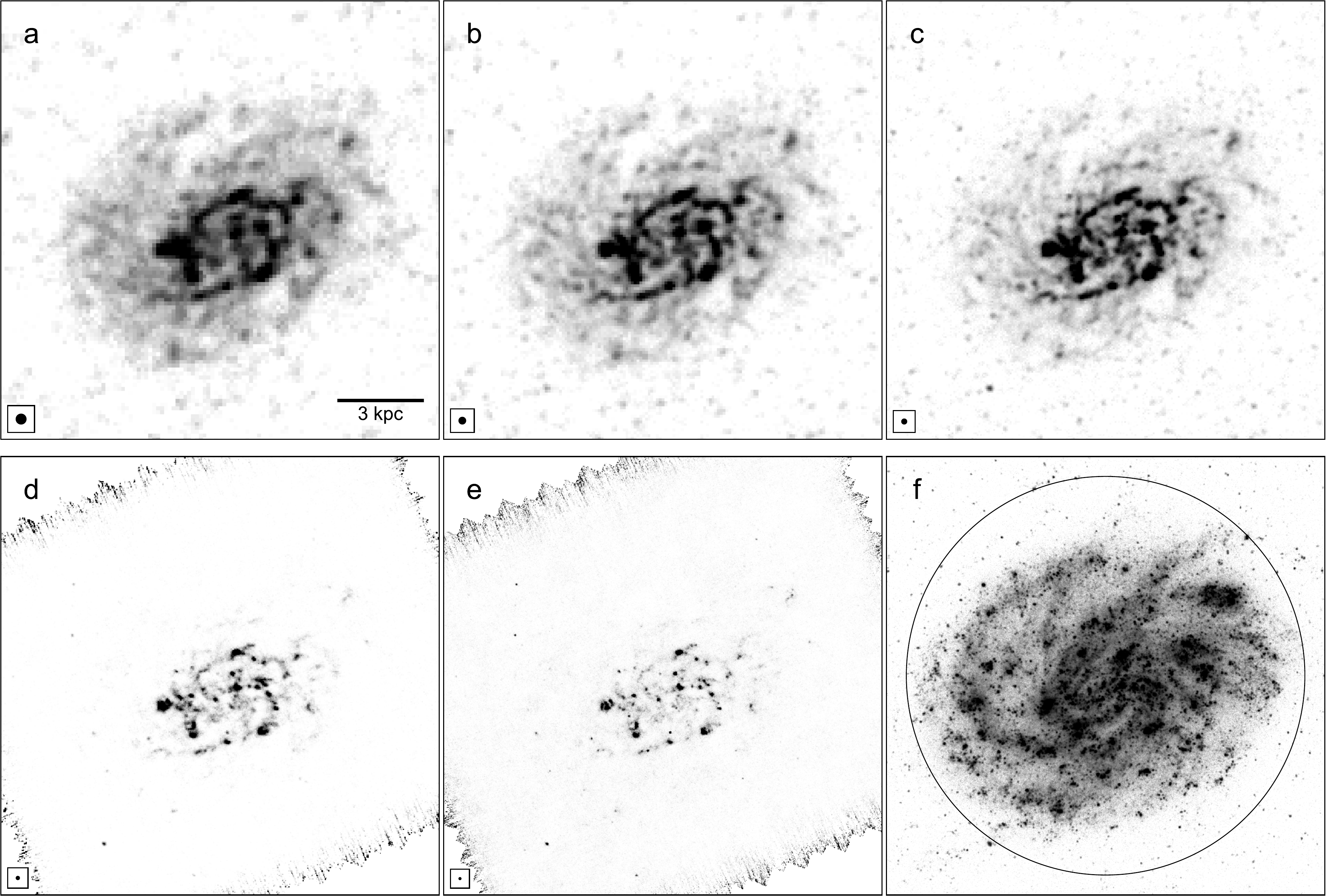}
\caption{Herschel maps and \textit{GALEX} FUV image of NGC~300. 
a) SPIRE-500, b) SPIRE-350, c) SPIRE-250, d) PACS-160, e) PACS-100, and f) \textit{GALEX} FUV. 
The beam sizes of the \textit{Herschel} images are indicated in the boxes on the lower left of the images. The black circle in the \textit{GALEX} FUV image indicates our chosen region of interest, inside which we searched for sources.}
\label{fig:herschel}
\end{figure*}

We obtained photometric observations of NGC~300 with the \textit{Herschel Space Observatory} (PI: Jan Forbrich) with two of its instruments, the Spectral and Photometric Imaging Receiver (SPIRE; \citealp{Griffin2010}) and the Photodetector Array Camera and Spectrometer (PACS; \citealp{Poglitsch2010}). 

The SPIRE observations took place on 2012 May 11 with a total exposure time of 4558 seconds; we obtained these \textit{Large map} observations in the \textit{Cross scan} pointing mode.
The three-band imaging photometer of SPIRE carried out broadband photometry in three spectral bands centred at approximately 250, 350, and 500~$\mu$m.
We refer to these images as SPIRE-250, SPIRE-350, and SPIRE-500, respectively.
In our further analysis, we used the absolute calibrated extended emission maps ("extdPxW"), for which the absolute flux offset of the maps was derived from \textit{Planck} all-sky maps%
\footnote{\label{spire-drg}\url{http://herschel.esac.esa.int/hcss-doc-13.0/load/spire_drg/html/ch06s10.html}}%
, as the flux estimates derived from these maps should be more reliable than the point source maps for our extended sources.

The two PACS observations were carried out on 2012 June 25 with a total exposure time of 3245 and 3803 seconds; the observation mode was \textit{Scan map} (\textit{largeScan} source mapping) with a \textit{Line scan} pointing mode and a medium scan map rate. 
For the PACS observations we chose the blue channel with the 85-125~$\mu$m filter in addition to the red channel with the 125-210~$\mu$m filter, corresponding to reference wavelengths of 100 and 160~$\mu$m.
We refer to these two images as PACS-100 and PACS-160, respectively.
For the PACS observations, we used maps generated with JScanam (HPPJSMAP), as these should be well-suited for our source detection purposes in NGC~300 (Herschel Science Center Helpdesk, personal communication).
All images were downloaded from the \textit{Herschel Science Archive} and processed through its automated pipeline (Standard Product Generation or SPG); the generator for the data products is \textit{SPG v13.0.0}.

Figure \ref{fig:herschel} shows all five \textit{Herschel} maps of NGC~300 and Table \ref{tbl:herschel} lists their most important parameters, including values for the FWHM spatial resolution in arcseconds  converted into physical units at the distance of NGC~300.

\begin{table}
\caption{Information on the \textit{Herschel} bands.\protect\footnotemark}
\centering
\small
\renewcommand{\arraystretch}{1.2}
\begin{tabular}{cccccc}
        \hline\hline
        Filter & Pixel size & Flux units & FWHM\tablefootmark{a} & FWHM\tablefootmark{b} \\
         & [arcsec] &  & [arcsec] & [pc] \\
        \hline
        PACS-100 & 1.6 & Jy/pixel & 7.0 & 67 \\ 
        PACS-160 & 3.2 & Jy/pixel & 11.5 & 110 \\
        SPIRE-250 & 6 & MJy/sr\tablefootmark{c} & 17.6 & 169 \\
        SPIRE-350 & 10 & MJy/sr\tablefootmark{c} & 23.9 & 229 \\
        SPIRE-500 & 14 & MJy/sr\tablefootmark{c} & 35.2 & 338\\
        \hline
\end{tabular}
\tablefoot{ 
\footnotesize 
\tablefoottext{a}{Adopted values after consultations with the \textit{Herschel} Science Center Helpdesk.}\\ 
\tablefoottext{b}{Estimated for a distance to NGC~300 of 1.98 Mpc.}\\
\tablefoottext{c}{Calibrated with \textit{Planck} (see Section \ref{sec:obs-herschel}).}
}
\label{tbl:herschel}
\end{table}

\subsection{Catalogue of HII~regions}
\label{sec:Hii}

\footnotetext{\url{http://herschel.esac.esa.int/twiki/pub/Public/%
DataProcessingWorkshop2015/Herschel_SPGV13.pdf}}

The most complete catalogue of HII~regions of NGC~300 up to the date of writing is from \citet[][hereafter D88]{Deharveng1988}. 
It is primarily based on an H$\alpha$ photographic plate taken by the ESO 3.6 m telescope with an exposure time of 90 minutes and further spectroscopic identifications for the smallest HII~regions \citepalias{Deharveng1988}.

We assessed the quality of this catalogue by overplotting it on a newer and more sensitive extinction corrected line-only H$\alpha$ map based on the ESO/WFI observations from \citetalias{Faesi2014} (see their $\S~2.2.$ for a detailed description on how this image was processed).
We found that the positions of the whole \citetalias{Deharveng1988} HII~region catalogue needed a linear shift to match the emission features from the \citetalias{Faesi2014} H$\alpha$ map. 
We excluded six HII~regions from our further analysis because we found no H$\alpha$ emission at their positions in the newer \citetalias{Faesi2014} H$\alpha$ map.
For 12 HII~regions we needed to make individual slight corrections to the \citetalias{Deharveng1988} position to make them match with HII~regions in the new \citetalias{Faesi2014} H$\alpha$ map. See Appendix \ref{cha:appendix_a} for a more detailed discussion on how we corrected the \citetalias{Deharveng1988} HII~region catalogue. 

We adopted the flux and size values of the HII~regions as given in \citetalias{Deharveng1988} without applying any corrections to their values; HII~regions for which no size estimate was given in \citetalias{Deharveng1988} are treated as point sources in our further discussion.
We also made no attempt at extending the catalogue with any previously uncatalogued HII~regions visible in the H$\alpha$ map of \citetalias{Faesi2014}. 

\subsection{Other ancillary data}
\label{sec:mpg-wfi}

Figure \ref{fig:herschel} also includes an image in the far ultraviolet (FUV) at an effective wavelength of 151.6~nm obtained by the \textit{Galaxy Evolution Explorer} (\textit{GALEX}). 
The \textit{GALEX} image is included in this comparison to show that the extent of the (visible) stellar disk is very well defined in the FUV. 
We therefore used this image for the definition of our region of interest (ROI), inside which we searched for GDC source candidates.

The European Southern Observatory (ESO) published an optical image of NGC~300, which we adopted for a qualitative comparison throughout this work. 
This image is a composite of individual observations with the Wide Field Imager (WFI) instrument of the MPG/ESO-2.2 m telescope at La Silla, Chile. 
The field of view of the image is $30.22\times 30.22$ arcminutes and the total exposure time amounted to around 50 hours.%
\footnote{\url{http://www.eso.org/public/images/eso1037a/}}
We downloaded the image as fullsize original file (TIFF) from the ESO homepage, after which we converted it into RGB FITS files with matching astrometry.

For the source selection we also used archival observations of NGC~300 in the optical by the Hubble Space Telescope.
We downloaded calibrated but unprocessed Level 0 images from the Mikulski Archive for Space Telescopes (MAST) and used the DrizzlePac software package%
\footnote{\url{http://drizzlepac.stsci.edu/}} %
to create combined images of exposures with the same filter, the same camera, and within the same visit. 
We corrected geometric distortions of the ACS/Wide Field Camera (WFC) images using the newest reference files%
\footnote{\url{http://www.stsci.edu/hst/acs/analysis/distortion}} %
and combined some of the images into mosaics using Montage (v3.3)%
\footnote{\url{http://montage.ipac.caltech.edu/}}. %
We created (false colour) RGB images if there were at least two exposures with different filters available for a region.  
In Appendix \ref{app:hubble} we present more details on the Hubble images we used.

\section{Source extraction}
\label{cha:sourceextraction}

\subsection{Algorithm}
\label{subsec:algo}

We used the multi-scale, multiwavelength source extraction algorithm \textit{getsources}%
\footnote{\label{getsources}
\url{http://www.herschel.fr/cea/gouldbelt/en/getsources/}}
 for the compilation of the source catalogues of NGC~300. 
We chose this algorithm because it was specifically created for the purpose of source extraction in \textit{Herschel} images.

The \textit{getsources} algorithm is primarily intended for FIR surveys of Galactic star-forming regions with \textit{Herschel}, in particular the detection of (point-like) protostars \citep{Men'shchikov2012}. 
However, it is also ideally suited for our purpose of detecting GMCs in nearby galaxies as most of these objects also exhibit a point-like structure due to our inability to resolve these sources in more detail, given the spatial resolution. 

The \textit{getsources} algorithm allows one to define which images should be used for the detection and on which images the combined multiwavelength extraction and measurements for the final catalogue should be performed. 
For example, it is possible to use all images for the detection and subsequently make multiwavelength extractions for various selections of the bands.

The algorithm also incorporates the multi-scale, multiwavelength filament extraction method \textit{getfilaments}, which detects and subtracts filamentary structures from the original images prior to the detection and measurement steps \citep{Men'shchikov2013}.
In our case, these filamentary structures correspond to diffuse extended emission throughout the stellar disk, which is particularly relevant for the SPIRE images (250-500~$\mu$m).

\subsection{Settings}
\label{subsec:settings}
 
For the source extraction with \textit{getsources}, we defined a circular ROI with a radius of $12'$ , centred at RA~=~$00^{h}~54^{m}~56.4432^{s}$ and DEC~=~$-37^{\circ}~40'~23.0016",$ inside which we searched for sources.
We based the definition of the ROI on the \textit{GALEX} FUV image, since the largest extent of the (visible) stellar disk is best seen at this wavelength (see Figure \ref{fig:herschel}). 
We wanted to incorporate the whole visible stellar disk while including only a minimum of background to minimize false detections due to background sources. 
Moreover, an upper limit for the size of the ROI was set by the PACS images whose field of view was smaller than for the SPIRE maps.

For the source detection we resampled all images to the resolution of the PACS~100 image ($1.6"/$pixel). 
We did not shift the resampled images in position as their relative position to each other was already satisfactory. 
We chose the upper limit in spatial scale for the image decomposition step as 1.5~times the FWHM of SPIRE-500, which corresponds to a physical scale of about 500~pc. 
Visual inspections confirmed that this value is sufficient to include even the biggest GDC structures seen in the SPIRE-500 image. 

The \textit{getsources} algorithm uses a global goodness (GOOD) value to classify the source candidates into reliable and tentative detections. 
This GOOD value incorporates a combination of the global detection significance over all wavelengths and the global S/N ratio. 
Sources classified as tentative are indicated by a negative GOOD value and are automatically discarded by \textit{getsources} for its final catalogue. 
However, we decided to include these tentative sources in our list of source candidates to check whether some of these were actually false negatives.

We performed two different multiwavelength extractions: one in which all five \textit{Herschel} bands were used (yielding 200 source candidates) and one that used only the three SPIRE images (yielding 156 source candidates). 
The latter approach should also be able to detect pre-star-forming GDCs that do not show significant emission at 100 and 160~$\mu$m.
For the most part, the two source extractions produced nearly identical results.
However, in the more crowded central region the extraction using all five \textit{Herschel} bands was able to decompose the extended objects into more individual regions. 
The inclusion of the PACS images also led to a decreased size estimate for the source candidates, as the extent of the sources could be better constrained.
Therefore, we decided to adopt the results of the source extraction performed for all \textit{Herschel} bands (200 source candidates), but added source candidates from the extraction limited to the SPIRE bands that did not show up in the SPIRE plus PACS run (six source candidates). 
We refer to this combined catalogue of 206 source candidates further on as the merged semifinal \textit{getsources} catalogue.
This catalogue includes 44 source candidates that \textit{getsources} labelled as tentative and would have discarded for its final catalogue. 

To verify the performance of \textit{getsources}, we compared its results to the source extraction with two other point source detection algorithms (\textit{SUSSEXtractor} and \textit{DAOPHOT}, both implemented in \textit{HIPE}, the Herschel Interactive Processing Environment). 
We found the results to be in good agreement regarding the detection of point sources (\textit{getsources} covers $\sim$ 80\% of the \textit{HIPE} point source candidates).
However, \textit{getsources} was superior in the overall detection of extended structures.

\section{Source catalogue}
\label{cha:sourcecat}

\subsection{Selection of the sources}
\label{sec:selection}

To analyse the merged semifinal \textit{getsources} catalogue in more detail, we created thumbnails to perform a visual inspection of each source to assess its significance. 
We overplotted the (wavelength-dependent) FWHM shapes of the 206 source candidates as determined by \textit{getsources} on the corresponding \textit{Herschel} maps. 
This allowed us to determine at which wavelengths the source candidates had a well-defined shape, that is whether they showed significant emission at that wavelength.
We adopted the shape of the FWHM at SPIRE-250 as a size estimate for the source candidates and additionally overplotted this shape on the \textit{GALEX} FUV, ESO/WFI, and Hubble images. 

We checked the ESO/WFI and Hubble images in particular for hallmark features of GDCs such as dark dust lanes seen in absorption or HII regions, and unmistakable signs of a false detection, such as a well-confined disky or elliptical structure indicating a background galaxy. 
This was especially significant for source candidates located between spiral arm features or positioned on the outskirts far away from the visible stellar disk.
Although we expect the majority of GDCs to be positioned inside a spiral arm feature, this was not a prerequisite for the final selection. 
\citet{Foyle2010}, for example, concluded in their analysis of three spiral galaxies that even in grand-design spirals star formation in interarm regions is significant. 

\begin{figure}
\centering
\includegraphics[width=\columnwidth]{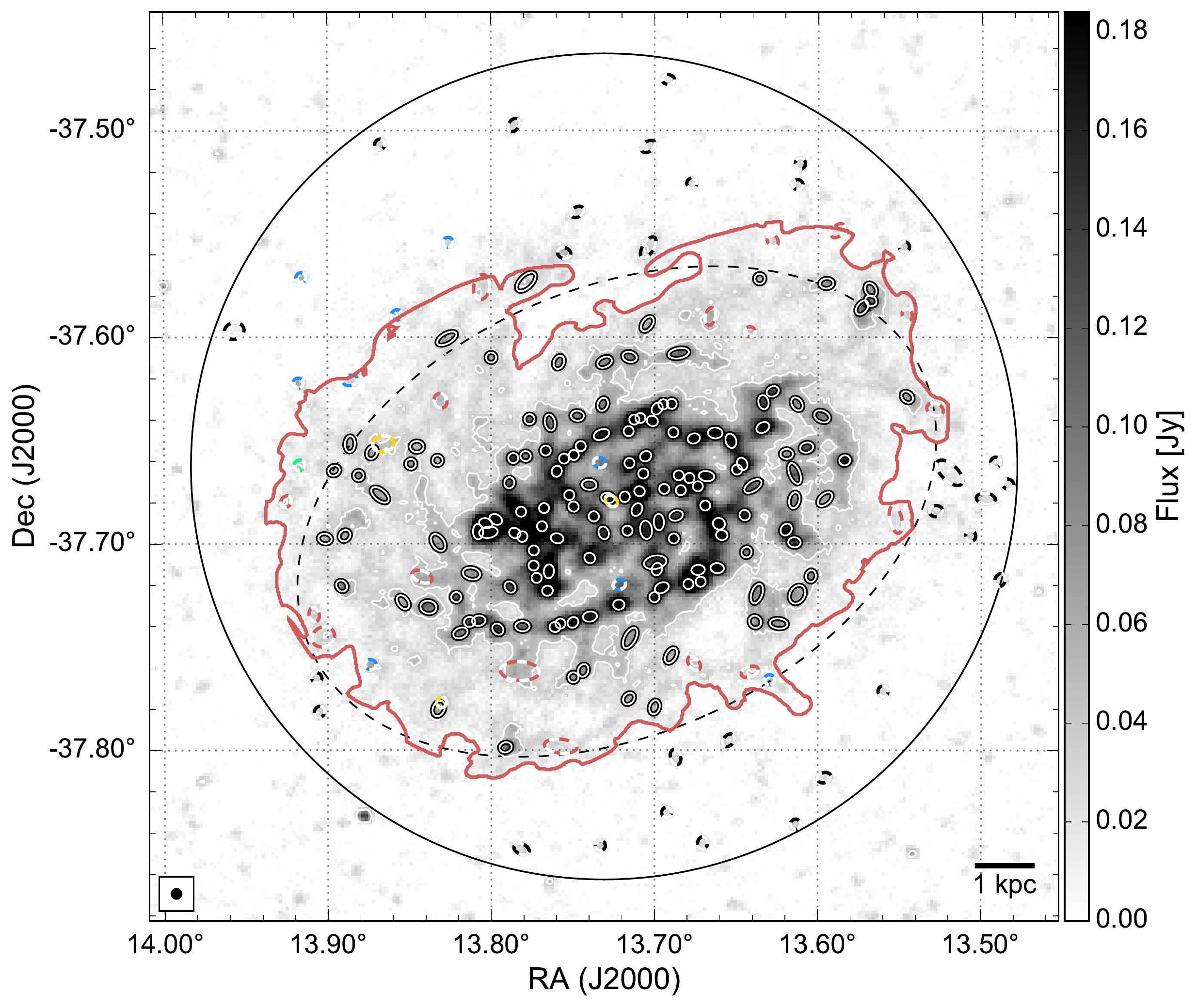}
\caption{Our catalogue of 206 source candidates overplotted on the SPIRE-250 image.
The small ellipses denote the SPIRE-250 FWHM of the 206 source candidates; the small dashed ellipses further indicate the 60 source candidates that we ultimately discarded (see text for details).
The white contour lines trace flux values of 0.04~Jy.
The large dashed black ellipse denotes the isophotal radius R$_{25}$ as defined in Table \ref{tbl:ngc300}.
The red contour indicates the area from which \textit{getsources} subtracted diffuse extended emission in addition to any background subtraction. 
The large black circle indicates our chosen ROI, inside which we searched for sources.}
\label{fig:src_can}
\end{figure}

Our most important criterion for the validation of a source was the existence of significant flux values in the \textit{Herschel} bands.
We required a S/N ratio $>$ 3 in at least two \textit{Herschel} bands for a source candidate to be considered as a valid source.
We especially checked the SPIRE bands for convincing emission features, as we expect the thermal emission of the cold dust of the GDCs to be most prominent at these wavelengths. 
In contrast, the existence of significant flux values in the PACS bands was not a necessary selection criterion, as some of the colder GDCs with only little or no ongoing star formation are not expected to show up brightly at 100 and 160~$\mu$m.
In particular in the central regions the increased density of the older stellar population contributes increasingly to the heating of the cold dust in the ISM of NGC~300 and thus also leads to higher values of diffuse emission in the SPIRE images.
We took into account that smaller GDCs located in a spiral arm near the central region might not stand out clearly from the surrounding high diffuse background emission.
Conversely, source candidates located near the edge of or beyond the visible stellar disk that show only a small contrast to their surroundings are much more likely to be background sources, since background objects begin to dominate the FIR emission outside the visible stellar disk, where there is only very little or no contribution from the galaxy itself (see Figure~\ref{fig:herschel}). 
Figure \ref{fig:src_can} shows the outermost contour of the diffuse extended emission of NGC~300 that was modelled by \textit{getfilaments}.
This structure traces the visible stellar disk very well, which is why one of our selection criteria for the GDC catalogue was that the source candidates had to be located inside this contour. 

If a source candidate was labelled as tentative by \textit{getsources} we required convincing evidence for its inclusion in the GDC catalogue.
We adopted the contrary approach for source candidates labelled as reliable: if there was no convincing evidence for a false or spurious detection or other problems such as blending effects, we kept the sources in our selection.

In total, we discarded 60 source candidates from the merged semifinal \textit{getsources} catalogue (marked in Figure \ref{fig:src_can} with dashed ellipses) for the following reasons:
\begin{itemize}
\item 
Source candidates identified as background galaxies (blue dashed ellipses). 
We visually identified nine source candidates as background galaxies, five of which are located inside the diffuse extended structure determined by \textit{getsources}. 
\item 
Source candidates blended with other sources (yellow dashed ellipses).
This applied to three source candidates, two of which were strongly blended. 
We added the flux of the two strongly blended discarded source candidates to their blended counterparts that we kept in the final catalogue.
\item 
Tentative source candidates positioned inside the diffuse extended structure determined by \textit{getfilaments}, but very likely spurious detections or background sources (red dashed ellipses).
This applied to 17 source candidates that were classified as tentative by \textit{getsources}; they have only weak fluxes in the \textit{Herschel} bands and no convincing GDC features (such as dark dust lanes, young massive star clusters, or HII~regions) were visible in the optical images.
Most of these source candidates were positioned near the outer edge of the fitted diffuse extended structure, where the likelihood of detecting background sources was higher.
\item 
Source candidates that did not fulfill our criterion of a positive detection (S/N $>$ 3) in at least two \textit{Herschel} bands (green dashed ellipses). 
This applied to one source candidate from the SPIRE-only source detection run with \textit{getsources,} which only had one positive detection in the SPIRE-250 band.
\item 
Source candidates positioned outside the diffuse extended structure determined by \textit{getfilaments}, which we adopted as a good tracer for the dust in the stellar disk (black dashed ellipses). 
This applied to 30 source candidates, of which 14 were labelled tentative by \textit{getsources}. 
Another two of these discarded sources are associated with the HII~regions \#1 and \#72 from \citetalias{Deharveng1988}.
Both of these regions were already targeted by \citetalias{Faesi2014} but do not show a detection in CO. 
These HII~regions also do not show up very brightly in the continuum-subtracted H$\alpha$ image from \citetalias{Faesi2014}.
\end{itemize}

The final catalogue automatically produced by \textit{getsources} would have consisted only of the 162 source candidates of the merged semifinal \textit{getsources} catalogue classified as reliable and would have neglected any sources labelled as tentative.
For our final catalogue of GDCs, we discarded 28 of those source candidates labelled as reliable, a third of which we could unambiguously identify as background galaxies.
However, we also chose to include 12 sources in our final catalogue that were classified as tentative by \textit{getsources} because they show convincing emission features in all SPIRE bands; five of these tentative sources also show associations with HII~regions from \citetalias{Deharveng1988}, which we took as additional evidence for their validity.
Based on these results we conclude that the catalogue automatically produced by \textit{getsources} would have had a false positive rate of 17.3\%, and 27.3\% of the objects neglected by \textit{getsources} as tentative sources would have been false negatives.

In total, we retained \selSrcAll ~of the 206 source candidates for our final catalogue of GDCs (see Figure \ref{fig:src_can}).
Table \ref{Tab:gdcs} in the Appendix lists the RA and DEC positions and the flux values for all five \textit{Herschel} bands as determined by \textit{getsources} for the \selSrcAll ~sources in the final catalogue. 
The flux values are the measurements of the total (background corrected) fluxes of the sources, for which the contributing flux inside the footprint of each source was summed up.
The footprint was determined by \textit{getsources} as the area containing non-negligible flux of a source and has a size that is about two times as large as its FWHM, although this can vary significantly depending on the source. 
In particular in the crowded central regions the footprints of the sources often overlap considerably, but the deblending algorithm contained in \textit{getsources} should be able to attribute correctly  the flux contribution to the respective sources to a very high degree. 
If the ratio of flux-to-flux error for a source at a certain wavelength was $<3.0$, we assumed that the total flux was not measurable at that wavelength. 
In such cases, we only give an upper limit of the flux as three times the estimate of the total flux error (indicated by '<' in front of the value). 
This affected mostly SPIRE-500 flux estimates. 

\begin{figure}
\centering
\includegraphics[width=\columnwidth]{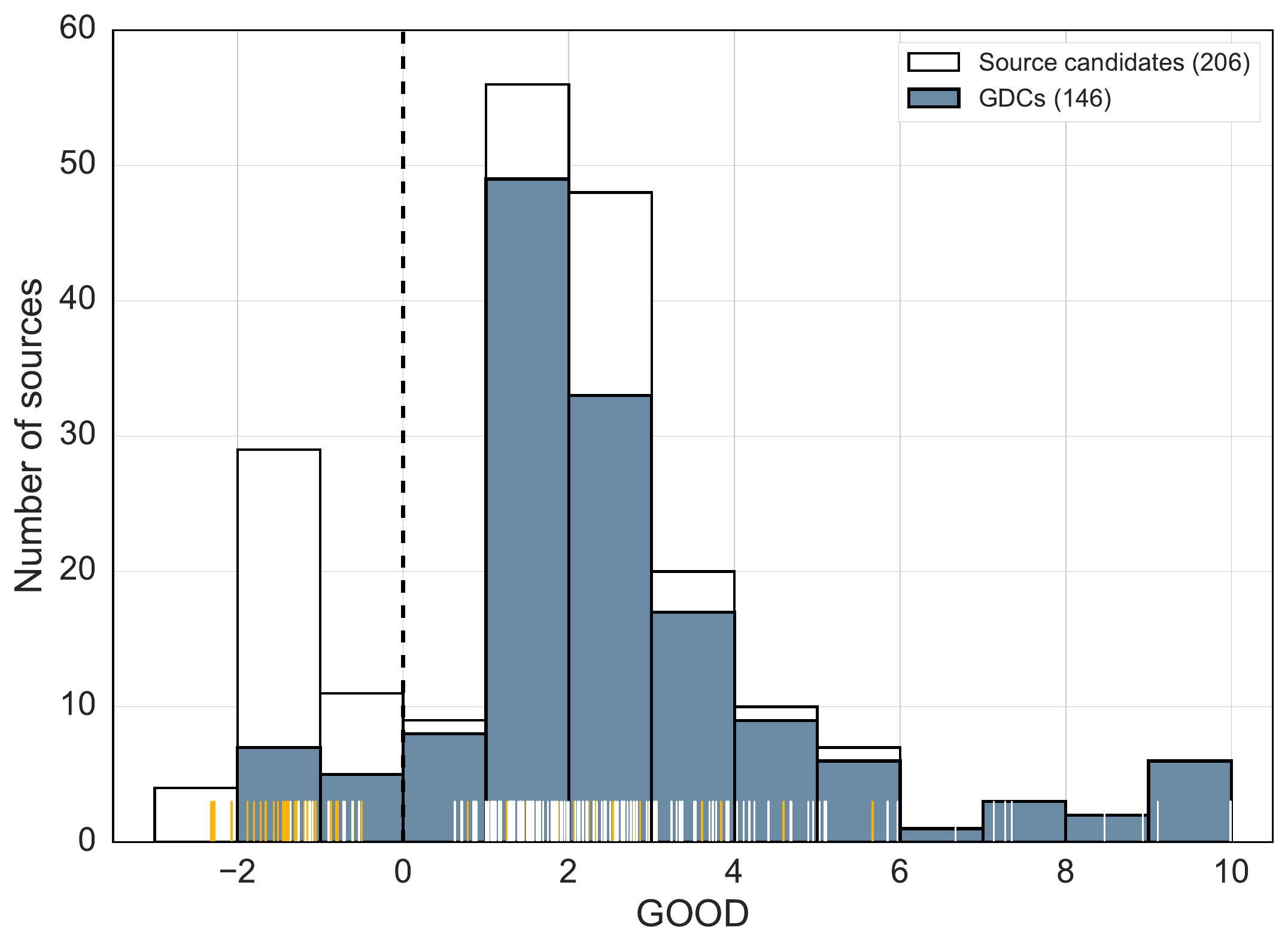}
\caption{Histogram showing the \textit{getsources} quality assessment ("GOOD") of the GDCs. 
The unfilled histogram includes all 206 source candidates from the merged semifinal \textit{getsources} catalogue; individual values of the source candidates are denoted by the ticks in the rug plot. 
The filled histogram shows the subsample of \selSrcAll ~GDCs selected for the final catalogue; individual values of these GDCs are indicated in white in the rug plot. 
The dashed vertical line denotes the threshold between reliable and tentative sources.}
\label{fig:hist_good}
\end{figure}

Figure \ref{fig:hist_good} shows a histogram of the quality assessment by \textit{getsources} for all source candidates of the merged semifinal \textit{getsources} catalogue and the subsample of sources selected for the final catalogue of GDCs. 
The majority of the source candidates and most of those that were ultimately discarded have a "GOOD" value below 3 (negative values just indicate that \textit{getsources} labelled them as tentative). 
According to additional information contained in the catalogues produced by \textit{getsources}, the user is advised to use only source candidates with "GOOD" values larger than one to three. 
As can be seen in Figure \ref{fig:hist_good}, this would have excluded most of our source candidates. 
However, since \textit{getsources} is intended mainly for source detection in Galactic star-forming regions, we argue that its application in an extragalactic context is expected to result in lower quality values due to the much poorer absolute spatial resolution and the detection of completely different physical objects (entire complexes of different star-forming regions instead of single protostars). 
In many cases, the modelling of the diffuse extended structure by \textit{getfilaments} could also have reduced the total flux values by an amount that is too much and would thus have had adverse consequences for the quality assessment.
The flux values of central sources embedded deep inside the fitted diffuse extended structure were reduced by a high amount of assumed (diffuse) background emission that could be more than twice as high as for sources located near the edge of the visible stellar disk, where the amount of flux allocated to diffuse extended emission drops to almost zero.

From the visual inspection of all source candidates in the ancillary data, we can almost fully rule out contamination of the catalogue by resolved background galaxies, as these would have been already identified in the ESO/WFI images and, where available, the Hubble images. 
In one problematic case (GDC \#24) a resolved background galaxy is situated close to a GDC, which shifted the central position of the GDC about 12" away from the centre of the dust emission peak.

\begin{figure}
\centering
\includegraphics[width=\columnwidth]{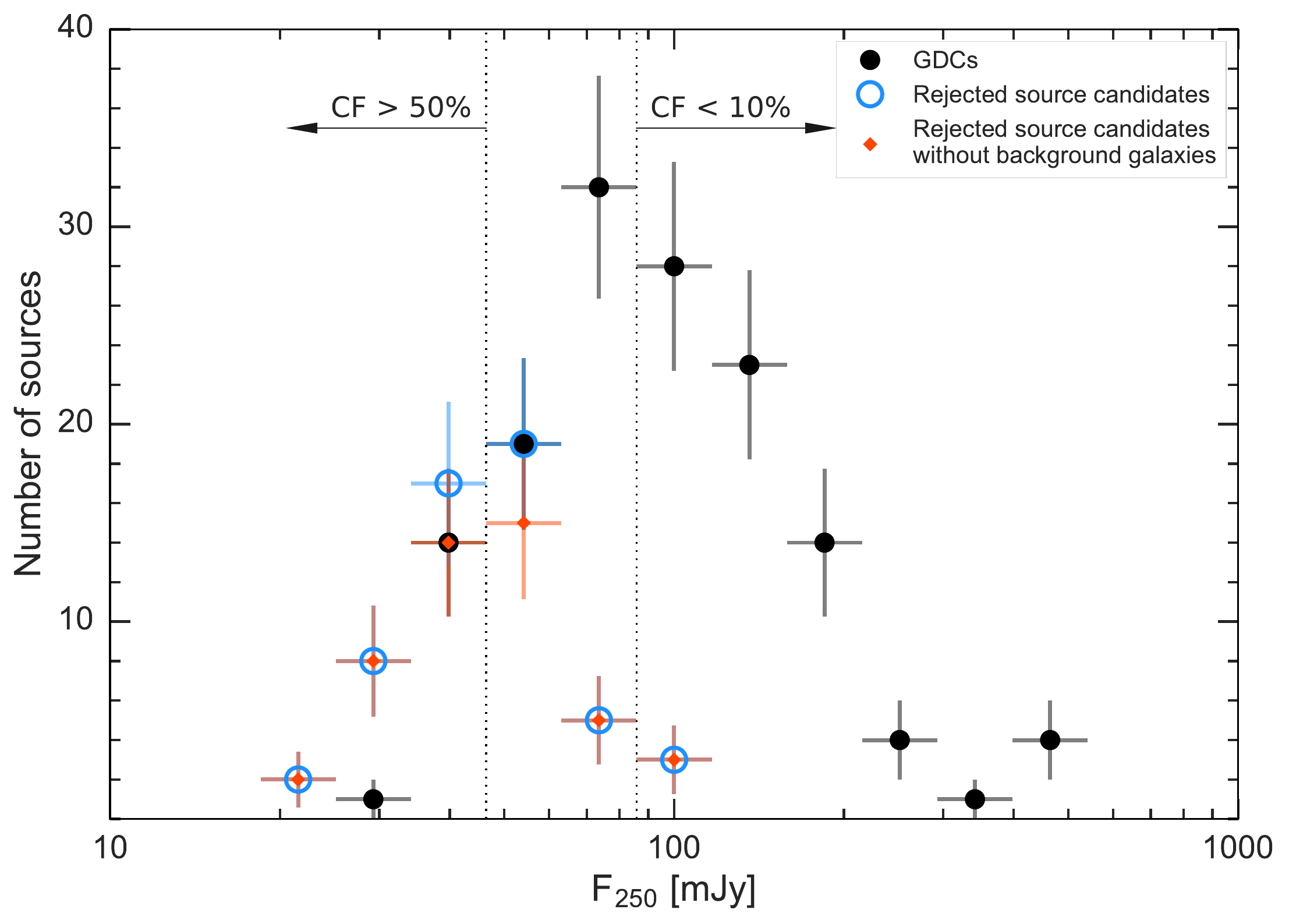}
\caption{Histogram of the \textit{getsources} SPIRE-250 total flux estimates for the GDCs in our final catalogue (black dots) and the discarded sources from the merged semifinal \textit{getsources} catalogue (blue open circles). The red diamonds show the discarded sources not including the nine resolved background galaxies.
The dotted vertical lines indicate thresholds for the contamination fraction (CF) of over 50\% (below 46~mJy) and less than 10\% (above 86~mJy).
The bin size is 0.125~dex, which corresponds to the median uncertainty of the total flux values.}
\label{fig:flux_limit}
\end{figure}

In Figure \ref{fig:flux_limit} we plot a histogram of the \textit{getsources} total SPIRE-250 flux estimates for the final GDC catalogue and for the rejected source candidates. 
We did not include sources with uncertain SPIRE-250 flux estimates in this plot, i.e. sources for which the total flux value did not exceed three times the flux uncertainty (this concerned six of the selected GDCs and six of the discarded source candidates).
The bin size of 0.125~dex corresponds to the median total flux error of the sources. 
We estimated the contamination fraction (CF) per flux bin as $N_{discarded}/(N_{real}+N_{discarded})$, i.e. the ratio of discarded sources to the total amount of source candidates (real and discarded) in the flux bin.
Our contamination fraction drops to below 10\% above a total flux value of 86~mJy and we therefore adopted this as our de facto sensitivity limit, beyond which we essentially detect only GDCs. 
Below this flux limit the number of rejected sources increases and reaches a peak around a total flux value of 46~mJy, below which our source candidates are dominated by background sources or spurious detections. 
All of the seven sources that we could resolve as background galaxies and that had reliable SPIRE-250 flux estimates are located around the 50\% CF threshold. 

\subsection{Deprojected galactocentric distances}
\label{sec:deprojdist}

\begin{figure}
\centering
\includegraphics[width=\columnwidth]{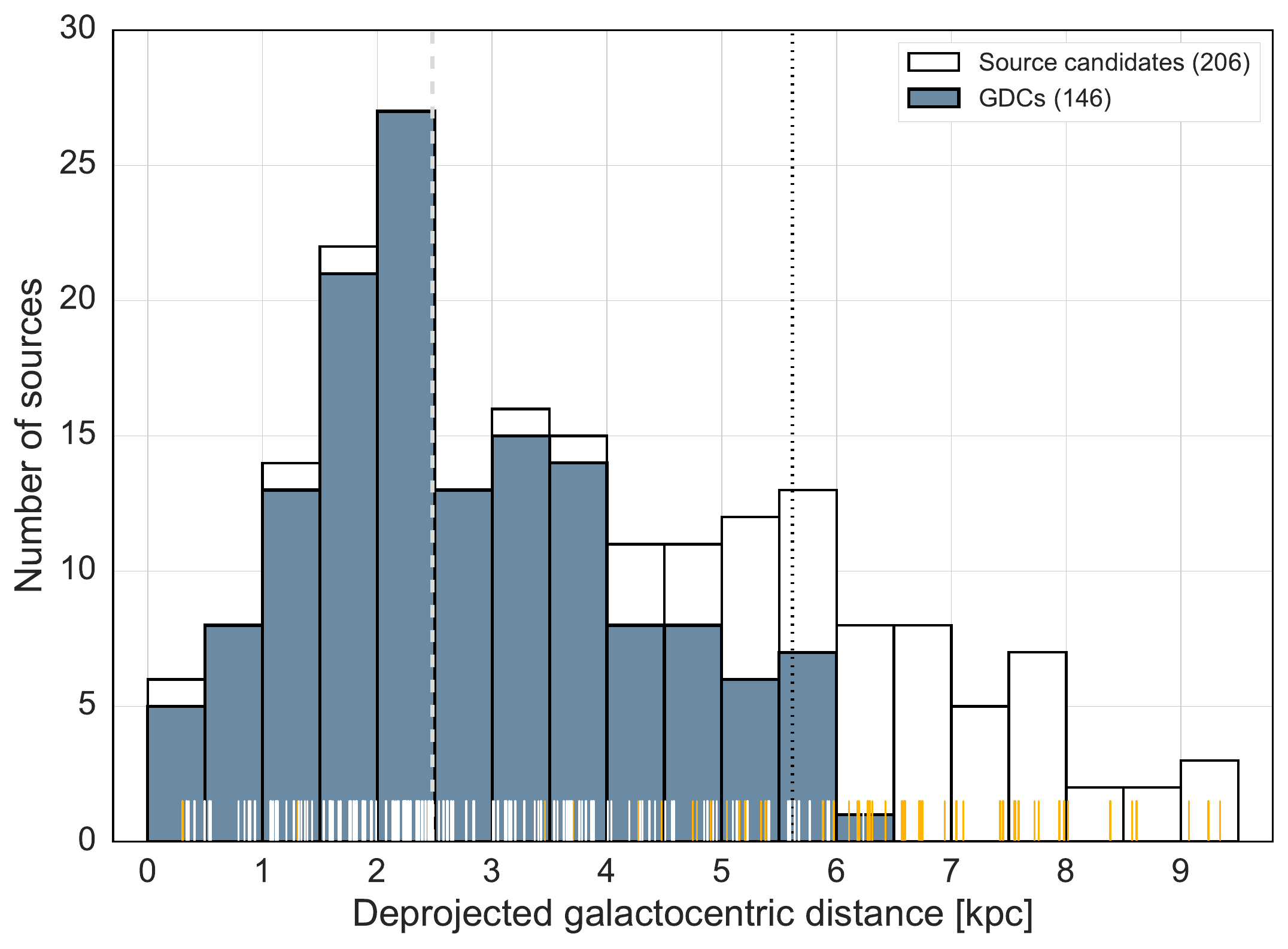}
\caption{Histogram of the deprojected galactocentric distances of the GDCs. 
The unfilled histogram includes all sources from the merged semifinal \textit{getsources} catalogue; individual values of these sources are indicated by the ticks in the rug plot. 
The filled histogram shows the subsample of GDCs selected for the final catalogue; individual values of these GDCs are indicated in white in the rug plot. 
The vertical lines indicate the R$_{25}$ distance of NGC~300 (dotted) and the median deprojected galactocentric distance of the selected GDCs (dashed).}
\label{fig:hist_Dist}
\end{figure}

We used a Python-script\footnote{\url{https://gist.github.com/jonathansick/9399842}} for the computation of the deprojected galactocentric distances of the GDCs. 
The values taken for the distance, position angle of the major axis, and inclination angle between the line of sight and polar axis are listed in Table \ref{tbl:ngc300}. 
Figure \ref{fig:hist_Dist} shows a histogram of the deprojected galactocentric distances of all source candidates detected by \textit{getsources} and the subsample of the final catalogue of GDCs. 
This shows that the sources we discarded were mostly situated outside the B-band 25$^{\text{th}}$ magnitude isophotal radius R$_{25}$. 

\subsection{Effective dust temperatures}
\label{sec:temp}

\begin{figure}
\centering
\includegraphics[width=\columnwidth]{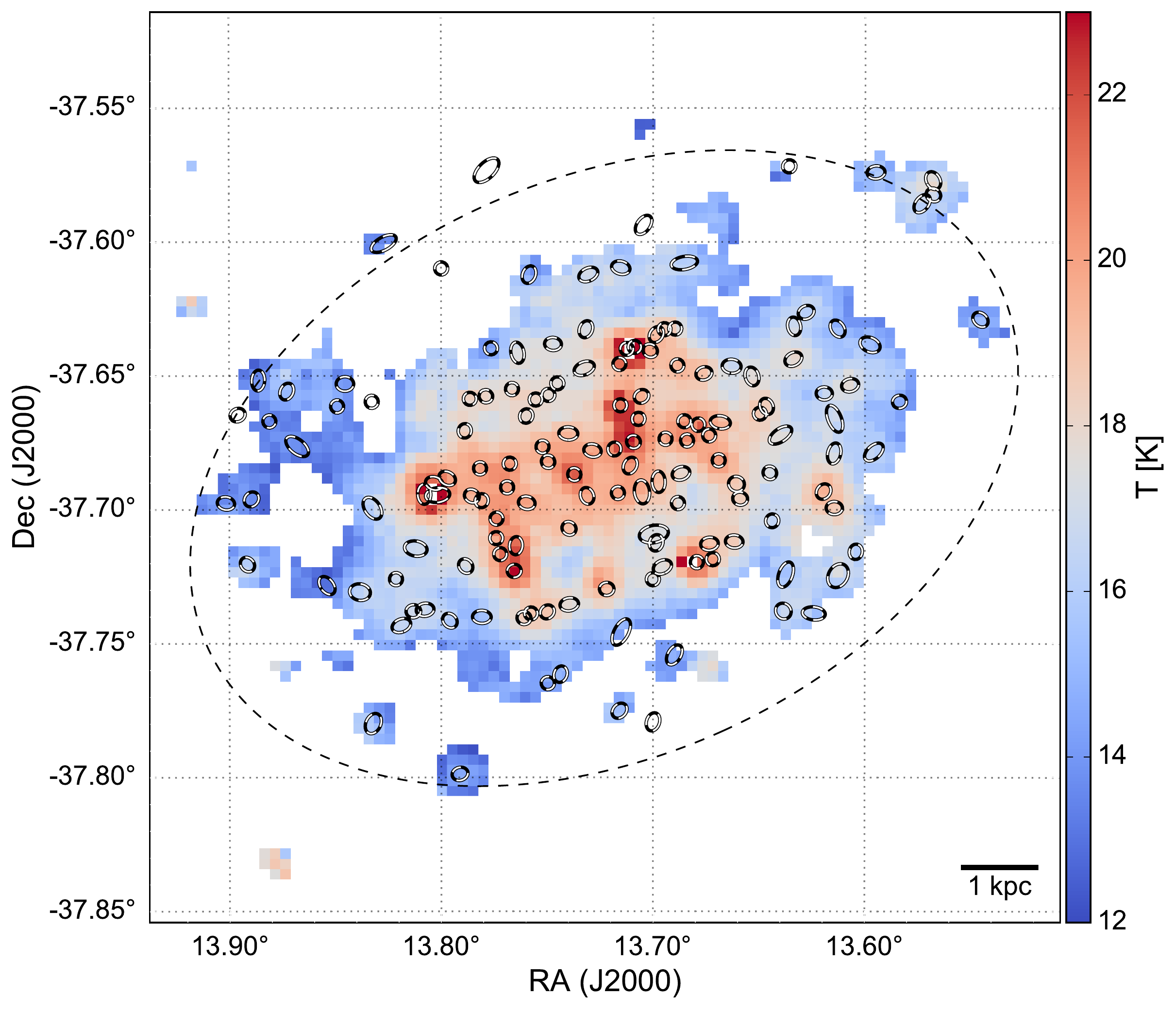}
\caption{Cold dust effective temperature map of NGC~300 overplotted with the final catalogue of \selSrcAll ~GDCs (small black dashed ellipses).
The large dashed black ellipse denotes the isophotal radius R$_{25}$ as defined in Table \ref{tbl:ngc300}.} 
\label{fig:temp_map}
\end{figure}

One common approach to estimate the dust temperature is to fit a single modified black body to the SED at FIR wavelengths (e.g. \citealp{Remy2013}; \citealp{Kirk2015}), which however is likely not the best method to characterize the physical causes responsible for the dust heating. 
\citet{Bendo2012} studied three nearby spiral galaxies and found that the evolved stellar population of a spiral galaxy is likely the dominant heating source for dust observed at wavelengths longer than 160~$\mu$m. These authors further concluded that dust models need to include two thermal components: one component whose SED peaks below 160~$\mu$m for the dust heated by star-forming regions and regions in which stochastically heated grain emission dominates (i.e. elevated 24-70~$\mu$ emission; e.g. in the Magellanic Clouds; \citealp{Bernard2008}), and one component with an SED that peaks at longer wavelengths for the cooler dust that is relatively unaffected by star formation activity and is heated mainly by the evolved stellar population. 

For NGC~300 we derived effective temperature maps using two-component black-body fits to seven photometric bands between 24 and 500~$\mu$m; \textit{Spitzer Space Telescope}’s
Multiband Imaging Photometer for \textit{Spitzer} (MIPS; \citealp{Rieke2004}) at 24 and 70~$\mu$m, PACS-100/160, and SPIRE-250/350/500).
We convolved all the images to the lowest resolution, i.e. the pixel scale of the SPIRE-500 image (see \S 2.4 of \citealp[][and references therein]{Galametz2012} for more information on the convolution). 
We used convolution kernels from \citet{Gordon2008} and performed the convolutions using the \textit{conv$\_$image} script%
\footnote{\url{http://dirty.as.arizona.edu/~kgordon/mips/conv_psfs/conv_psfs.html}}.
We used the resulting maps to construct the 24-500~$\mu$m SED for each individual pixel.
We then fit each pixel's SED with a two-component modified black body using the \textit{MPFIT} package%
\footnote{\url{http://cow.physics.wisc.edu/~craigm/idl/mpfittut.html}}.
This algorithm performs a Levenberg-Marquardt least-squares minimization to the data points for the following assumed two-component model (see also Eq. 3 of \citealp{Galametz2012}): 

\begin{equation}
F_{\nu} = A_{W} \left[ \nu^{~\beta_{W}} B_{\nu}\left( T_{W} \right) \right]  + A_{C} \left[ \nu^{~\beta_{C}} B_{\nu}\left( T_{C} \right) \right] ,
\label{eqn:fit}
\end{equation} 

\noindent where $B_{\nu}\left( T_{C} \right)$ and $B_{\nu}\left( T_{W} \right)$ are the Planck functions of the cold and warm dust effective temperatures, A$_{C}$ and A$_{W}$ are the cold and warm dust component amplitudes (incorporating optical depth and normalization frequency for $\beta$), and $\beta_{C}$ and $\beta_{W}$ are the power law exponents of the dust emissivities. 
Based on theoretical expectations \citep{Li2001} we decided to fix $\beta _{W}$ to a value of 2. 
We fixed $\beta _{C}$ to a value of 1.7, which was motivated by results from the \textit{Planck} mission that found values for the spectral index at FIR wavelengths of $1.78 \pm 0.08$ for local clouds \citep{PlanckXXV2011} and $1.62 \pm 0.10$ over the whole sky \citep{PlanckXI2013}. 
Our fit thus contains four free parameters: T$_{C}$, T$_{W}$, A$_{C}$, and A$_{W}$.
To produce reliable effective temperature estimates, we required that each pixel have intensity $\geq$ 3$\sigma$ in at least six of the seven bands.
 
Our comparisons show that the dust mass in our GDCs is completely dominated by the cold dust contribution, where the total mass of dust in the cold component is 100-1000 times higher than the total mass of dust in the warm component. 
For this reason we focus only on the cold dust effective temperature in our further discussion, as its thermal emission completely dominates the dust mass in our GDCs at the wavelength regime of 250-500~$\mu$m. 

Figure \ref{fig:temp_map} shows our cold dust effective temperature map for NGC~300.
We calculated mean cold dust effective temperatures for our GDCs by averaging across the temperature values inside the elliptical structure that we adopted for their size estimate.

A single fixed value of $\beta$ is likely not the best choice for all of our GDCs (Faesi et al., in prep.).
The emission and absorption properties of the dust grains depend on their composition (e.g. \citealp{Zubko2004}; \citealp{Demyk2017}) and the dust emissivity varies with the environment \citep[e.g.][]{Paradis2009}.
\citet{Tabatabaei2014}, for example, studied the variation of $\beta$ across M33 and found that it declines from close to 2 in the centre to about 1.3 in the outer disk; these authors conclude that this decrease is likely due to reduced metallicity. 
Our estimate for the uncertainty of the dust effective temperature values reflects this possible variation of $\beta$ throughout the disk of NGC~300. 
For the error estimate, we compared the cold dust effective temperature map derived for $\beta$ = 1.7 with cold dust effective temperature maps for $\beta$ values of 1.5 and 2. 
We calculated the difference in cold dust effective temperature for each pixel, which yielded a median value of $\pm$1.5~K that we adopt as our uncertainty for the cold dust effective temperature.

\subsection{Mass determination}

We estimated the dust masses of the GDCs, assuming that the dust is optically thin, i.e.
\begin{equation}
M_{d} = \dfrac{F_{\nu} D^{2}}{B_{\nu}\left( T_{d} \right) \kappa_{\nu}},
\label{eqn:dustmass}
\end{equation} 

\noindent where $F_{\nu}$ is the flux at frequency $\nu$, $B_{\nu}\left( T_{d} \right)$ is the Planck function at frequency $\nu$ for dust effective temperature $T_{d}$, $\kappa_{\nu}$ is the mass absorption coefficient that expresses the effective surface area for extinction per unit mass, and $D$ is the distance. 

We estimated $\kappa_{\nu}$ from computations of a model for interstellar dust consisting of a mixture of carbonaceous grains and amorphous silicate grains with R$_{V}=3.1$ \citep{Weingartner2001}.
We fitted their values of $\kappa_{\nu}$ from $245.471$~$\mu$m to $512.861$~$\mu$m in log-log space with the lowest order polynomial that yielded a good match%
\footnote{The polynomial we used was of the form y(x)~$=0.1131$~x$^{3}~-~0.08277$~x$^{2}~-~3.869$~x$~+~8.798$.}
to interpolate the values of $\kappa_{\nu}$ at 250, 350, and 500~$\mu$m as 4.023, 1.915, and 0.949~cm$^{2}$/g, respectively.

We based our mass determinations on the SPIRE-250 flux measurements by \textit{getsources} and calculated the uncertainty of the mass values by error propagation of the individual parameters.%
\footnote{For the error propagation calculations we used \textit{Uncertainties}, a Python package for calculations with uncertainties, developed by Eric O. Lebigot: 
\url{http://pythonhosted.org/uncertainties/}}
For the uncertainty of the distance, we adopted a value of $\pm$0.06~Mpc, which is the 1$\sigma$ error in fractional distance for NGC~300 from the Cosmicflows-2 catalogue \citep{Tully2013}. 
We used an error of $\pm 1.5$~K for the cold dust effective temperature (see previous section). 
Owing to the exponential term in the Planck function, this temperature uncertainty would translate to upper and lower bounds for $B_{\nu}\left( T_{d} \right)$ that differ by a factor of about 1.1 to 1.3. 
To simplify the error calculation, we took the bigger uncertainty value as its symmetric error, which leads to slight overestimates for the lower uncertainty limits of the dust mass values.
The flux uncertainties we used are given in Table \ref{Tab:gdcs}.
We calculated no masses for GDCs for which the flux uncertainty exceeded the flux values and for which we only list upper limits for the flux. 
We assume no uncertainty for the mass absorption coefficient, but caution that for a different assumption of the dust grain composition its value could be higher by a factor of about two (\citealp{Fanciullo2015}, \citealp{Demyk2017}).
Moreover, in its outskirts the metallicity of NGC~300 is similar to that of the Large Magellanic Cloud (LMC; \citealp{Kudritzki2008}, \citealp{Bresolin2009}), for which \citet{Galliano2011} showed that the standard grain composition (graphite and silicate) used for the Milky Way does not give physically realistic results. 
Therefore, we caution that our mass estimates may vary within a factor of two depending on the detailed and spatially varying dust composition in NGC~300.

We compared the GDC masses based on the SPIRE-250 fluxes to the masses obtained using the two other SPIRE bands, for which we used the corresponding flux values and mass absorption coefficients for 350 and 500~$\mu$m. 
For the total flux values of SPIRE-350/500 \textit{getsources} determined bigger footprints, inside which the flux was summed up. 
Thus the deblending got more unreliable, especially in the more crowded central regions of NGC~300 and the number of reliable mass estimates decreased.
The number of GDCs for which we could compare SPIRE-250 mass estimates with SPIRE-350 or SPIRE-500 mass estimates was thus reduced to 106 and 33 objects, respectively. 
The SPIRE-350 mass values are on average about 30\% higher than the SPIRE-250 mass values, but about 93\% are still compatible within the uncertainties.
The 33 SPIRE-500 mass values are on average a factor of 2.5 higher than the SPIRE-250 mass values and only 30\% are compatible within the uncertainties. 
However, the number of GDCs for which we could calculate mass estimates using the SPIRE-500 band is very low (only about 24\% of the total sample of GDCs), which effectively excluded this wavelength for the mass estimation.

Finally, we also checked how lower or higher values of the spectral index $\beta$ would affect our dust mass estimate.
For $\beta = 1.3$ and $\beta = 1.5$ the mass estimates would be on average about 35\% and 19\% lower, respectively. For $\beta = 2$ the mass estimates would be on average about 39\% higher. 
However, almost all of those values are compatible within the uncertainties to our mass determinations for $\beta = 1.7$ with an assumed error for the cold dust effective temperature of $\pm 1.5$~K. 

\subsection{Flux at 24~$\mu$m}
\label{sec:flux24mu}

We used the \textit{Spitzer}/MIPS~24~$\mu$m image from the \textit{Spitzer} Local Volume Legacy Survey\footnote{\url{http://irsa.ipac.caltech.edu/data/SPITZER/LVL/}} \citep{Dale2009} to estimate flux values of our GDCs at 24~$\mu$m.
In addition to the background subtraction that was already performed on the image we corrected for an additional background flux due to the stellar disk of the galaxy.
We determined a median flux value inside the ROI of 0.001~mJy, which we subtracted from all the pixels.
We used the \textit{aperture$\_$photometry} task of the \textit{photutils}\footnote{\url{https://photutils.readthedocs.io}} package to sum up all positive flux values inside the SPIRE-250 FWHM ellipses, which we adopted as the effective contour sizes of our GDCs. 
For the uncertainty values we compared these flux values to aperture photometry results for GDC FWHM minor and major axes multiplied by a factor of 0.9 and 1.1 and took the bigger flux difference as its symmetric error.  
We made no corrections for blended GDCs.
Table \ref{Tab:gdcs} lists the flux values at 24~$\mu$m for all GDCs. 
In section \ref{sec:halpha_emission} we use the 24~$\mu$m fluxes to correct the H$\alpha$ measurements for the effects of extinction.

\subsection{H$\alpha$ flux and correlation with HII~regions}
\label{sec:hii_correl}

We used the line-only H$\alpha$ map from \citetalias{Faesi2014} to calculate H$\alpha$-fluxes for our GDCs. 
We performed the aperture photometry again with the \textit{photutils} package.
We used its \textit{Background2D} task and determined a median background value per pixel of $0.34\cdot10^{-18} erg~s^{-1} cm^{-2}$, which we subtracted from all the pixels.
Similar to the flux determination at 24~$\mu$m, we summed up all positive flux values inside the SPIRE-250 FWHM ellipses that we adopted as the effective contour sizes of our GDCs. 
For the uncertainty values we compared these flux values to aperture photometry results for GDC FWHM minor and major axes multiplied by a factor of 0.9 and 1.1 and took the bigger flux difference as its symmetric error.  
We made no corrections for blended GDCs.
Table \ref{Tab:gdcs} lists the H$\alpha$ flux values for all GDCs. 

We also correlated our GDCs with the slightly modified catalogue of HII~regions from \citetalias{Deharveng1988} (see $\S~\ref{sec:Hii}$). 
We adopted the size estimates from \citetalias{Deharveng1988} and treated all HII~regions for which no size estimate existed as point sources at the resolution of the ESO/WFI image. 
We slightly increased the effective contour size (i.e. both the semi-minor and semi-major axis) of the GDCs to 1.1~times the area of their SPIRE-250 FWHM size to account for HII~regions with no size estimate situated just outside the SPIRE-250 FWHM extent of the GDC.
We established an association between a GDC and an HII~region if their areas intersected; the point source-like HII~regions had to be located inside the association area that we adopted for the GDCs.

Column (17) in Table \ref{Tab:gdcs} lists the HII~regions associated with each GDC. 
The superscripted letters in brackets indicate whether APEX observations by \citetalias{Faesi2014} of these HII~regions exist and whether they showed a detection (D), marginal detection (M), or non-detection (N) in CO(J=2-1). 

We find that \selAllHii ~of our GDCs (about 62\% of the final catalogue) are associated with at least one HII~region from \citetalias{Deharveng1988}. 
We note that this result does not sensitively depend on the effective contour size we adopted for the GDCs: even when we double the semi-minor and semi-major axes of the SPIRE-250 FWHM size about 30\% of the GDCs still show no association with any \citetalias{Deharveng1988} HII~regions according to our criterion.
For 54 GDCs an associated \citetalias{Deharveng1988} HII~region was already targeted by \citetalias{Faesi2014}, 31 of which showed a CO detection (D) and 17 showed no detection in CO (N).
We can thus confirm that the \citetalias{Faesi2014} sample is a reasonably unbiased sampling of the GDC distribution in NGC~300 because by targeting its HII~regions for associated molecular gas they also sampled the distribution of GDCs, in particular the dynamic range in cold dust effective temperature and dust masses, very well (see Table \ref{Tab:gdcs}).

Altogether, 117 of the 166 HII~regions  ($\sim$ 70\%) from our revised \citetalias{Deharveng1988} catalogue show an association with at least one of our GDCs.
Even though in our source selection we saw the presence of an HII~region as an affirmative sign for the correct detection of a GDC with ongoing star formation, this was not a primary selection criterion and our catalogue should thus not be biased towards GDCs associated with HII~regions.

We compared these results based on the \citetalias{Deharveng1988} HII~region catalogue to two more recent H$\alpha$ observations of NGC~300. 
\citet{Soffner1996} observed three fields in the central region of NGC~300 with an H$\alpha$ filter at the 3.5 m NTT telescope. 
Each of these fields had an extent of 2.2 $\times$ 2.2 arcminutes, an exposure time of 900 and 1800 seconds, and a resolution of 0.13 arcsec/pixel; this allowed \citet{Soffner1996} to identify many more HII~regions (88) than \citetalias{Deharveng1988} (33) for the same area. 
If we associate our GDCs in the fields observed by \citet{Soffner1996} with their catalogue of HII~regions instead of that by \citetalias{Deharveng1988}, the number of GDCs from our catalogue correlated with HII~regions increases by a third (from 21 to 28). 
If the whole stellar disk of NGC~300 were mapped in H$\alpha$ similar to the study of \citet{Soffner1996}, we would thus also expect our number of GDCs associated with HII~regions to increase by up to a third (or about 30). 

We also qualitatively checked the line-only H$\alpha$ map from \citetalias{Faesi2014} and the ESO/WFI images in Fig. \ref{footprints} for HII~regions not catalogued by \citetalias{Deharveng1988} and get a similar result of an additional 26 GDCs associated with at least one small HII~region.

If we consider these two estimates for GDCs associated with HII~regions not catalogued by \citetalias{Deharveng1988}, the percentage of GDCs from our catalogue not associated with any HII~region decreases to 18-20\%.

\subsection{Comparison with APEX observations}
\label{sec:apexcomp}

In this section we would like to examine the relation between the dust emission and molecular gas in NGC~300, as it is well-known from work on other nearby galaxies that the flux in the FIR correlates with the CO line emission (e.g. \citealp{Corbelli2012}, \citealp{Wilson2012}, \citealp{Grossi2016}).

Since at the time of writing there are no galaxy-wide molecular gas observations available for NGC~300, we have to restrict our comparison to the 76 pointed CO(J=2-1) APEX observations of \citetalias{Faesi2014}.
The APEX observations have a FWHM beam size of 27“, which corresponds to a spatial resolution of about 250~pc at the distance of NGC~300.
The goal of \citetalias{Faesi2014} was to get CO measurements for a representative sample of star-forming regions throughout the stellar disk of NGC~300, for which they targeted HII~regions for their associated molecular gas.

\begin{figure}
\centering
\includegraphics[width=\columnwidth]{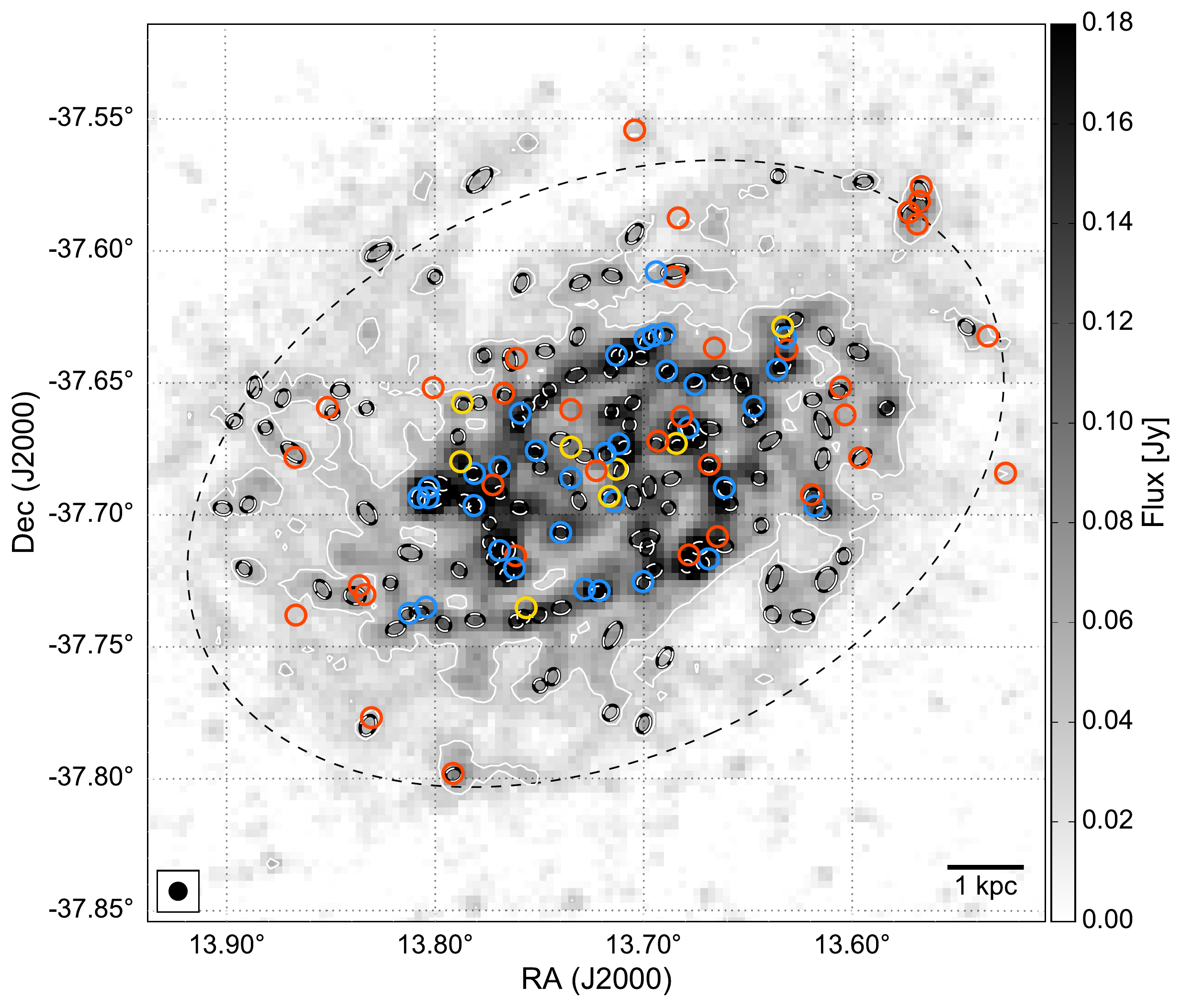}
\caption{SPIRE-350 map of NGC~300 overplotted with the final catalogue of \selSrcAll ~GDCs (small black dashed ellipses) and the APEX CO(J=2-1) pointed observations from \citetalias{Faesi2014} (remaining small circles; the colours indicate whether there was a detection (blue), marginal detection (yellow), or non-detection (red) in CO(J=2-1)).
The beam size of the \textit{Herschel} image is indicated in the box in the lower left of the image.
The white contour lines trace flux values of 0.04~Jy.
The large black dashed ellipse indicates the isophotal radius R$_{25}$ as defined in Table \ref{tbl:ngc300}.} 
\label{fig:apex_gdc_map}
\end{figure}

Figure \ref{fig:apex_gdc_map} shows a comparison between the catalogue of GDCs (yellow ellipses) and the APEX CO(J=2-1) pointed observations from \citetalias{Faesi2014}. 
For 33 of the 42 CO detections in \citetalias{Faesi2014} (shown as blue and yellow circles), the footprints of our GDCs overlap more than 50\% with the APEX beam area. 
The remaining 9 APEX pointings that yielded a detection in CO either partly overlap with a GDC or at least have a GDC close by.
There are also 15 GDCs whose footprint overlaps more than 50\% with the APEX beam area of observations that yielded no CO detections. 
The percentage of APEX CO detections coinciding with GDCs is thus much higher (79\%) than the percentage of APEX CO non-detections coinciding with GDCs (44\%).
It is clear from this comparison that dust structures are much more likely to be found where molecular gas is present.
Of the 15 GDCs associated with APEX CO non-detections, 11 are situated at a galactocentric distance beyond 2.81~kpc ($=0.5\cdot$R$_{25}$), i.e. they are located in the outer part of the galaxy, where the lower metallicity values might explain the CO non-detections. 
It is also possible that in these regions the dust is associated with HI instead of H$_{2}$ traced by CO.
Another explanation might be that these GDCs are associated with evolved HII~regions in which much of the molecular gas has been dissociated by long-term exposure to the intense radiation field present there. 
This would also explain the four GDCs located in the inner part of the galaxy, which are associated with non-detections in CO.

Any comparisons between CO observations to our GDCs need to account for some limitations of our catalogue.
Owing to the fairly large beam size at 250~$\mu$m ($\sim$ 170~pc) compact dust peaks with sizes below the \textit{Herschel} resolution that spatially correspond to peaks in CO emission might not be present in our catalogue, especially if the dust is heated by active star-forming regions and the bulk of the dust emission is shifted to the mid-infrared.
The sizes we adopted for our GDCs (their FWHM shape at 250~$\mu$m) is of course also tied to the spatial resolution of the SPIRE-250 image and, in cases in which larger dust complexes were segmented into individual GDCs, the deblending and size attribution might have been compromised.

A direct comparison between the GMC complexes (GMCCs) detected with APEX and the GDCs from our catalogue is further difficult because the APEX CO measurements represent single observations towards the central position of the associated bright HII~region.
Follow-up observations of a subsample of the APEX pointings with ALMA confirm that in many cases the brightest CO source is indeed offset from the central APEX pointing (Faesi et al., in prep.).
We thus restricted our comparison to a sample of 21 GDCs, where the APEX beam fully encompassed the elliptical shape we adopted as the size estimate for our GDC. 

\begin{figure}
\centering
\includegraphics[width=\columnwidth]{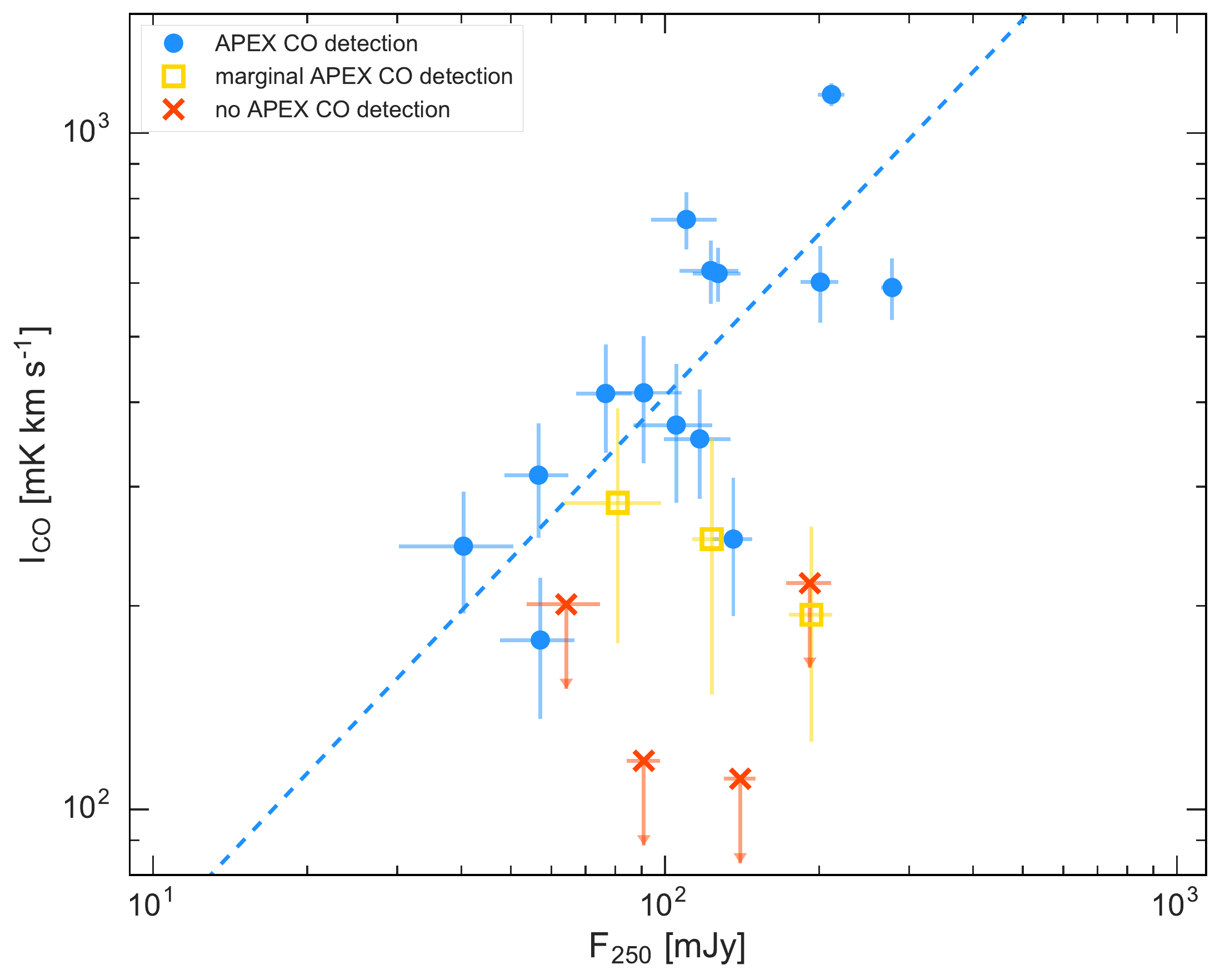}
\caption{Total flux at 250~$\mu$m for 21 of our GDCs plotted against the integrated CO(J=2-1) intensity of the associated APEX pointings from \citetalias{Faesi2014}.
The colours indicate whether \citetalias{Faesi2014} detected CO (blue circles and yellow squares) or not (red crosses: upper limits in I$_{CO}$).
The blue dashed line shows a fit only through the blue data points (orthogonal distance regression also taking the error bars into account); the slope of this fit is $0.8\pm0.2$.
} 
\label{fig:apex_herschel_flux}
\end{figure}

Figure \ref{fig:apex_herschel_flux} shows a plot of the total flux of the GDCs at 250~$\mu$m against the integrated CO(J=2-1) intensity of the corresponding GMC complexes as determined by \citetalias{Faesi2014}.
Neglecting the CO upper limits and marginal detections, we see a good correlation between the CO intensity and the dust emission with a Pearson correlation coefficient of 0.63. 
A fit through only the data points that have CO detections (blue circles) yields a slope of $0.8\pm0.2$.
Only two of the CO non-detections are located at larger galactocentric radii (3.3 and 5.4~kpc, with CO upper limits $<$118~mK km s$^{-1}$), while all the other GDCs associated with CO detections, marginal detections, and non-detections are located at galactocentric distances less than 3.1~kpc. 

\begin{figure}
\centering
\includegraphics[width=\columnwidth]{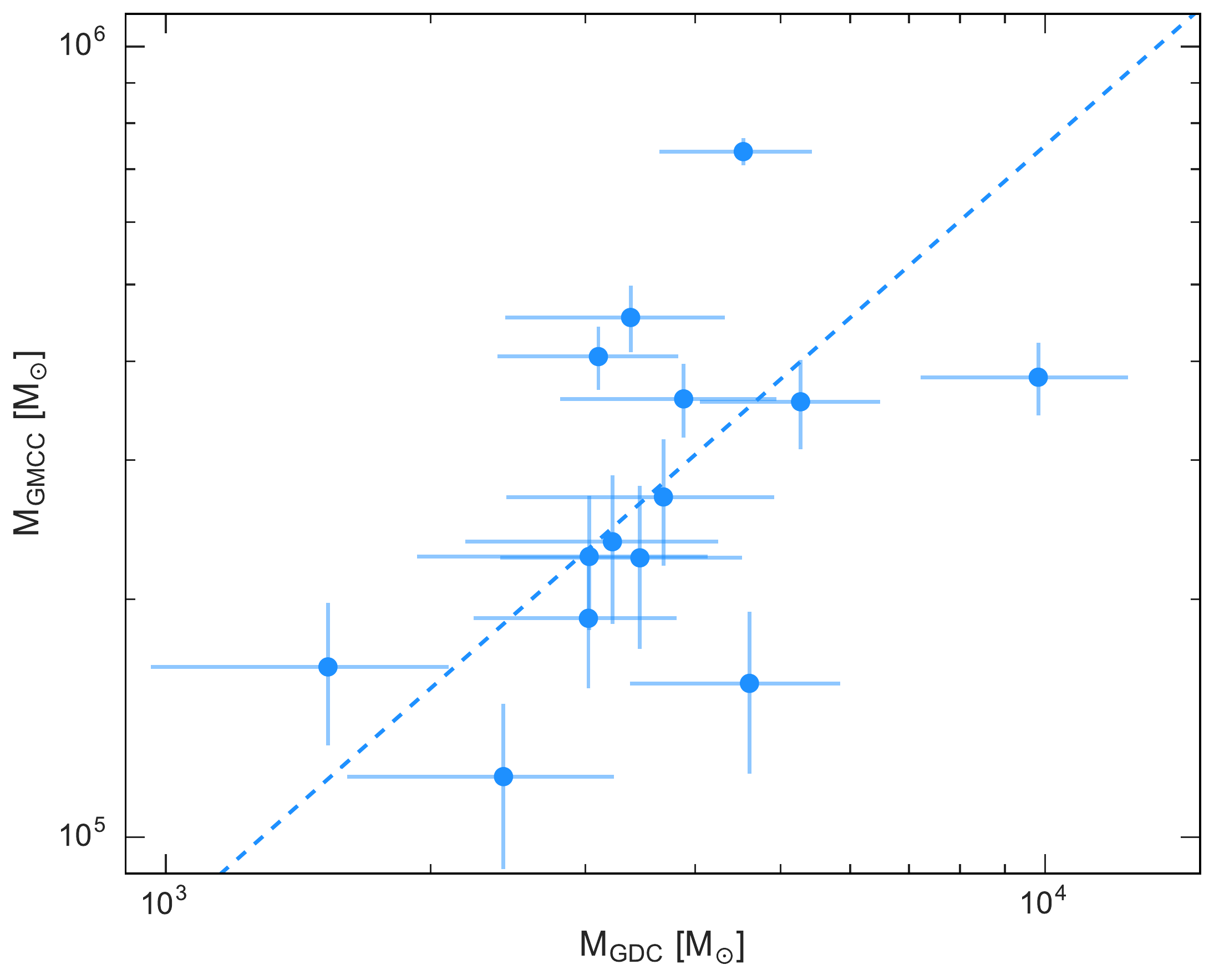}
\caption{Dust masses vs. molecular gas masses for the 14 of our GDCs coinciding with GMCCs having CO(2-1) detections from \citetalias{Faesi2014}.
The dashed line shows a fit through the data points (orthogonal distance regression also taking the error bars into account); the slope of this fit is $1.0\pm0.3$.
} 
\label{fig:apex_herschel_mass}
\end{figure}

In Figure \ref{fig:apex_herschel_mass} we compare the dust masses of 14 GDCs with the molecular gas masses of coinciding GMC complexes that show a detection in CO(2-1) as determined by \citetalias{Faesi2014}.
Since \citetalias{Faesi2014} used different values for the distance to NGC~300 and its inclination, we recalculated the GMCC masses for the values we used throughout this paper (see Table \ref{tbl:ngc300}).
Notwithstanding the rather low Pearson correlation coefficient of 0.35, most of the 14 GDCs in Figure \ref{fig:apex_herschel_mass} are consistent with a general trend that regions with higher dust masses also have higher molecular gas masses.
A fit through those 14 data points yields a slope of $1.0\pm0.3$.
Even though the scatter is substantial and we are limited to low-number statistics, Figure \ref{fig:apex_herschel_mass} thus nonetheless serves as a valuable starting point for the discussion of the gas-to-dust ratio in NGC~300. 
However, we have to caution that both mass estimates are subject to many assumptions and uncertainties.
For the molecular gas mass estimates from \citetalias{Faesi2014}, systematic uncertainties might be due to the CO-to-H$_{2}$ conversion factor and possible variations of the CO(2-1)-to-CO(1-0) line ratio throughout the disk of the galaxy. 
For our calculation of the total dust mass, systematic uncertainties might have been introduced by our adopted values for the mass absorption coefficient and spectral emissivity index.
Even though the 14 GDCs plotted in Figure \ref{fig:apex_herschel_mass} are covered by an APEX beam, most of these GDCs are not centred in the beam. 
There could thus be a systematic bias towards lower APEX gas masses in this comparison, if we assume that the dust emission peaks are for the most part spatially coincident with the brightest CO emission peaks.
The fraction of molecular mass residing in CO-dark gas might also be significant and its importance increases in the outer regions of galaxies in which the metallicity decreases and the amount of dust shielding gets lower \citep{Wolfire2010}.
The fit through the data points in Figure \ref{fig:apex_herschel_mass} that corresponds almost exactly to a constant gas-to-dust ratio of 80 acts thus as a strict lower limit on the molecular gas-to-dust ratio. 

We note again that \citetalias{Faesi2014} only calculated the molecular gas mass (including helium), not the total mass including atomic hydrogen.
However, since the neutral ISM is much more extended than the molecular ISM and GMCs are typically clumpy and less than 100~pc in size \citep[e.g.][]{Murray2011}, we expect that at the size scales of the \citetalias{Faesi2014} GMCCs ($\sim$250~pc) HI contributes more significantly to the total gas mass than at smaller scales and might even be a dominating factor.
\citet{Mizuno2001} found that most of the GMCs in the LMC are distributed in dense parts of HI gas.
In their study of GMCs in M33, \citet{Imara2011} also found that the majority of clouds coincide with a local peak in the surface density of HI with mean values for $\Sigma_\mathrm{HI}$ of about 10~\msun ~pc$^{-2}$ near the GMCs.
The average value of $\Sigma_\mathrm{HI}$ in NGC~300 is about 7~\msun ~pc$^{-2}$ within R$_{25}$ \citep{Westmeier2011}, which is slightly higher than the average molecular gas surface densities of about 6~\msun ~pc$^{-2}$ for the GMCCs with CO detections from \citetalias{Faesi2014}.
If we assume a constant HI surface density for the GMCCs with a value equal to the average $\Sigma_\mathrm{HI}$ we would expect HI to contribute about 3.5$\cdot$10$^{5}$~\msun ~to the total gas mass of a structure with radius 125~pc.
This estimate for the HI mass exceeds the average molecular gas mass of the 14 GMCCs included in Figure \ref{fig:apex_herschel_mass} by about 13\%, so on a first approximation we would expect the average value of the total gas-to-dust ratio to be about 170, which is slightly more than double the amount of the molecular gas-to-dust ratio.
Given that the atomic-to-molecular transition typically seems to occur at an HI surface density of 10~\msun ~pc$^{-2}$ \citep[e.g.][]{Leroy2008} this would yield an upper limit for the HI mass contribution to the GMCCs of about 4.9$\cdot$10$^{5}$~\msun ~and gives us an upper limit of about 210 for the average total gas-to-dust ratio of the 14 GMCCs in Figure \ref{fig:apex_herschel_mass}.
\citet{Leroy2011} empirically found a metallicity-dependent total gas-to-dust ratio in nearby galaxies of 
$\mathrm{log}_{10}(\delta_\mathrm{GDR}) =  9.4 - 0.85\cdot  (12+\mathrm{log}_{10}(\mathrm{O}/\mathrm{H}))$.
For a characteristic oxygen abundance of $12+\mathrm{log}_{10}(\mathrm{O}/\mathrm{H}) = 8.41$ at 0.4$\cdot$R$_{25}$ in NGC~300 \citep{Bresolin2009}, this yields a gas-to-dust ratio of about 180, which is in agreement with our estimate above.  
Because of the metallicity gradient in NGC~300, the oxygen abundance value ranges from about 8.57 in the centre to 8.16 at R$_{25}$ \citep{Bresolin2009}, which translates to a prediction for the gas-to-dust ratio in NGC~300 from about 130 near the centre to about 300 in the outskirts of the stellar disk. 
For the morphologically similar galaxy M33, \citet{Braine2010} estimated the total gas-to-dust ratio from derived dust cross-sections and found that it ranges from $\sim$125 in the inner parts $\sim$200 near R$_{25}$.
Compared to M33, our predicted values of the total gas-to-dust ratio in NGC~300 are almost identical in the central part but higher near R$_{25}$, which would also be expected from the similar but slightly steeper slope of the metallicity gradient of NGC~300 as measured from O abundances of HII~regions \citep{Magrini2016}.

\section{Discussion}
\label{cha:discuss}

\subsection{Cold dust effective temperature}

\begin{figure}
\centering
\includegraphics[width=\columnwidth]{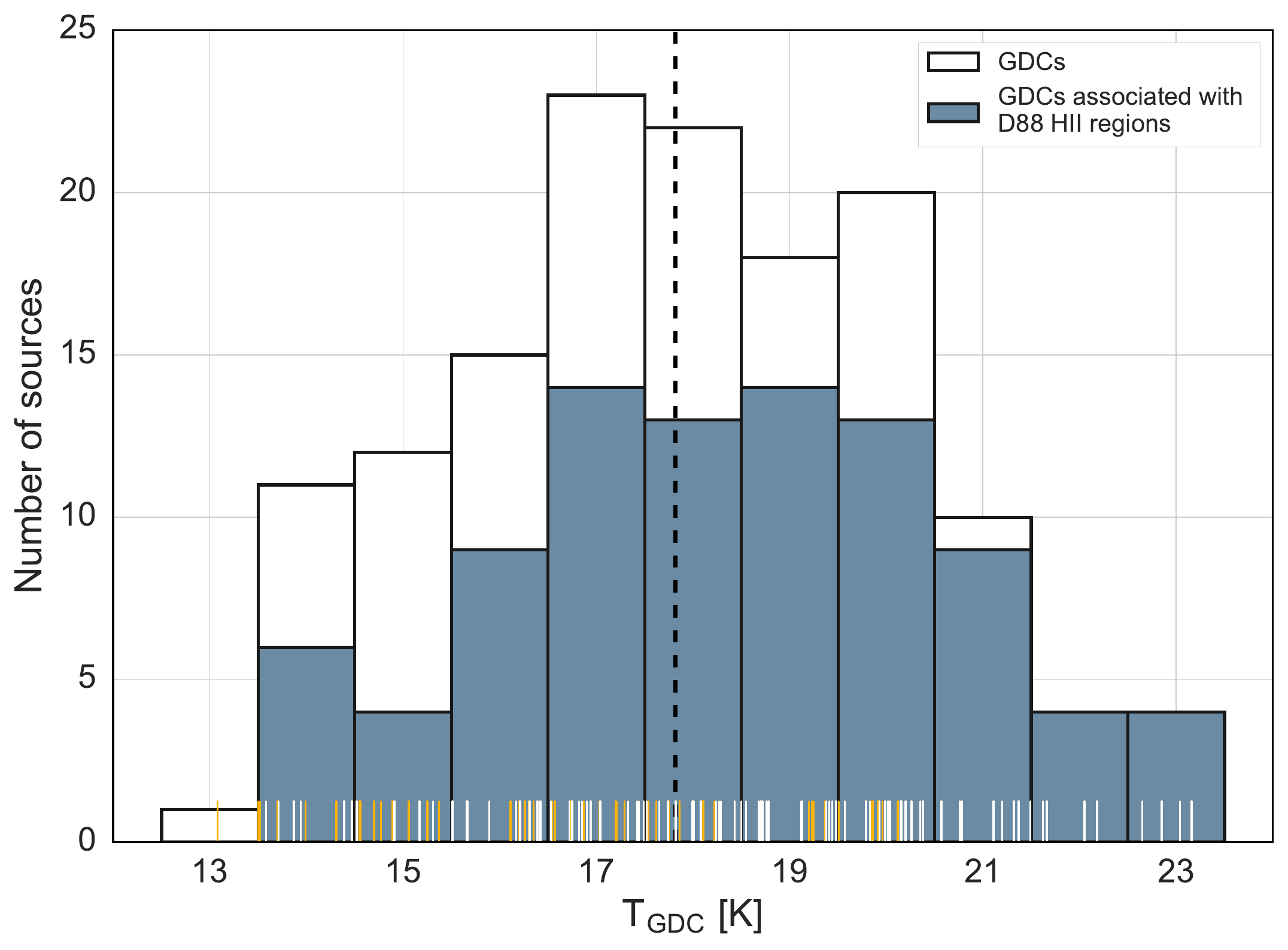}
\caption{Histogram of the cold dust effective temperature for the \tempAll ~of our GDCs with dust effective temperature measurements; individual temperature values of the GDCs are indicated by the ticks in the rug plot. 
The shaded bars indicate the subsample of \tempD ~GDCs associated with HII~regions from \citetalias{Deharveng1988}; individual values of these GDCs are indicated in white in the rug plot. 
The dashed vertical line shows the median temperature (\tempMedGdc) of the entire sample.}
\label{fig:hist_Tcold}
\end{figure}

Figure \ref{fig:hist_Tcold} shows the histogram of the cold dust effective temperatures for \tempAll ~GDCs from the final catalogue (for the remaining six sources no temperature values could be determined).
This figure shows that the GDCs probe a wide range of effective temperatures from $\sim$~13 to 23~K. 
As one might expect, GDCs that are associated with known HII~regions have in general a slightly higher cold dust effective temperature. 
All the GDCs with a cold dust effective temperature above the median value of \tempMedGdc ~that show no association with any \citetalias{Deharveng1988} HII~region are situated in the inner part of the galaxy; these GDCs have a deprojected galactocentric distance smaller than 2.81~kpc ($=0.5\cdot$R$_{25}$). 
One likely explanation for these increased temperatures is the radiation field produced by the dense older stellar population near the centre leading to higher stellar surface densities in the inner parts of the galaxy \citep{Westmeier2011}, which is reflected in the radial gradient in cold dust effective temperature in Figure \ref{fig:temp_map}. 

Table \ref{Tab:gdcs} in the Appendix confirms that the GDCs with the highest cold dust effective temperature values nearly all correlate with giant HII~regions%
\footnote{An exception is GDC \#107, which has a high dust effective temperature but correlates only with a medium-sized HII~region.},
which indicates that a physical association is at least partly responsible for this distribution.
However, our approach of estimating the effective temperature is subject to many uncertainties, such as our implicit assumptions that the SED can be properly modelled by the two-component black-body fit and that the dust effective temperature along a given line of sight is constant.

\begin{figure}
\centering
\includegraphics[width=\columnwidth]{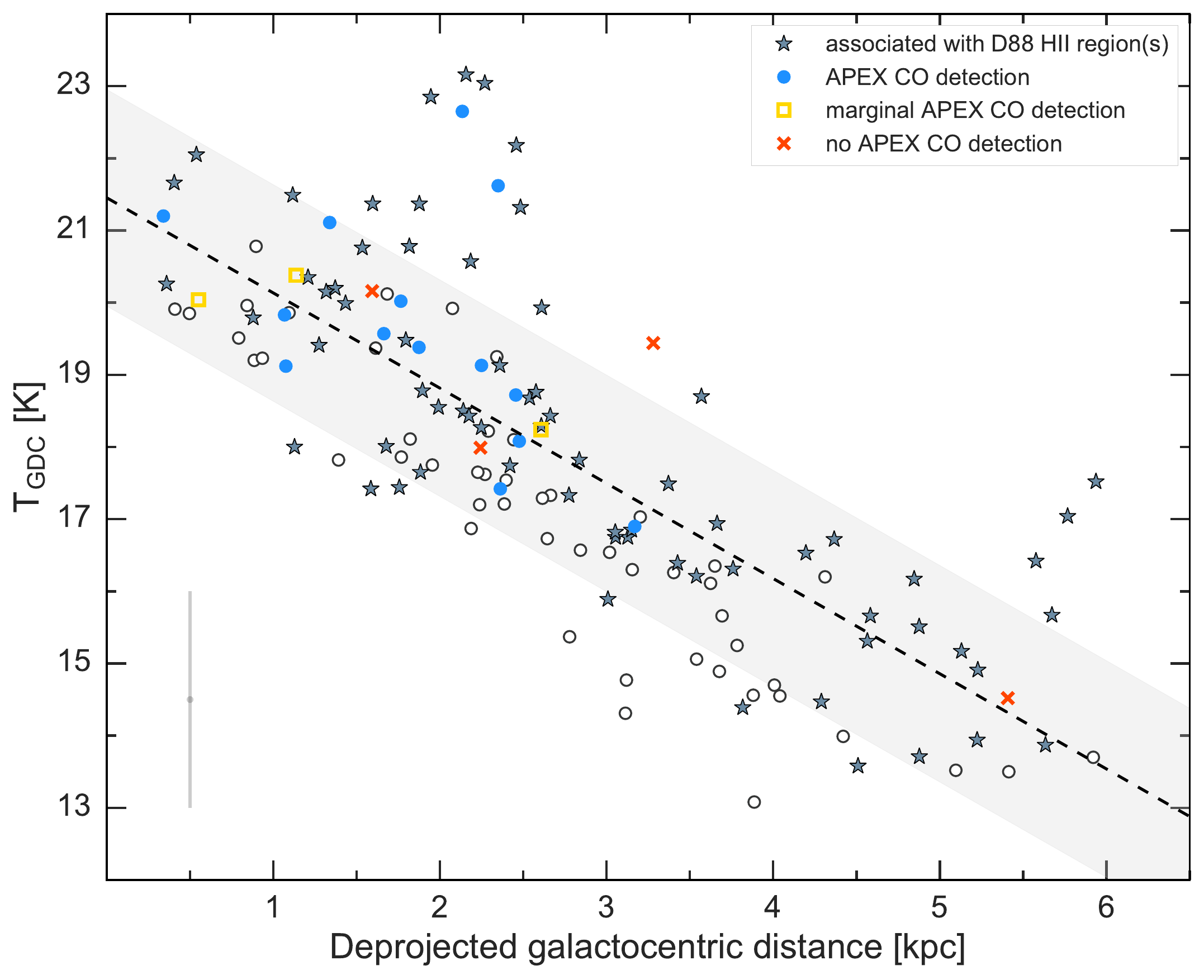}
\caption{Cold dust effective temperature vs. deprojected galactocentric distances for the \tempAll ~of our GDCs with temperature measurements.
The coloured symbols all indicate GDCs that are associated with \citetalias{Deharveng1988} HII~regions.
The GDCs that also coincide with regions that were observed by APEX (see section \ref{sec:apexcomp}) are indicated by blue filled circles (CO detections), yellow unfilled squares (marginal CO detections), and red crosses (no CO detections).
The vertical error bar in the lower left indicates the uncertainty in temperature that we assume for all GDCs.
The dashed line is a fit through the data points and has a slope of $-1.3\pm 0.1$.
The shaded region indicates the typical temperature uncertainty interval of $\pm$1.5~K.
}
\label{fig:dist_vs_temp}
\end{figure}

In Figure \ref{fig:dist_vs_temp} we plot the estimate of the cold dust effective temperature for each GDC against their deprojected galactocentric distances. 
There is no obvious accumulation at a certain temperature value and the high negative Pearson correlation coefficient of -0.77 suggests a relatively smooth radial cold dust effective temperature gradient in the disk of NGC~300; a fit through the data points yields a slope of $-1.3\pm 0.1$.
This gradient in cold dust effective temperature ranges from $\sim$22~K in the centre to $\sim$14~K at a galactocentric distance of 6~kpc, which is similar to the cold dust effective temperature gradient found in M33 (\citealp{Kramer2010}; \citealp{Xilouris2012}).
There are some deviations from the general trend, most notably the GDCs with very high cold dust effective temperature at a galactocentric distance of $\sim$~$2 - 2.5$~kpc (GDCs \#42, \#65, \#68, \#102, \#123, \#124, and \#126) and $\sim$~6~kpc (GDCs \#2, \#3, and \#4); however, they are all associated with big HII~region complexes, which may explain their increased dust effective temperature. 
The general trend in decreasing cold dust effective temperature with increasing galactocentric distance is likely due to the radial decrease in stellar density \citep{Westmeier2011} and thus interstellar radiation field. 
The amount of heating of the cold dust is thus mainly a function of the galaxy's exponential surface brightness profile, i.e. the radiation field produced by the total stellar population of the galaxy. 

\subsection{GDC dust masses}

\begin{figure}
\centering
\includegraphics[width=\columnwidth]{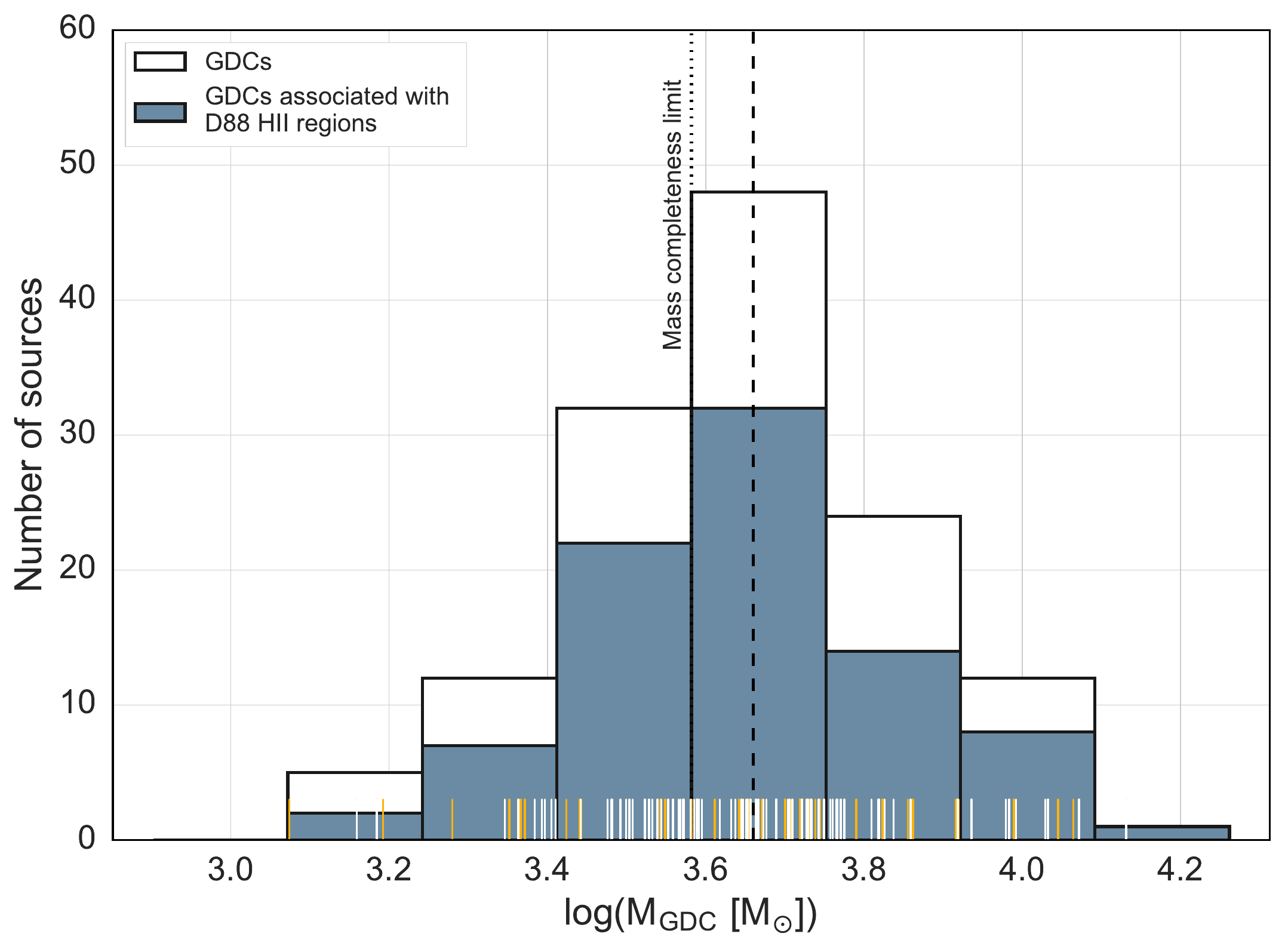}
\caption{Histogram of the dust mass distribution for the \massAll ~of our GDCs with reliable mass measurements; individual mass values of the GDCs are indicated by the ticks in the rug plot.
The bin size of \massBin ~corresponds to our median mass uncertainty.
The shaded bars indicate the subsample of \massD ~GDCs associated with HII~regions from \citetalias{Deharveng1988}; individual values of these GDCs are indicated in white in the rug plot. 
The dashed vertical line indicates the median mass value of \massMed ~for the entire sample and the vertical dotted line indicates our mass completeness limit of \massLimit .}
\label{fig:hist_mass}
\end{figure}

Figure \ref{fig:hist_mass} shows the dust mass distribution of all GDCs from the final catalogue for which reliable mass estimates could be determined from the SPIRE-250 flux values (\massAll ~sources in total). 
The total dust mass in the sample is \massTotDust ~with a median mass of \massMed.
Assuming our completeness estimate from Figure \ref{fig:flux_limit} we used the median cold dust effective temperature of our GDC sample (\tempMedGdc) to translate the flux value of 86~mJy (above which the contamination fraction drops below 10\%; see Fig. \ref{fig:flux_limit}) to an approximate mass completeness limit of \massLimit.
Our sample of GDCs covers a range in dust masses of about \massMin ~to \massMax.

The likelihood that a GDC is associated with an HII~region appears to be independent of its mass.
Of the 86 GDCs above the mass completeness limit 65\% are associated with a \citetalias{Deharveng1988} HII~region. 
The GDCs \#5, \#30, and \#45 are the most massive objects of our catalogue that do not show any association with known HII~regions and, thus, are of particular interest for further studies, as they might be pre-star-forming environments. 
For the remaining GDCs below the mass completeness limit 64\% show a correlation with a \citetalias{Deharveng1988} HII~region. 

\begin{figure}
\centering
\includegraphics[width=\columnwidth]{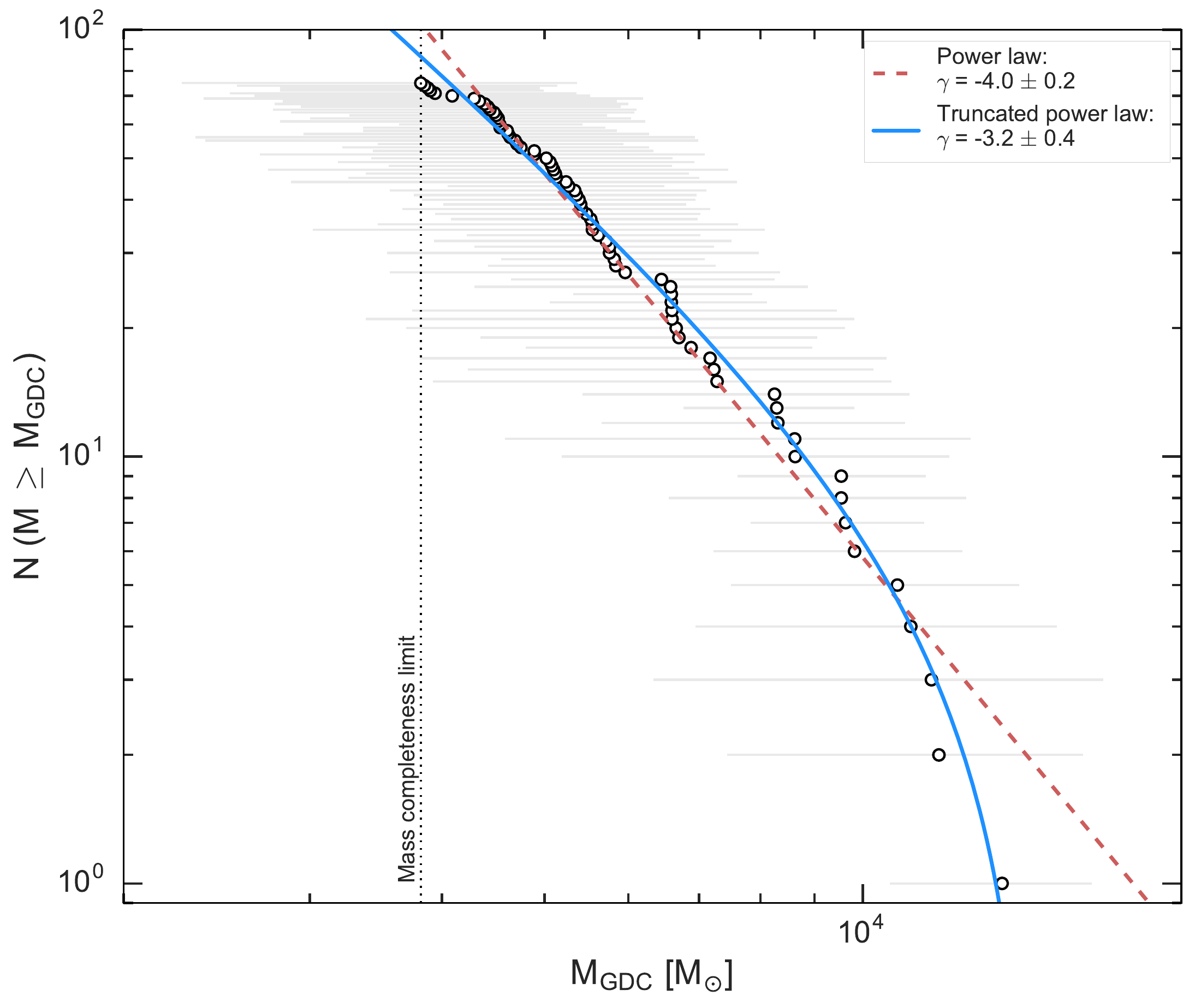}
\caption{Cumulative dust mass distribution for the \cumMassNr ~GDCs with mass measurements above the mass completeness limit of \massLimit ~(vertical dotted line). 
The horizontal error bars indicate the uncertainties in the dust masses.
The red dashed line and blue solid line show a power law and a truncated power law fit to the data points, respectively. 
}
\label{fig:diff_mass}
\end{figure}

In Figure \ref{fig:diff_mass} we plot the cumulative mass distribution for \cumMassNr ~GDCs that have dust masses above the mass completeness limit of \massLimit ~and fractional mass uncertainties less than 0.5. 
We fit truncated and non-truncated power law functions to the data using the \textit{mspecfit}\footnote{\url{https://github.com/low-sky/idl-low-sky/blob/master/eroslib/mspecfit.pro}} routine of \citet{Rosolowsky2005}, which is designed to fit cumulative distributions and implements a maximum-likelihood algorithm to account for uncertainties in both the mass values and the number distribution.
The non-truncated power law distribution is defined as

\begin{equation}
N\left(M' > M \right) = \left( \dfrac{M}{M_{0}}\right) ^{\gamma + 1},
\label{eqn:powerlaw}
\end{equation} 

with the power law index $\gamma$ and the maximum mass in the distribution $M_{0}$.

The truncated power law distribution is defined as

\begin{equation}
N\left(M' > M \right) = N_{0}\left[\left( \dfrac{M}{M_{0}}\right) ^{\gamma + 1} - 1\right],
\label{eqn:truncpowerlaw}
\end{equation} 

with the additional parameter $N_{0}$ indicating the point where the distribution 
shows a significant deviation from a non-truncated power law\footnote{A more physical interpretation of $N_{0}$ is obtained by setting $N$ to $N_{0}$ in Eq. \ref{eqn:truncpowerlaw} and solving for $M$;
$N_{0}$ thus gives us the number of clouds with mass greater than $M = 2^{1/(\gamma + 1)}M_{0}$.}. 

For the non-truncated power law fit we obtained values for $\gamma$ and $M_{0}$ of -4.0$\pm$0.2 and 1.80$\cdot$10$^{4}$~\msun ~($\pm$0.14$\cdot$10$^{4}$~\msun), respectively.
The values of $\gamma$, $M_{0}$, and $N_{0}$ for the truncated power law fit are -3.2$\pm$0.4, 1.45$\cdot$10$^{4}$~\msun ~($\pm$0.14$\cdot$10$^{4}$~\msun), and 5.1$\pm$4.7, respectively.

The truncated power law distribution is a much better fit to the data points in Figure \ref{fig:diff_mass} than the non-truncated power law distribution.
Although the value of the power law index for the truncated power law distribution ($\gamma = -3.2\pm 0.4$) is higher than the value of the power law index that \citetalias{Faesi2014} derived from their mass spectrum for resolved molecular clouds ($\gamma= -2.7\pm 0.5$), both of which are in agreement considering the uncertainty intervals.

\begin{figure}
\centering
\includegraphics[width=\columnwidth]{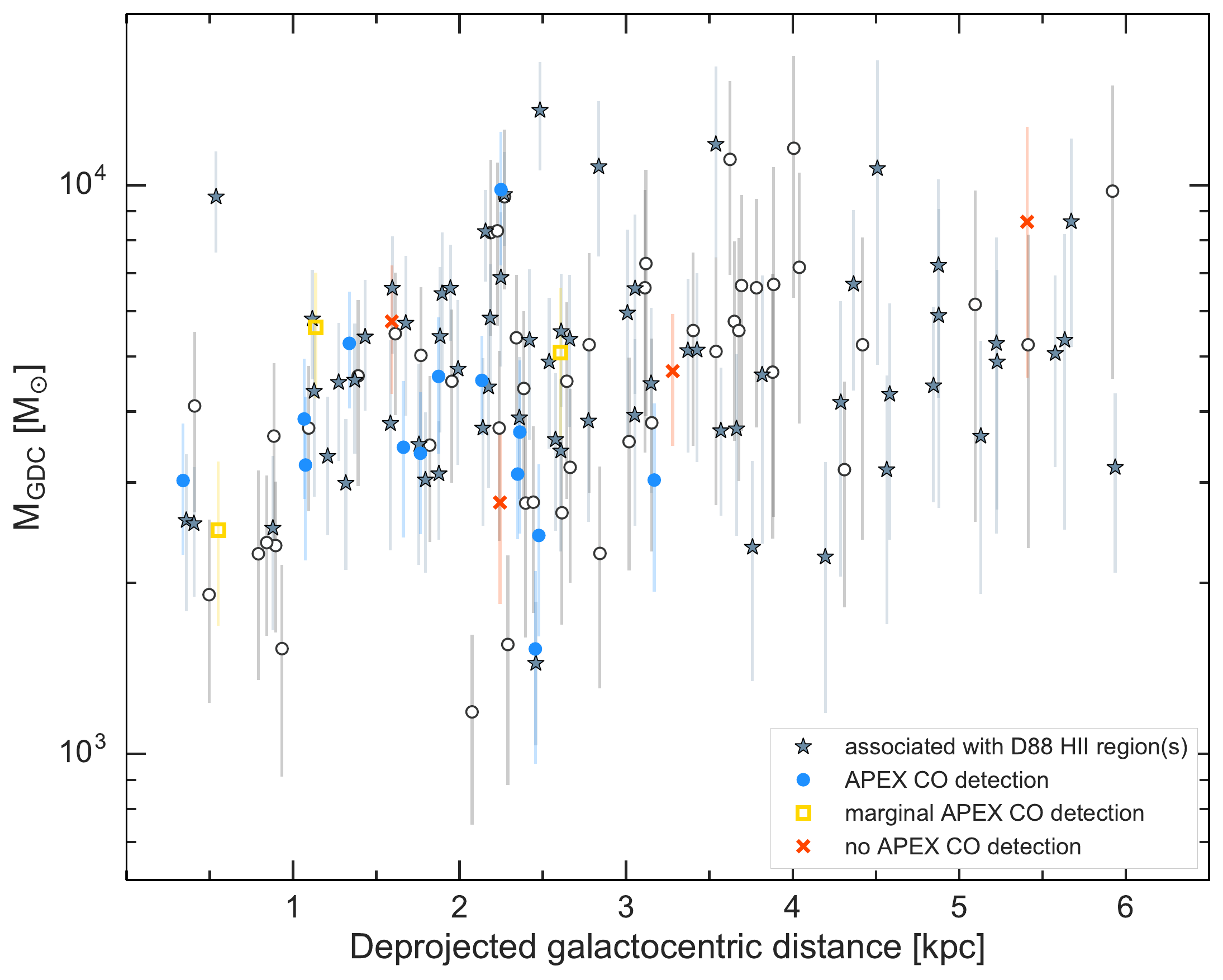}
\caption{Plot of the dust mass estimates against the deprojected galactocentric distances for \massAll ~of our GDCs with reliable mass measurements.
The coloured symbols all indicate GDCs that are associated with \citetalias{Deharveng1988} HII~regions.
The GDCs that also coincide with regions that were observed by APEX (see section \ref{sec:apexcomp}) are indicated by the blue filled circles (CO detections), yellow unfilled squares (marginal CO detections), and red crosses (no CO detections).
The error bars denote the uncertainty in the mass estimates.
}
\label{fig:dist_vs_mass}
\end{figure}

Figure \ref{fig:dist_vs_mass} shows GDC dust mass as a function of deprojected galactocentric distance. 
At a Pearson correlation coefficient of 0.29, the weak trend of GDC mass estimates to increase with galactocentric radius does not appear to be significant in view of the uncertainties.
A likely explanation for the apparently lower masses of the GDCs located below a galactocentric distance of 1~kpc could be that the subtracted flux attributed to diffuse emission was highest in the central region of the galaxy and that the GDCs in that location are also more crowded and thus might be more affected by the deblending, which could lead to possible underestimations of the dust mass values of some of these GDCs. 
Since our mass estimate is highly dependent on the value of the cold dust effective temperature, with lower values leading to a significant increase in the computed mass estimates, it is possible that we overestimate the masses for some of the GDCs located far away from the centre for which we could only determine uncertain temperature estimates (these GDCs are indicated in Table \ref{Tab:gdcs}).
However, the similar scatter in dust masses of GDCs located at $\sim$2 to 6~kpc is most likely a real effect, since \citet{Casasola2017} also found that the dust mass surface density gradient in NGC~300 is almost flat out to $\sim$ 5~kpc, which is the outermost distance they consider for their dust mass surface density calculations, and cannot be fit by an exponential profile. 
Moreover, Figure \ref{fig:dist_vs_temp} and Figure \ref{fig:dist_vs_mass} show that both GDCs with and without associated HII~regions are found in the entire galaxy and across the full mass range.

\subsection{Size distribution}

\begin{figure}
\centering
\includegraphics[width=\columnwidth]{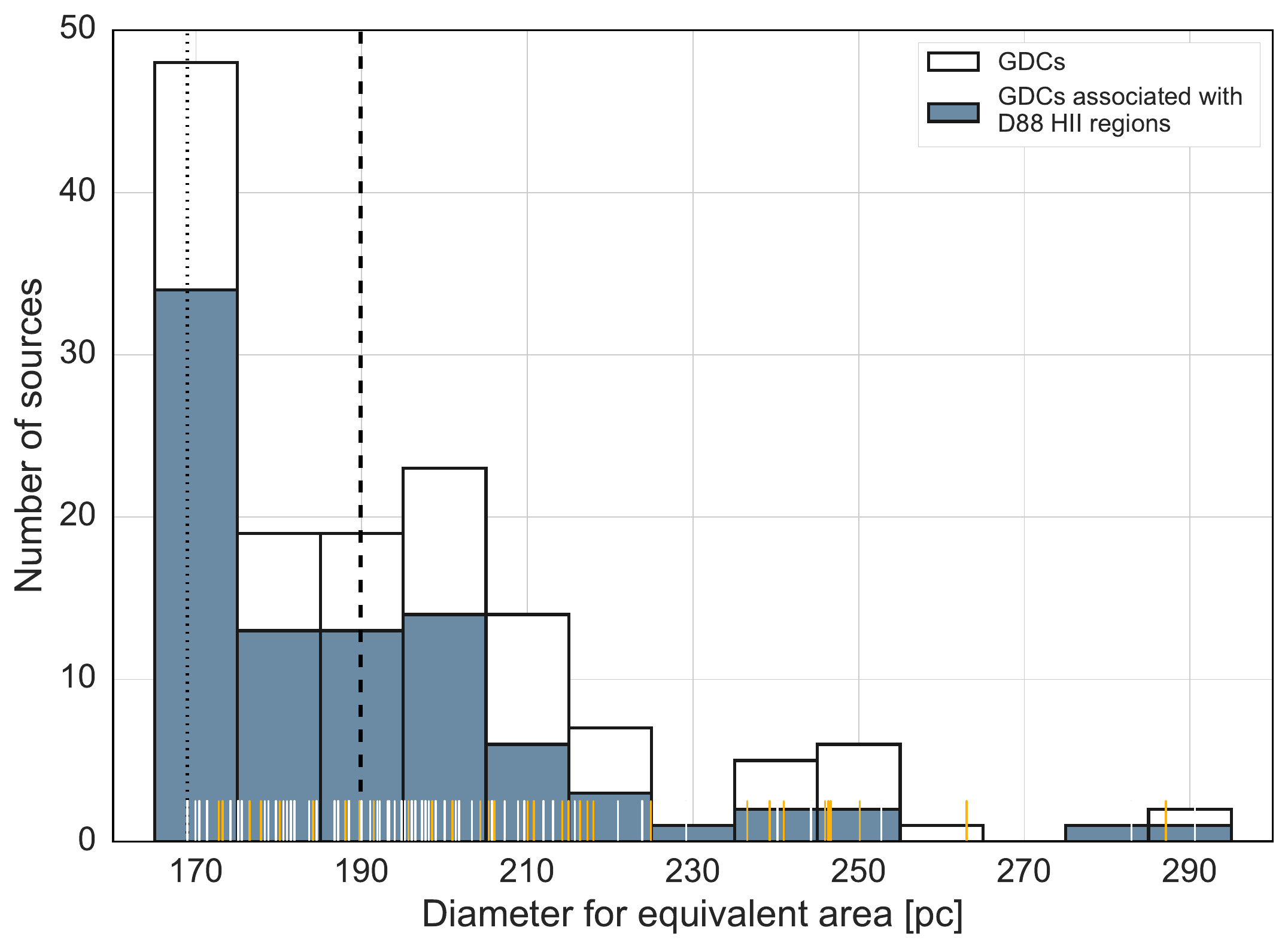}
\caption{Histogram of the size distribution for the GDCs. The diameter of an equivalent area is proxy for the real elliptical size; individual size values of the GDCs are indicated by the ticks in the rug plot.
The shaded bars indicate the subsample of \sizeD ~GDCs associated with HII~regions from \citetalias{Deharveng1988}; individual values of these GDCs are indicated in white in the rug plot. 
The vertical lines indicate the FWHM beam size (169~pc) at 250~$\mu$m (dotted) and the median size value of \sizeMed~(dashed) for our entire sample.}
\label{fig:hist_Size}
\end{figure}

We adopted the elliptical shape of the FWHM at 250~$\mu$m as determined by \textit{getsources} for our size estimate of the GDCs. 
Figure \ref{fig:hist_Size} shows the distribution of diameter values for the equivalent circular areas.  
About a third of the GDCs are situated at the lower limit at 170~pc, which is defined by the beam size at 250~$\mu$m.
It is likely that many of these GDCs could not be resolved in the SPIRE-250 image and are actually much smaller, even though some could be individual large dust structures.
For the GDCs bigger than about 180~pc in size, we assume that these are indeed complexes or associations of several dust clouds, as their extent matches or exceeds the size of even the largest known individual clouds in our Milky Way \citep[e.g.][]{Murray2011}.

The average size of the GDCs increases slightly at larger galactocentric distances.
For GDCs located below a galactocentric distance of 2.81~kpc ($=0.5\cdot$R$_{25}$) the median diameter for an equivalent circular area is 178~pc, whereas for GDCs at larger galactocentric distances the median diameter is 198~pc.
For about 35\% of the 87 GDCs located at a galactocentric distance of less than 2.81~kpc, the sizes are the same as the beam size of the SPIRE-250 image, which means that these are likely unresolved and thus more compact objects. 
For the 59 GDCs located beyond a galactocentric distance of 2.81~kpc, only 10\% have sizes identical to the SPIRE-250 beam size. 

However, we have to caution that for some of the GDCs in the outskirts of the galaxy that show only weak fluxes at 250~$\mu$m the size estimate as determined by \textit{getsources} might be too high; we indicate the GDCs that have uncertain size estimates in Table \ref{Tab:gdcs}.

\subsection{Association with H$\alpha$ emission}
\label{sec:halpha_emission}

\begin{figure}
\centering
\includegraphics[width=\columnwidth]{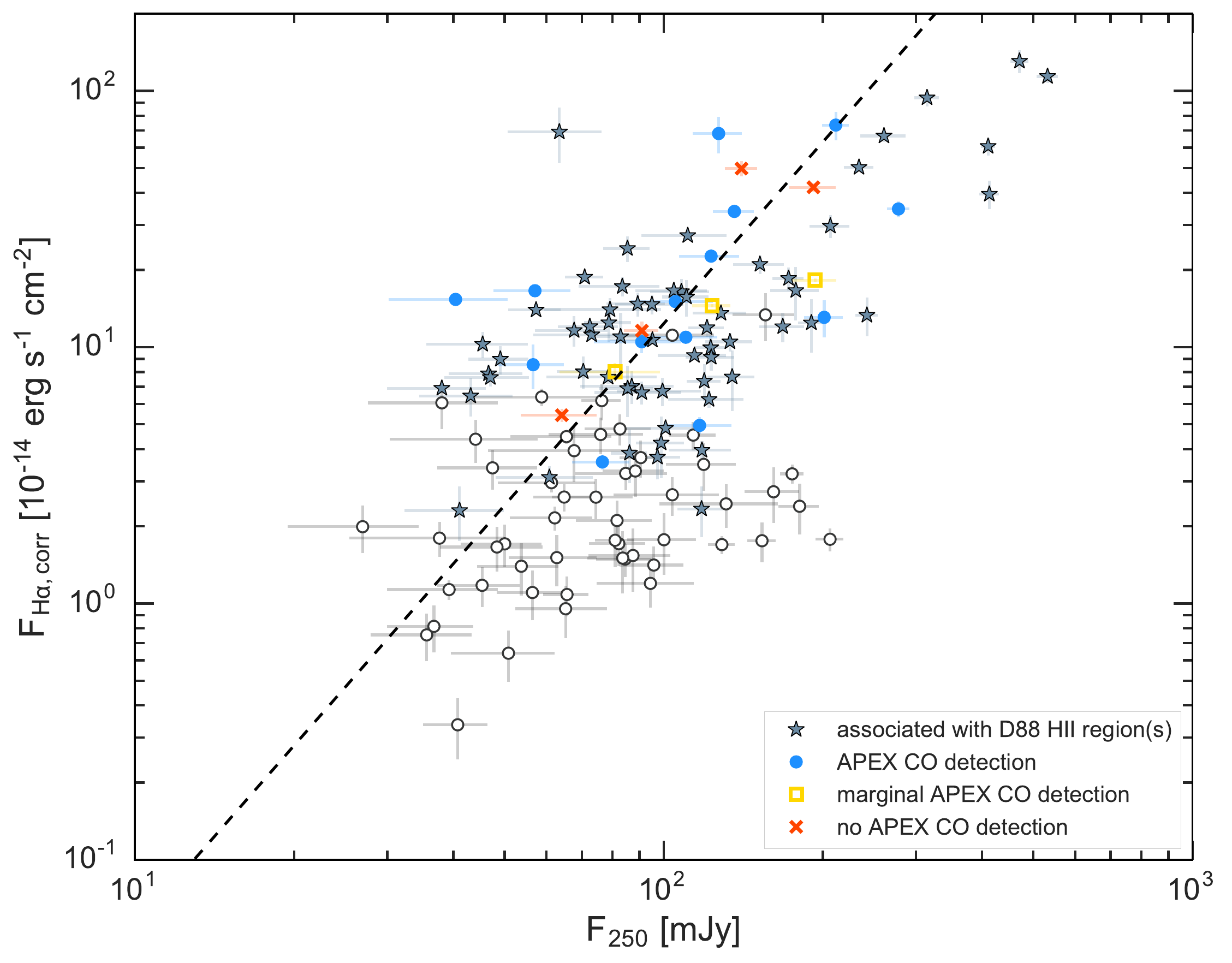}
\caption{Total flux at 250~$\mu$m vs. extinction-corrected H$\alpha$ emission for 134 of our GDCs with reliable flux measurements (S/N~$>3$) at both wavelengths.
The coloured symbols all indicate GDCs that are associated with \citetalias{Deharveng1988} HII~regions.
The GDCs that also coincide with regions that were observed by APEX (see section \ref{sec:apexcomp}) are indicated by the blue filled circles (CO detections), yellow unfilled squares (marginal CO detections), and red crosses (no CO detections).
The dashed line shows a fit through the GDCs associated with \citetalias{Deharveng1988} HII~regions (orthogonal distance regression also taking the error bars into account); the slope of this fit is $2.4\pm0.3$.
}
\label{fig:flux250_vs_halpha}
\end{figure}

Another interesting property of our GDC sample is their associated H$\alpha$ emission.
For GDCs that are in a pre-star-forming or early star-forming stage we would expect plenty of dust emission in the FIR but only little associated H$\alpha$ emission.
For GDCs in an intermediate star-forming stage we would still expect high dust emission in addition to an increased H$\alpha$ emission if the dust is associated with GMC complexes that formed massive stars that in turn created HII~regions.
However, the presence of an HII~region and higher H$\alpha$ emission can also indicate cloud destruction and thus a decrease of FIR emission for regions at a later evolutionary star-forming stage. 
In any case, we would assume that the heating from the massive stars of the HII~regions also leads to higher dust effective temperatures for the associated GDCs.
As discussed in section \ref{sec:temp} there are two dust temperature components and within the HII~regions and in their photodissociation regions there is associated hotter dust.
However, for the GDCs the column density and mass of this hot dust component is negligible compared to the cold dust component, which is why we only focus on the latter in this discussion.

Since the effect of dust extinction on the H$\alpha$-emission is non-negligible we corrected for it using the flux values determined at 24~$\mu$m, assuming that this wavelength is a good tracer for the absorbed H$\alpha$-emission that gets re-emitted by the dust at mid-infrared wavelengths. 
We calculated the H$\alpha$-extinction according to Eq. (7a) in \citetalias{Faesi2014}, which is based on the results from \citet{Calzetti2007}, and used it to correct the H$\alpha$-flux values that we report in table \ref{Tab:gdcs}.

Figure \ref{fig:flux250_vs_halpha} shows the total flux at 250~$\mu$m versus the associated extinction-corrected H$\alpha$ emission for the GDCs for which we could determine the flux values. 
The GDCs associated with \citetalias{Deharveng1988} HII~regions show an increase in H$\alpha$ emission with increasing flux at 250~$\mu$m, yielding a Pearson correlation coefficient of 0.75.
A fit through the GDCs associated with \citetalias{Deharveng1988} HII~regions yields a slope of $2.4\pm0.3$.
The majority of the 21 GDCs that coincide with regions observed in CO(2-1) by APEX have an H$\alpha$ emission that is higher than the median H$\alpha$ flux (6.1$\cdot$10$^{-14}$~erg~s$^{-1}$~cm$^{-2}$) of all GDCs plotted in Figure \ref{fig:flux250_vs_halpha}. 
This is not that surprising, given that the APEX observations were targeting HII~regions. 
Two of the GDCs coinciding with CO non-detections (\#17 and \#36) show some of the highest H$\alpha$ flux values of the whole sample, which makes it possible that these are regions in which the associated GMCCs were already disrupted by the giant HII~regions they formed. 

\begin{figure}
\centering
\includegraphics[width=\columnwidth]{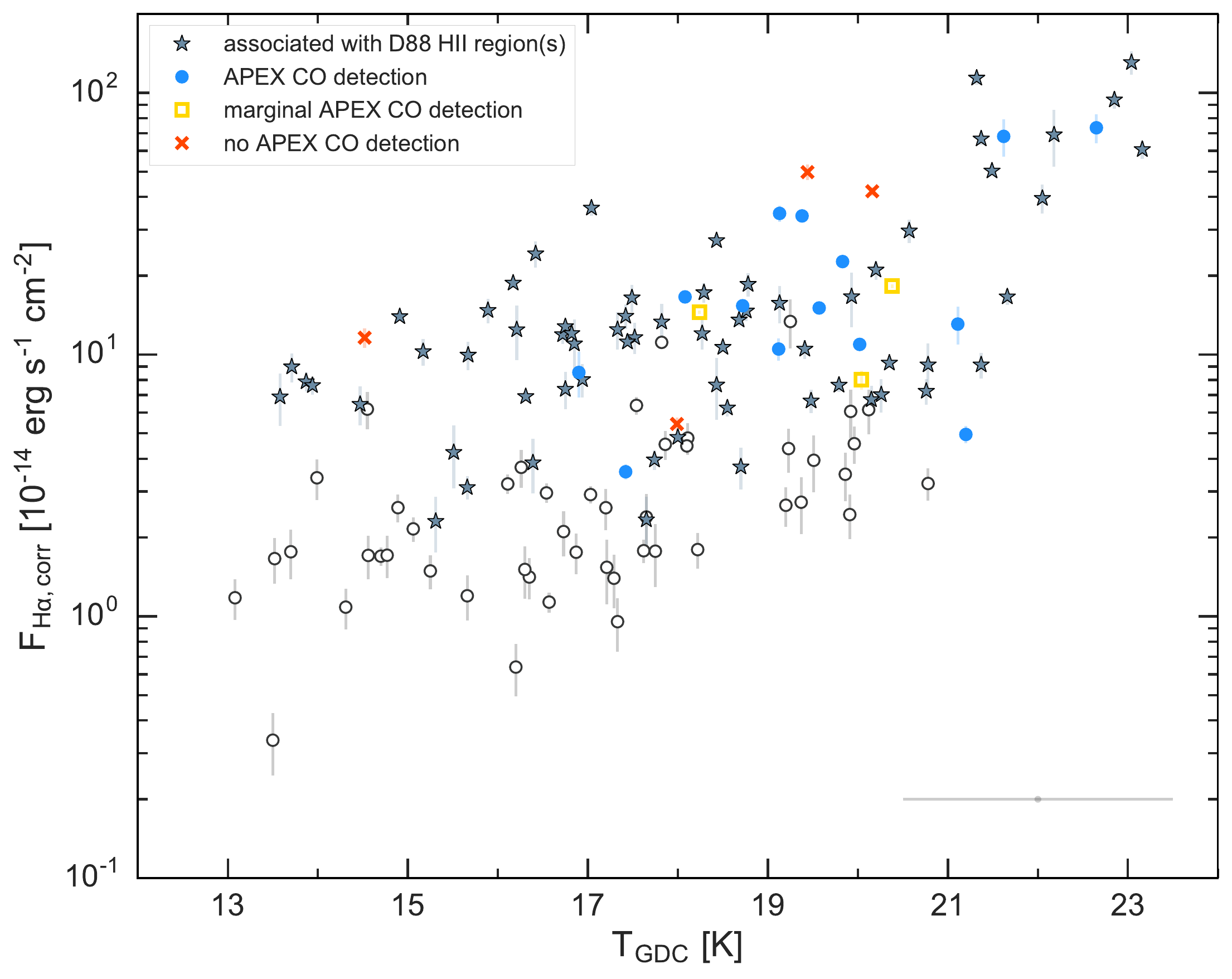}
\caption{Cold dust effective temperature vs. extinction-corrected H$\alpha$ emission for 134 of our GDCs with temperature measurements and reliable flux measurements (S/N~$>3$).
The coloured symbols all indicate GDCs that are associated with \citetalias{Deharveng1988} HII~regions.
The GDCs that also coincide with regions that were observed by APEX (see section \ref{sec:apexcomp}) are indicated by the blue filled circles (CO detections), yellow unfilled squares (marginal CO detections), and red crosses (no CO detections).
The horizontal error bar in the lower right indicates the uncertainty in temperature that we assume for all GDCs. 
}
\label{fig:temp_vs_halpha}
\end{figure}

Figure \ref{fig:temp_vs_halpha} shows the cold dust effective temperature versus their associated extinction-corrected H$\alpha$ emission for the GDCs for which we could determine temperature and H$\alpha$ flux values.
The overall shallow increasing trend of higher cold dust effective temperature values with brighter associated H$\alpha$ emission (with a Pearson correlation coefficient of 0.56) likely reflects that the massive stars that created the HII~regions are also heating the dust.
Below a cold dust effective temperature value of 21~K both GDCs associated with HII~regions and GDCs not associated with HII~regions or associated only with very little or compact H$\alpha$ emission cover about the same range in dust effective temperature values, whereas GDCs with dust effective temperatures above 21~K are all associated with HII~regions. 
Figure \ref{fig:flux250_vs_halpha} and Figure \ref{fig:temp_vs_halpha} also again show that by targeting HII~regions the parameter range in flux and effective cold dust temperature of the associated GDC population can be  sampled well.

\subsection{Total dust mass of NGC~300}

We used the aperture photometry method of the \textit{photutils} package to derive total background subtracted fluxes of NGC~300 at 250~$\mu$m.
We summed up the flux inside the region of interest that we also used for \textit{getsources}, and estimated a median value for the subtracted background per pixel from an annulus around the region of interest (inner radius: 7.2~kpc; outer radius: 8.8~kpc).
We get a total flux value of 97.46~Jy at 250~$\mu$m for NGC~300.
We compared this to the sum of the total flux values of the GDCs from our catalogue (15.76 $\pm$ 1.91~Jy); we adopted the upper limit of three times the error as flux values for the GDCs where the uncertainty exceeded the flux estimate. We find that about 16\% of the total flux of NGC~300 in SPIRE-250 belongs to the GDCs of our catalogue. 

\begin{figure}
\centering
\includegraphics[width=\columnwidth]{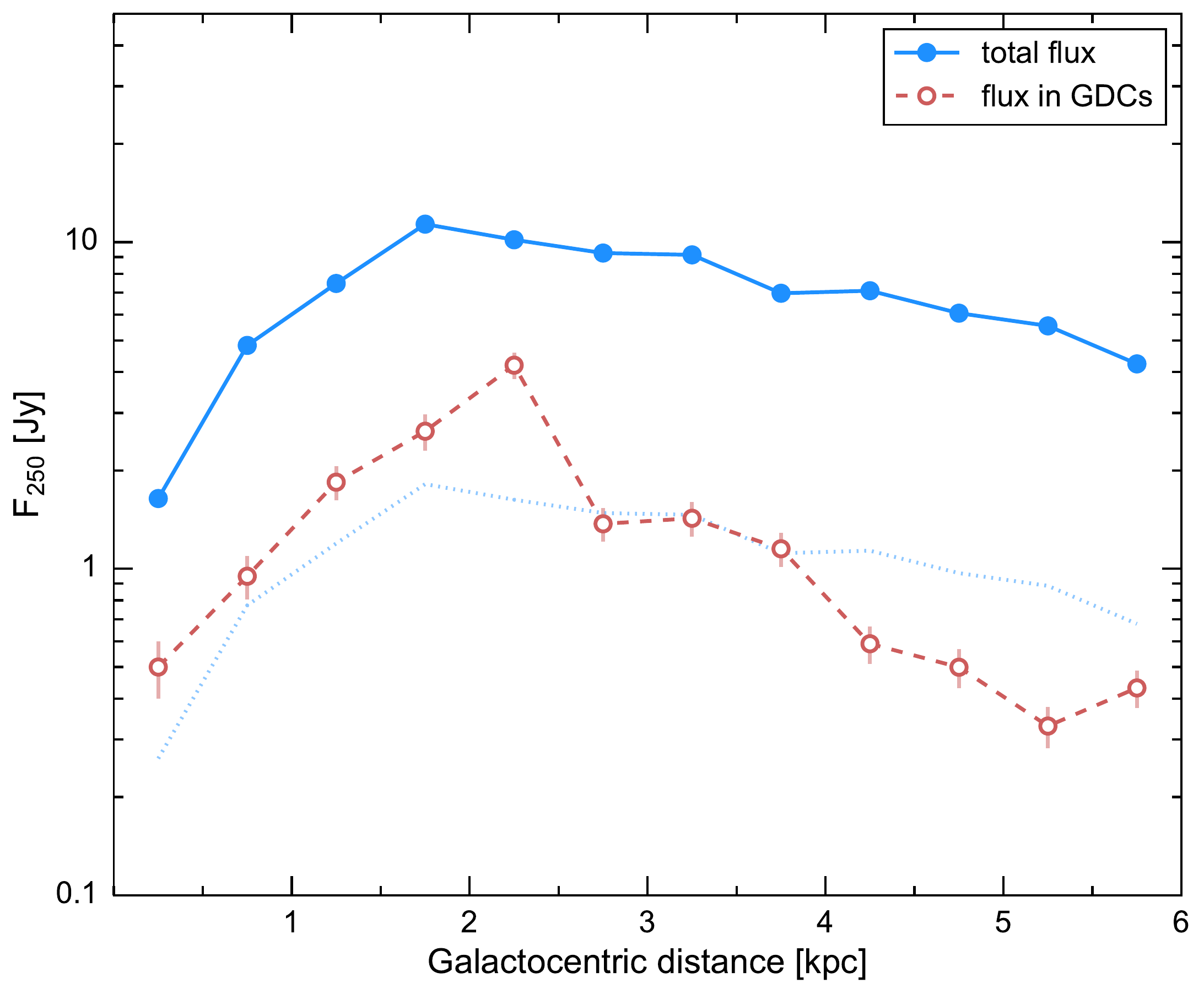}
\caption{Comparison of the fraction of the total flux residing in GDCs as a function of galactocentric distance. 
The blue filled circles and solid line show the total background-subtracted flux at 250~$\mu$m binned in radial distances of 0.5~kpc.
The red open circles and dashed line show the flux at 250~$\mu$m due to GDCs in the same radial bins. 
The blue dotted line indicates 16\% of the total flux, which is the average fraction of the the total flux residing in GDCs. 
}
\label{fig:total_vs_gdc_flux}
\end{figure}

Figure \ref{fig:total_vs_gdc_flux} shows the comparison between the total flux of the galaxy (blue filled circles and solid line) versus the flux that is in GDCs (red open circles and dashed line) for the SPIRE-250 band. 
For the total flux values, we performed aperture photometry for annuli at every 0.5~kpc and subtracted the background as detailed above. 
To get the fraction of the total flux that is in GDCs, we summed up the flux values at 250~$\mu$m of all GDCs whose central positions were located inside the annuli.
For the GDCs with a S/N ratio below 3 we used the upper limits as given in Table \ref{Tab:gdcs}.
In Figure \ref{fig:total_vs_gdc_flux} we also indicate the 16\% average of the total flux that is due to the GDCs (blue dotted line).
In the inner part of the galaxy ($<$ 2.75~kpc) the fraction of the total flux in GDCs is higher (with a maximum of 41\% in the radial bin at 2.25~kpc) than in the outer part, where the fraction of total flux in GDCs drops to as low as 6\% in the radial bin at 5.25~kpc. 

Assuming the median cold dust effective temperature of $16.7\pm 1.5$~K we derived for NGC 300, we find a total dust mass of 5.4$\cdot$10$^{6}$~\msun ~($\pm$1.8$\cdot$10$^{6}$~\msun). 
Given a total stellar mass of 2.1$\cdot$10$^{9}$~\msun ~(see Table \ref{tbl:ngc300}) this yields a dust-to-stellar mass ratio of about 0.25\%.

\subsection{Comparisons to other nearby galaxies}

We compared our results for NGC~300 to other nearby galaxies that offer the advantage of higher spatial resolution, in particular the morphologically similar M33.
Given an estimate of the stellar mass of M33 of about $3-6\cdot$10$^{9}$~\msun~\citep{Corbelli2003} and total dust mass estimates for M33 of about $1\cdot$10$^{7}$~\msun ~\citep{Kramer2010} to $1.67\cdot$10$^{7}$~\msun ~\citep{Hermelo2016}, the ratio of the dust mass to the stellar mass is very similar in M33 and NGC~300. 
The global SFR of M33 is $0.45\pm0.10$~\msun~yr$^{-1}$ \citep{Verley2009}, which is about 3-5 times higher than the SFR values estimated for NGC~300 \citep{Helou2004}.
However, the total HI~mass seems to be very similar in M33 \citep{Gratier2010} and NGC~300 \citep{Westmeier2011}, which makes predictions about the total molecular gas mass and the global dust-to-gas ratio in NGC~300 in relation to M33 difficult at this point. 
Reliable estimates of the total molecular gas mass in NGC~300 thus have to await further galaxy-wide molecular observations for this galaxy. 

For M31, the ratio of the total dust mass ($5.4\cdot$10$^{7}$~\msun, \citealp{Draine2014}) to the stellar mass ($10-15\cdot$10$^{10}$~\msun, \citealp{Tamm2012}) is about 5-6 times lower than in NGC~300 and M33. 
If we neglect the contribution of the bulge to the total stellar mass ($\sim$30\%; \citealp{Tamm2012}) and the dust mass (0.5\%; \citealp{Groves2012}), the total dust mass to stellar mass ratio in M31 is still by a factor of about 3 lower than in NGC~300 and M33.

At the time of writing we are aware of two other studies that used \textit{getsources} in an extragalactic context to derive FIR source catalogues based on \textit{Herschel} observations. 
\citet{Foyle2013} used \textit{getsources} to produce a list of compact dust sources for the Southern Pinwheel Galaxy (M83; distance of 4.5~Mpc).
These authors found 90 sources with dust masses in the range of 10$^{4}$ to 10$^{6}$~\msun ~and they associate most of these sources with giant molecular associations (GMAs). 
In another study, \citet{Natale2014} analysed \textit{Herschel} observations of M33 (distance of 0.859~Mpc) and created a catalogue of 183 clumpy FIR-bright sources with dust masses from $\sim$10$^{2}$ to 10$^{4}$~\msun ~that they associate with GMCs and GMAs.
Since the spatial scales we probe in NGC~300 are intermediate to those of \citet{Foyle2013} and \citet{Natale2014}, it is more likely that our GDCs will trace complexes of GMCs rather than individual clouds. 

In M33, the CO emission generally seems to follow the overall dust emission \citep{Braine2010} and dust sources detected in CO show higher dust luminosities \citep{Natale2014}.
In M31, the peaks of the dust continuum are also traced well by CO emission and the CO luminosity of GMCCs is correlated with the GMCC masses derived from \textit{Herschel} data \citep{Kirk2015}.
A relationship between the amount of dust and CO intensity was likewise found for regions across a range of metallicity in the Milky Way and in the Magellanic Clouds \citep{Lee2015}.
These results are in agreement with the general trend of increasing CO emission with higher dust emission that we find for our comparison of GDCs with GMCCs from \citetalias{Faesi2014} (see section \ref{sec:apexcomp}).
However, we find that GDCs that have about the same dust mass are not equally well detected in CO emission; this is similar to the results of \citet{Natale2014} for M33, who attribute this difference in CO detection mostly to an evolutionary effect of advanced gas dissipation for more evolved clouds.
Even though for NGC~300 our small sample size of GDCs that are directly comparable with CO observations does not allow us to address this result more conclusively, we note that in NGC~300 the metallicity gradient and the corresponding radial increase of the CO-to-H$_{2}$ conversion factor \citepalias{Faesi2014} is likely the most important reason for the difference in CO detections for similar GDCs. 

In M33, \citet{Verley2010} found a good correlation between the 250~$\mu$m emission and associated H$\alpha$ emission of compact dust sources; these authors found that the FIR emission delineates shell structures around HII~regions.
\citet{Anderson2012} also found that the associated FIR emission of Galactic HII~regions is dominated by the cold dust in their photodissociation regions. 
These results are in agreement with our findings in section \ref{sec:hii_correl} that the majority of our GDCs in NGC~300 are associated with HII~regions, which in turn implies that those GDCs are also associated with complexes of molecular clouds. 

\citet{Gratier2012} found that clouds in M33 that are associated with massive star formation have higher CO luminosities, which is in agreement with the results of \citet{Natale2014}, who found that FIR sources in M33 that were detected in CO have on average higher dust temperatures and higher H$\alpha$ luminosities. 
However, \citet{Lee2015} found that for regions in the LMC with higher dust temperatures the CO emission for a given amount of dust is smaller, which they attribute to environmental effects such as increased photodissociating radiation fields.
Even though GDCs that are associated with HII~regions in NGC~300 generally also have higher dust temperatures (see Figure \ref{fig:dist_vs_temp} and Figure \ref{fig:temp_vs_halpha}), we do not find any significant trend of dust temperature or H$\alpha$ luminosity with CO intensity for the 21 GDCs that we could directly compare with GMCCs from \citetalias{Faesi2014}.
However, all the \citetalias{Faesi2014} GMCCs are linked to high-mass star formation and follow-up molecular gas observations of GDCs without associated HII~regions are necessary for a more conclusive comparison.

\subsection{Implications for the formation of molecular gas in NGC~300}

\citet{Krumholz2009} found that dust shielding and H$_{2}$ self-shielding are both nearly equally important for the formation of molecular clouds in essentially all galactic environments.
\citet{Krumholz2009} also found that the molecular fraction in a galaxy is nearly independent of the strength of the interstellar radiation field and mainly determined by its column density and to a lesser degree by its metallicity. 
In their study of the relationship between dust and CO intensity in the Milky Way and the Magellanic Clouds, \citet{Lee2015} suggest that dust shielding is the dominant factor determining the distribution of bright CO emission. 

\begin{figure}
\centering
\includegraphics[width=\columnwidth]{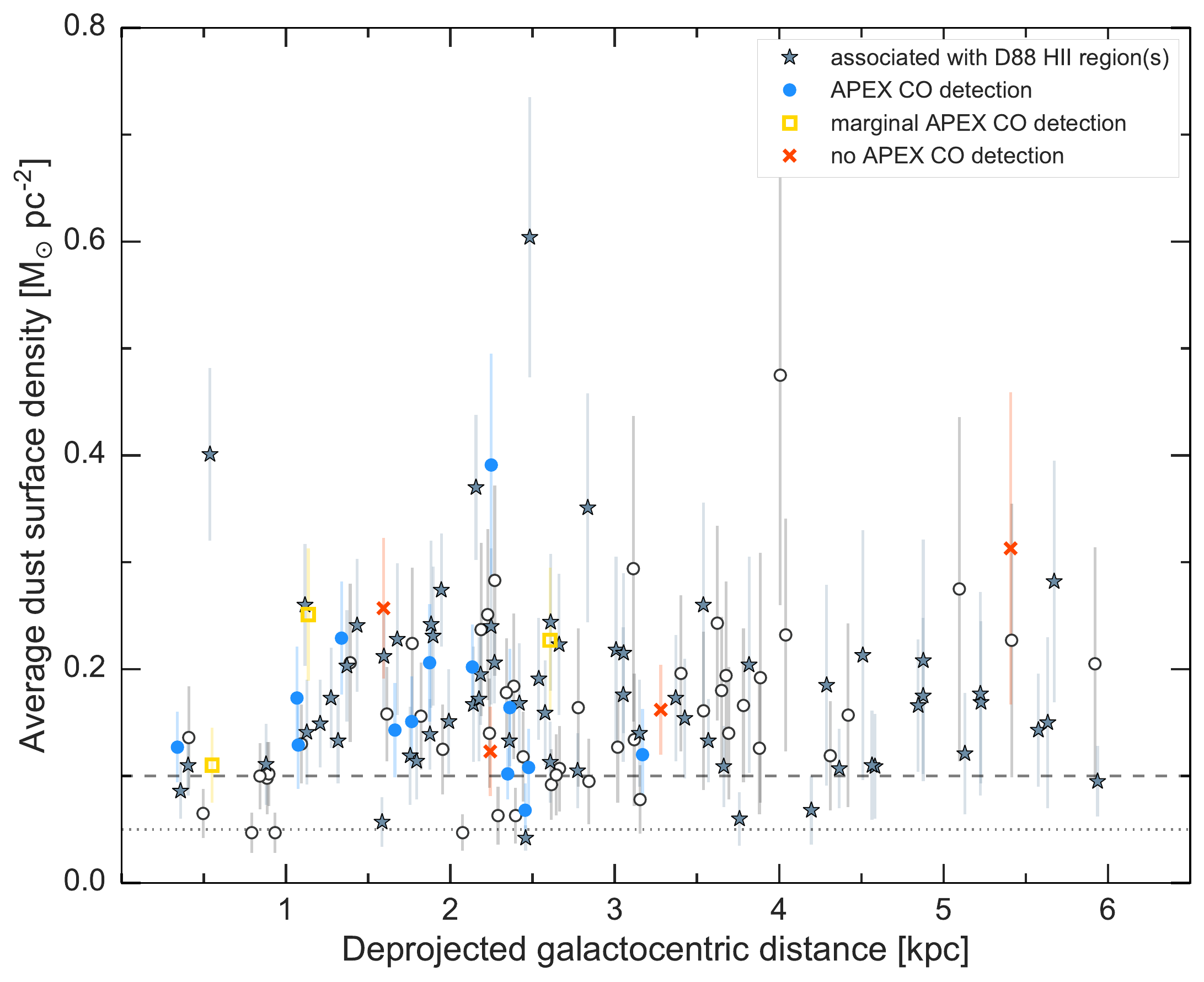}
\caption{Plot of the average dust surface density against the deprojected galactocentric distances for \massAll ~of our GDCs with reliable mass measurements.
The coloured symbols all indicate GDCs that are associated with \citetalias{Deharveng1988} HII~regions.
The GDCs that also coincide with regions that were observed by APEX (see section \ref{sec:apexcomp}) are indicated by the blue filled circles (CO detections), yellow unfilled squares (marginal CO detections), and red crosses (no CO detections).
The error bars denote the uncertainty in the dust surface density estimates.
The dotted and dashed horizontal lines indicate constant dust surface density values of 0.05 and 0.1~\msun~pc$^{-2}$, respectively.
At least 87\% of the sample have dust surface densities sufficient for the formation of molecular gas (see the text).
}
\label{fig:dist_vs_surf_dens}
\end{figure}

To assess the impact of dust shielding on the formation of molecular gas in NGC~300, we calculated the dust surface density of our GDCs by simply dividing their dust masses through their elliptical footprint defined at the SPIRE-250 image.
Figure \ref{fig:dist_vs_surf_dens} shows these average dust surface density values of the 134~GDCs for which we have mass estimates plotted against their galactocentric distance. 
For 97\% of these GDCs the average dust surface density exceeds 0.05~\msun~pc$^{-2}$ and for about 87\% it exceeds a value of 0.1~\msun~pc$^{-2}$.
\citet{Roman-Duval2014} found that the atomic-to-molecular transition in the Magellanic Clouds is located at dust surface densities of about 0.03 to 0.05~\msun~pc$^{-2}$.
Given that the metallicity in the outskirts of NGC~300 is comparable to the metallicity values of the Magellanic Clouds \citep{Bresolin2009}, we would expect that the transition from the atomic to molecular phase also occurs at similar values of the dust surface density, suggesting, in turn, that the vast 
majority of GDCs in the outer regions of NGC~300 harbour the potential to form molecular clouds.
Towards the centre of NGC~300, the metallicity reaches values intermediate to the LMC and the Sun \citep{Bresolin2009}.
For solar metallicity the transition from atomic to molecular gas occurs at a characteristic shielding column of $\Sigma_\mathrm{HI}$ $\sim$ 10 \msun ~pc$^{-2}$ \citep{Krumholz2009}, which compares favourably with the average HI surface density  in NGC~300 found by \citet{Westmeier2011} and suggests that most of the GDCs in the central part of NGC~300 are also harbouring molecular gas.

The points discussed in this and the previous section together with the comparison in section \ref{sec:apexcomp} gives us confidence that most of the GDCs of our catalogue will indeed be associated with GMCCs.
However, new galaxy-wide CO observations centred on the peaks of the FIR dust emission are required for a more detailed and conclusive comparison between the dust and molecular gas in NGC~300.


\section{Summary}

We present the first comprehensive study of the giant dust cloud (GDC) population throughout the spiral galaxy NGC~300 based on images obtained by the \textit{Herschel Space Observatory} covering a wavelength range from 100-500~$\mu$m.
For the source detection we used \textit{getsources}, a multiwavelength source extraction algorithm developed for \textit{Herschel} observations.
We based the final selection of GDCs for our catalogue on a visual comparison of the \textit{getsources} source candidates with additional archival observations in the optical (by the MPG/ESO 2.2 m telescope and the Hubble Space Telescope) and the FUV (by \textit{GALEX}).
Using the H$\alpha$ observations by the MPG/ESO-2.2 m telescope, we slightly improved the currently most complete catalogue of HII~regions in NGC~300 by \citet{Deharveng1988} and spatially correlated its HII~regions with GDCs from our catalogue.
We derived cold dust effective temperatures for the GDCs from a two-component SED-fitting from 24 to 500~$\mu$m wavelength range, assuming a value of 1.7 for the power law exponent $\beta$ of the dust emissivity.
We used the cold dust effective temperature to derive dust mass estimates based on the SPIRE-250 total fluxes. 

We used aperture photometry measurements and the median dust effective temperature throughout NGC~300 (16.7~K) to derive an estimate of its total dust mass. 

In the following, we summarize our main results and conclusions. 

   \begin{enumerate}
      \item We compiled a catalogue of \selSrcAll ~GDCs, for which we give the positions, total flux estimates in all five \textit{Herschel} bands, size estimates, cold dust effective temperatures, dust mass estimates, flux values at 24~$\mu$m, H$\alpha$ emission, and correlation with HII~regions from the catalogue of \citet{Deharveng1988} in Table \ref{Tab:gdcs}. 

      \item The GDCs in our catalogue cover a cold dust effective temperature range of $\sim$~13-23~K, showing a distinct temperature gradient from the centre to the outer parts of the stellar disk.
      
      \item We found the total dust mass of the galaxy to be 5.4$\cdot$10$^{6}$~\msun ~($\pm$1.8$\cdot$10$^{6}$~\msun), of which about 16\% of this mass is attributable to GDCs of our catalogue. 
      We determined individual dust masses for \massAll ~of the GDCs. 
      These masses range from $\sim$~\massMin ~to \massMax, and the median mass is \massMed .
      The GDC masses are not significantly increasing with galactocentric radius.
      
      \item The power law index of the truncated power law fit to the cumulative dust mass distribution of our GDCs ($\gamma = 3.2\pm 0.4$) is statistically indistinguishable from the power law index of the mass spectrum obtained by \citet{Faesi2014} for giant molecular cloud complexes (GMCCs) in NGC~300 ($\gamma = 2.7\pm 0.5$).
      
      \item We found that about 62\% of our GDCs are associated with HII~regions from \citet{Deharveng1988}. 
      The GDCs with associated HII~regions are found in the entire stellar disk of NGC~300 and across the full range of dust masses that we probe.
      Our results thus suggest that targeting HII~regions in nearby galaxies for associated GDC structures is a good way to sample the total GDC population of the galaxy.
      
      \item For GDCs associated with HII~regions, the 250~$\mu$m and extinction-corrected H$\alpha$ fluxes are reasonably well correlated. 
      We also found a shallow trend of higher cold dust effective temperature values with brighter associated H$\alpha$ emission that is likely due to the heating of the dust by the massive stars of the HII~regions.
      
      \item We compared a subsample of our GDCs to previous pointed APEX CO (2-1) observations from \citet{Faesi2014} and found that GDCs with brighter dust emission are in general also associated with brighter CO emission.
Based on the average dust mass surface densities of the GDCs, we argue that at least 87\% of these GDCs provide enough shielding to harbour GMCs. 
Additional comparisons with the results from other nearby galaxies also suggests that most of our GDCs are associated with GMC complexes.
New galaxy-wide molecular gas observations will help to further elucidate the important relationship between dust structures and molecular clouds in NGC~300. 
      
   \end{enumerate}

\begin{acknowledgements}
      This work is based in part on observations made with Herschel, a European Space Agency Cornerstone Mission with significant participation by NASA. 
      Support for this work was provided by NASA through an award issued by JPL/Caltech through contract RSA:1472984.
      This project has also received funding from the European Union’s Horizon 2020 research and innovation programme under grant agreement No 639459 (PROMISE). 
      This paper is partly based on the master thesis of M. Riener.
      This work made use of Montage, funded by the National Science Foundation under Grant Number ACI-1440620, and previously funded by the National Aeronautics and Space Administration's Earth Science Technology Office, Computation Technologies Project, under Cooperative Agreement Number NCC5-626 between NASA and the California Institute of Technology; NASA's Astrophysics Data System Bibliographic Service; SAOImage DS9, developed by Smithsonian Astrophysical Observatory.
      We wish to thank the anonymous referee for helpful comments that improved this paper.
      We would also like to thank Alexander Men'shchikov for his help with questions about the source extraction with \textit{getsources} and the evaluation of its results.
      
      \\\textbf{Code bibliography}: Astropy \citep{astropy}; Matplotlib \citep{Hunter2007}; APLpy \citep{APLpy}; Photutils \citep{Photutils}.
\end{acknowledgements}

\bibliographystyle{aa} 
\bibliography{ngc300_gathering_dust} 

\begin{appendix} 

\section{Astrometric correction of the D88 HII region catalogue}

\label{cha:appendix_a}

\begin{figure*}
\centering
\includegraphics[width=0.9\textwidth]{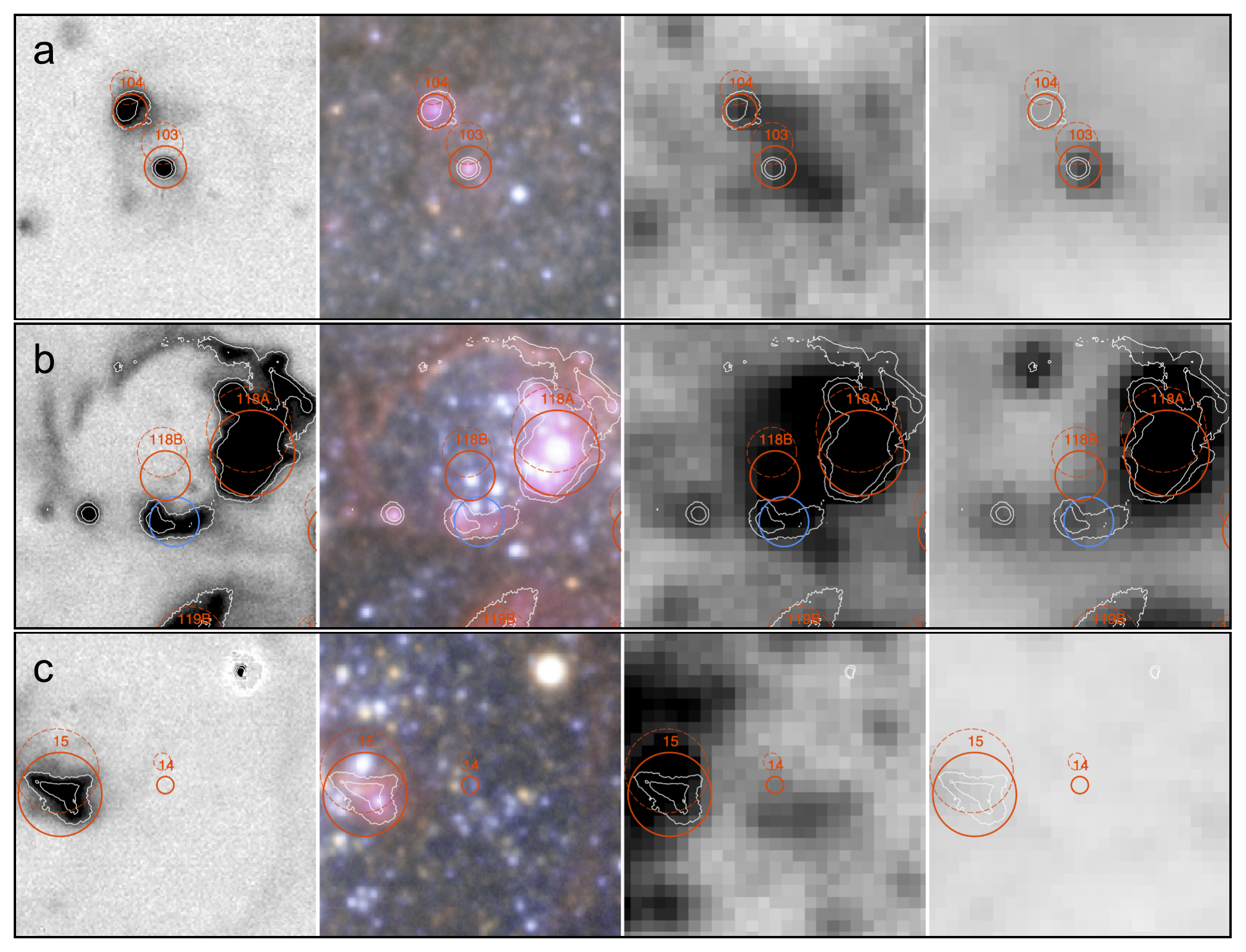}
\caption{(\textit{Left to Right:}) WFI H$\alpha$ map \citepalias{Faesi2014}; ESO/WFI image (Blue: 475~nm and OIII; Green: 605~nm; Orange: SII; Red: H$\alpha$. Credit: ESO); \textit{GALEX} FUV image; and \textit{Spitzer}/MIPS~24~$\mu$m image.
a) Example of the astrometrical shift adopted for all HII regions. 
b) Example of an additionally manually shifted HII~region. 
Red lines indicate the position of the source before (\textit{dashed}) and after (\textit{solid}) the general astrometric shift, the blue line indicates the final adopted position.
c) Example of a discarded HII region showing no analogues in the newer observations.
White contours denote flux thresholds in H$\alpha$.}
\label{fig:appHii}
\end{figure*}

Here we discuss the corrections we made to the \citetalias{Deharveng1988} HII~region catalogue.
To verify the validity of the \citetalias{Deharveng1988} HII~regions, we overplotted these regions on a newer and more sensitive extinction corrected line-only H$\alpha$ map based on the ESO/WFI observations, an extinction corrected \textit{GALEX} FUV image, and a \textit{Spitzer}/MIPS~24~$\mu$m image (all three images are from \citetalias{Faesi2014}; see their $\S~2.2.$, $\S~2.3.$ and $\S~2.4.$ for details on how these images were processed). 
The massive stars responsible for creating the HII~regions emit a significant fraction of their flux at higher energies and should therefore also be visible in the \textit{GALEX} FUV image. 
The \textit{Spitzer}/MIPS~24~$\mu$m image can be used to check whether the UV flux may be absorbed by too much dust in the line of sight. 
We found that after transferring the positions of the HII~regions given in \citetalias{Deharveng1988} into the FK5 system an additional linear shift of $-0.034^{s}$ in right ascension and $-3.03"$ in declination was necessary to yield a good agreement to the newer H$\alpha$ map from \citetalias{Faesi2014} (see Figure \ref{fig:appHii}a).
Twelve sources needed an additional correction in their position to yield a good correspondence to emission features visible in the WFI H$\alpha$ map (see Figure \ref{fig:appHii}b). We list the final position of these 12 HII~regions in Table \ref{Tbl:hii-shift}.
We excluded six HII~regions identified by
\citetalias{Deharveng1988} because we could not find any emission features at their position in the \citetalias{Faesi2014}
H$\alpha$ map and FUV image (see Figure \ref{fig:appHii}c). 
The original \citetalias{Deharveng1988} catalogue numbers of these discarded HII~regions are 14, 22, 44, 89 NUCLEUS, 99, and 105.

\begin{table}[h!]
\small
\begin{center}
\renewcommand{\arraystretch}{1.2}
\caption{Final positions of the additionally shifted \citetalias{Deharveng1988} HII regions.}
\begin{tabular}{ccc}
\hline\hline
\# D88 & RA (J2000) & DEC (J2000) \\
\hline
57 & 0:54:44.500 & -37:36:40.56 \\
62 & 0:54:45.429 & -37:38:42.03 \\
69 & 0:54:48.026 & -37:43:33.21 \\
70 & 0:54:48.389 & -37:39:48.14 \\
73 & 0:54:49.461 & -37:44:52.91 \\
97 & 0:54:56.611 & -37:45:55.54 \\
102 & 0:54:57.364 & -37:44:02.01 \\
118B & 0:55:04.451 & -37:42:58.15 \\
121 & 0:55:03.919 & -37:46:26.14 \\
140 & 0:55:15.186 & -37:44:15.48 \\
149 & 0:55:27.579 & -37:40:55.82 \\
156 & 0:55:30.387 & -37:41:16.47 \\\hline
\end{tabular}
\label{Tbl:hii-shift}
\end{center}
\end{table}

\section{Catalogue of GDCs}
\label{app:catalogue}

Table \ref{Tab:gdcs} lists all \selSrcAll ~sources that we identified as GDCs. (01) gives the catalogue number of the sources: the footnotes indicate sources that (a) were labelled as tentative by \textit{getsources}, (b) were strongly deblended, (c) show an incorrect shift in their position, (d) were added from the SPIRE only source extraction, (e) are sources for which the effective temperature estimate is very uncertain, or (f) for which the given size estimate is likely too high (as the FWHM at 250~$\mu$m is larger than the FWHM at 350~$\mu$m). 
Column (02) and (03) give the central position of the GDCs and (04 - 08) gives the total flux measurements in all  five \textit{Herschel} bands. 
For sources for which the flux value was lower than three times the error only upper limits (defined as three times the error and indicated by a '<') are given. Columns
(09) and (10) give the size estimates of the major and minor axis of the source, respectively (corresponding to the SPIRE-250 FWHM); the rotation angle (east of the vertical axis) of the major axis is listed in column (11). 
Column (12) gives the deprojected galactocentric distance of the source (see $\S$ \ref{sec:deprojdist}), column (13) gives the size of the GDC as a diameter of an equivalent circular area, and column (14) lists the effective temperature estimate of the cold dust component of the sources. 
Column (15) is the mass estimate for the source calculated from the total flux in SPIRE-250 (07) and the cold dust effective temperature (14). 
Column (16) lists the flux values obtained from the \textit{Spitzer}/MIPS 24~$\mu$m image (see $\S$~\ref{sec:flux24mu}).
Column (17) gives the non-extinction-corrected H$\alpha$ fluxes determined from the line-only H$\alpha$ map from \citetalias{Faesi2014} (see $\S$~\ref{sec:hii_correl}). 
For GDCs where the flux value was lower than three times the uncertainty value only upper limits (defined as three times the uncertainty value and indicated by a '<') are given.
Column (18) lists the \citetalias{Deharveng1988} HII~regions associated with our GDCs (see $\S$~\ref{sec:hii_correl}). 
The superscripted letters in brackets indicate whether the HII~regions were observed by \citetalias{Faesi2014} with APEX and whether they show a detection (D), marginal detection (M), or non-detection (N) in CO(J=2-1). 

\onecolumn
\begin{landscape}
\scriptsize
\setlength\tabcolsep{5pt}\renewcommand{\arraystretch}{1.2}
\begin{longtable}{cccccccccccccccccc}
\caption {Source list of the GDCs in NGC 300}\label{Tab:gdcs}\\\hline\hline
\endfirsthead
\caption* {\textbf{Table \ref{Tab:gdcs} continued:} Source list of the GDCs in NGC 300}\\\hline\hline
$\#$ & RA & DEC & F$_{100}$ & F$_{160}$ & F$_{250}$ & F$_{350}$ & F$_{500}$ & A & B & $\Theta$ & D & d$_{\textrm{eq}}$ & T$_{\textrm{cold}}$& Dust mass & F$_{24}$ & F$_{\textrm{H}\alpha}$ & D88 HII \\
 & (J2000) & (J2000) & [mJy] & [mJy] & [mJy] & [mJy] & [mJy] & ["] & ["] & [$^{\circ}$] & [kpc] & [pc] & [K]& [10$^{3}$ M$_{\odot}$] & [mJy] & [10$^{-14}$ erg/s/cm$^{2}$] & regions \\
 (01) & (02) & (03) & (04) & (05) & (06) & (07) & (08) & (09) & (10) & (11) & (12) & (13) & (14)& (15) & (16) & (17) & (18)\\
\hline
\endhead
\LongtableFooter
\endfoot
\hline
\endlastfoot
$\#$ & RA & DEC & F$_{100}$ & F$_{160}$ & F$_{250}$ & F$_{350}$ & F$_{500}$ & A & B & $\Theta$ & D & d$_{\textrm{eq}}$ & T$_{\textrm{cold}}$& Dust mass & F$_{24}$ & F$_{\textrm{H}\alpha}$ & D88 HII \\
 & (J2000) & (J2000) & [mJy] & [mJy] & [mJy] & [mJy] & [mJy] & ["] & ["] & [$^{\circ}$] & [kpc] & [pc] & [K]& [10$^{3}$ M$_{\odot}$] & [mJy] & [10$^{-14}$ erg/s/cm$^{2}$] & regions \\
 (01) & (02) & (03) & (04) & (05) & (06) & (07) & (08) & (09) & (10) & (11) & (12) & (13) & (14)& (15) & (16) & (17) & (18)\\
\hline
1 & 13.54564 & -37.62894 & 85.2 $\pm7.7$ & 91.3 $\pm11.1$ & 57.4 $\pm6.5$ & $<$ 17.8 & 44.0 $\pm8.9$ & 22.7 & 17.6 & 48.3 & 5.23 & 192 & 14.9 & 4.9 $\pm2.2$ & 3.2 $\pm0.2$ & 12.7 $\pm0.1$ & 3$^{}$ \\
2 & 13.56791 & -37.58238 & $<$ 24.8 & 83.6 $\pm11.9$ & $<$ 21.3 & 62.6 $\pm8.7$ & 44.5 $\pm8.4$ & 20.3 & 17.6 & 57.0 & 5.77 & 181 & 17.0 & -- & 4.0 $\pm0.5$ & 34.8 $\pm2.3$ & 6$^{ N}$ \\
3\tablefootmark{a} & 13.56803 & -37.57702 & 87.6 $\pm14.0$ & 172.6 $\pm16.8$ & 67.7 $\pm10.6$ & 46.5 $\pm9.7$ & $<$ 24.0 & 25.2 & 18.5 & 28.8 & 5.94 & 207 & 17.5 & 3.2 $\pm1.1$ & 3.5 $\pm0.5$ & 10.3 $\pm1.4$ & 5$^{ N}$ \\
4\tablefootmark{f} & 13.57323 & -37.58564 & 126.9 $\pm7.7$ & 115.6 $\pm14.5$ & 85.4 $\pm8.6$ & 68.5 $\pm8.3$ & 48.5 $\pm7.7$ & 27.7 & 17.6 & 142.3 & 5.58 & 212 & 16.4 & 5.1 $\pm1.9$ & 2.5 $\pm0.4$ & 23.3 $\pm2.6$ & 6$^{ N}$, 9$^{ N}$ \\
5 & 13.58374 & -37.65966 & 40.3 $\pm7.7$ & 72.5 $\pm11.5$ & 128.8 $\pm7.4$ & 105.9 $\pm6.5$ & 60.8 $\pm14.1$ & 19.2 & 17.6 & 103.7 & 4.01 & 176 & 14.7 & 11.6 $\pm5.3$ & 1.1 $\pm0.2$ & 1.3 $\pm0.1$ & -- \\
6\tablefootmark{b} & 13.59496 & -37.57401 & -- & -- & 122.7 $\pm8.4$ & 68.1 $\pm9.6$ & 29.4 $\pm6.5$ & 24.0 & 17.6 & 95.0 & 5.67 & 197 & 15.7 & 8.6 $\pm3.4$ & 2.5 $\pm0.3$ & 9.0 $\pm1.1$ & 11$^{}$, 12$^{}$ \\
7\tablefootmark{a,f} & 13.5958 & -37.67845 & 59.4 $\pm11.3$ & 83.1 $\pm13.2$ & 38.0 $\pm8.0$ & $<$ 16.1 & $<$ 21.4 & 30.1 & 17.6 & 132.5 & 3.76 & 221 & 16.3 & 2.3 $\pm1.0$ & 1.8 $\pm0.2$ & 6.2 $\pm0.3$ & 10$^{}$, 13$^{ N}$ \\
8 & 13.59769 & -37.63838 & 76.8 $\pm19.6$ & 112.1 $\pm16.7$ & 84.5 $\pm10.1$ & 50.3 $\pm5.7$ & $<$ 34.9 & 28.3 & 19.4 & 66.7 & 3.78 & 225 & 15.2 & 6.6 $\pm2.9$ & 1.3 $\pm0.2$ & 1.0 $\pm0.1$ & -- \\
9 & 13.60428 & -37.71571 & 61.9 $\pm15.9$ & 119.7 $\pm15.2$ & 50.9 $\pm11.3$ & 36.2 $\pm8.8$ & $<$ 31.3 & 20.8 & 17.6 & 169.3 & 4.31 & 184 & 16.2 & 3.2 $\pm1.4$ & 0.7 $\pm0.1$ & 0.4 $\pm0.1$ & -- \\
10 & 13.60705 & -37.6535 & 153.0 $\pm13.6$ & 159.9 $\pm14.6$ & 108.0 $\pm13.8$ & 48.7 $\pm8.8$ & $<$ 23.7 & 23.2 & 17.6 & 103.8 & 3.37 & 194 & 17.5 & 5.1 $\pm1.7$ & 3.0 $\pm0.4$ & 15.3 $\pm1.8$ & 17$^{ N}$ \\
11\tablefootmark{f} & 13.61263 & -37.72462 & $<$ 41.0 & 76.9 $\pm15.5$ & 120.7 $\pm9.3$ & 79.3 $\pm7.6$ & 45.3 $\pm9.2$ & 33.8 & 25.7 & 150.4 & 4.37 & 283 & 16.7 & 6.7 $\pm2.4$ & 2.5 $\pm0.5$ & 11.0 $\pm0.9$ & 18$^{}$, 19$^{}$ \\
12\tablefootmark{a} & 13.61298 & -37.63236 & $<$ 57.3 & $<$ 22.9 & 62.2 $\pm11.1$ & 33.3 $\pm5.8$ & $<$ 19.6 & 24.9 & 17.6 & 33.6 & 3.54 & 201 & 15.1 & 5.1 $\pm2.4$ & 0.9 $\pm0.1$ & 1.8 $\pm0.2$ & -- \\
13 & 13.61434 & -37.66604 & 452.7 $\pm45.3$ & $<$ 153.0 & $<$ 91.7 & 67.6 $\pm14.3$ & $<$ 30.3 & 38.3 & 18.1 & 22.1 & 3.13 & 253 & 16.8 & -- & 3.1 $\pm0.3$ & 11.6 $\pm0.7$ & 20$^{}$ \\
14 & 13.61444 & -37.69937 & 161.7 $\pm13.8$ & 126.3 $\pm17.9$ & 97.3 $\pm10.0$ & 66.8 $\pm9.9$ & 50.8 $\pm14.1$ & 21.9 & 17.6 & 87.7 & 3.57 & 188 & 18.7 & 3.7 $\pm1.1$ & 2.4 $\pm0.4$ & 2.8 $\pm0.5$ & 21$^{}$ \\
15 & 13.61452 & -37.67897 & 49.8 $\pm14.0$ & $<$ 63.1 & $<$ 42.4 & $<$ 25.6 & $<$ 22.3 & 28.5 & 17.6 & 166.9 & 3.20 & 215 & 17.0 & -- & 1.3 $\pm0.2$ & 2.4 $\pm0.2$ & -- \\
16\tablefootmark{a} & 13.61924 & -37.65659 & $<$ 22.9 & $<$ 29.7 & 61.3 $\pm12.8$ & $<$ 18.1 & $<$ 21.6 & 21.8 & 17.6 & 87.9 & 3.02 & 188 & 16.5 & 3.5 $\pm1.4$ & 1.2 $\pm0.2$ & 2.5 $\pm0.2$ & -- \\
17 & 13.6195 & -37.69322 & 147.2 $\pm9.7$ & 177.4 $\pm14.2$ & 140.3 $\pm9.9$ & 88.3 $\pm9.9$ & 58.6 $\pm14.0$ & 21.9 & 18.3 & 143.2 & 3.28 & 192 & 19.4 & 4.7 $\pm1.2$ & 5.4 $\pm0.6$ & 47.6 $\pm3.2$ & 24$^{ N}$ \\
18\tablefootmark{b} & 13.62406 & -37.73873 & -- & -- & 60.8 $\pm12.7$ & 41.5 $\pm9.2$ & 45.8 $\pm10.6$ & 30.9 & 17.6 & 84.2 & 4.58 & 224 & 15.7 & 4.3 $\pm1.9$ & 1.0 $\pm0.1$ & 2.7 $\pm0.3$ & 28$^{}$ \\
19 & 13.62783 & -37.6263 & 57.1 $\pm8.5$ & 70.8 $\pm12.3$ & 90.5 $\pm7.9$ & 46.7 $\pm5.6$ & $<$ 28.2 & 22.2 & 17.6 & 116.8 & 3.40 & 190 & 16.3 & 5.5 $\pm2.1$ & 1.9 $\pm0.3$ & 3.0 $\pm0.5$ & -- \\
20 & 13.63348 & -37.63153 & 37.1 $\pm8.7$ & 112.0 $\pm13.6$ & 82.9 $\pm10.0$ & 50.6 $\pm6.5$ & $<$ 19.8 & 25.1 & 17.6 & 11.4 & 3.15 & 202 & 16.9 & 4.5 $\pm1.6$ & 2.8 $\pm0.5$ & 9.9 $\pm2.4$ & 30$^{ D}$, 31$^{ M}$ \\
21 & 13.63383 & -37.64379 & 261.0 $\pm15.9$ & 288.6 $\pm17.4$ & 242.2 $\pm11.7$ & 136.2 $\pm10.0$ & 89.0 $\pm11.1$ & 24.1 & 17.6 & 118.7 & 2.84 & 198 & 17.8 & 10.8 $\pm3.3$ & 5.1 $\pm0.8$ & 11.4 $\pm2.0$ & 32$^{}$, 34$^{ D}$ \\
22\tablefootmark{e} & 13.63586 & -37.57177 & $<$ 23.0 & 63.5 $\pm14.1$ & 40.8 $\pm5.7$ & 18.5 $\pm5.5$ & $<$ 15.6 & 18.1 & 17.6 & 122.9 & 5.41 & 171 & 13.5 & 5.2 $\pm2.9$ & 0.4 $\pm0.1$ & 0.2 $\pm0.1$ & -- \\
23\tablefootmark{f} & 13.63739 & -37.72423 & 132.2 $\pm24.5$ & 95.5 $\pm20.4$ & 94.4 $\pm19.7$ & 59.1 $\pm14.8$ & $<$ 43.3 & 35.4 & 18.6 & 160.3 & 3.69 & 246 & 15.7 & 6.7 $\pm3.0$ & 1.3 $\pm0.2$ & 0.7 $\pm0.1$ & -- \\
24\tablefootmark{d} & 13.63845 & -37.73788 & 66.3 $\pm9.8$ & 82.2 $\pm13.7$ & 38.3 $\pm12.0$ & $<$ 18.6 & $<$ 26.7 & 22.1 & 20.3 & 38.4 & 4.20 & 203 & 16.5 & 2.2 $\pm1.0$ & 1.5 $\pm0.2$ & $<$ 7.5 & 33$^{}$ \\
25 & 13.63987 & -37.67216 & $<$ 44.8 & 113.7 $\pm19.0$ & 58.8 $\pm16.5$ & $<$ 28.1 & $<$ 24.0 & 34.5 & 17.6 & 124.0 & 2.40 & 237 & 17.5 & 2.8 $\pm1.2$ & 2.1 $\pm0.4$ & 5.6 $\pm0.4$ & -- \\
26 & 13.64376 & -37.70416 & 76.7 $\pm12.3$ & 69.2 $\pm13.2$ & 39.3 $\pm9.3$ & $<$ 36.8 & $<$ 25.2 & 18.7 & 17.6 & 158.2 & 2.84 & 174 & 16.6 & 2.3 $\pm0.9$ & 1.3 $\pm0.2$ & 0.6 $\pm0.1$ & -- \\
27 & 13.64488 & -37.68625 & 126.7 $\pm18.8$ & 116.0 $\pm16.6$ & 87.5 $\pm15.4$ & 47.7 $\pm14.2$ & $<$ 47.0 & 18.7 & 17.6 & 50.1 & 2.39 & 174 & 17.2 & 4.4 $\pm1.6$ & 1.8 $\pm0.4$ & 0.8 $\pm0.2$ & -- \\
28 & 13.64645 & -37.66143 & 349.6 $\pm23.1$ & 265.1 $\pm28.9$ & 167.9 $\pm16.4$ & 101.0 $\pm11.8$ & 71.2 $\pm15.8$ & 22.5 & 17.6 & 24.1 & 2.25 & 191 & 18.3 & 6.9 $\pm2.1$ & 7.3 $\pm1.1$ & 9.2 $\pm1.2$ & 37$^{ D}$ \\
29 & 13.64959 & -37.66416 & 129.1 $\pm11.8$ & 138.7 $\pm18.5$ & 95.2 $\pm15.9$ & $<$ 36.2 & $<$ 40.7 & 17.6 & 17.6 & 153.7 & 2.14 & 169 & 18.5 & 3.7 $\pm1.2$ & 5.4 $\pm0.7$ & 8.6 $\pm0.3$ & 38$^{}$ \\
30\tablefootmark{f} & 13.65335 & -37.65024 & 116.7 $\pm18.3$ & 236.3 $\pm26.5$ & 206.3 $\pm12.5$ & 119.6 $\pm10.9$ & 49.1 $\pm14.0$ & 25.6 & 18.2 & 17.1 & 2.27 & 207 & 17.6 & 9.5 $\pm3.0$ & 3.3 $\pm0.5$ & 0.5 $\pm0.1$ & -- \\
31 & 13.6589 & -37.69589 & 187.5 $\pm10.4$ & 202.7 $\pm19.6$ & 111.0 $\pm20.3$ & 59.8 $\pm16.8$ & $<$ 58.4 & 20.2 & 17.6 & 80.5 & 2.17 & 181 & 18.4 & 4.4 $\pm1.5$ & 12.3 $\pm1.1$ & 22.5 $\pm0.5$ & 39$^{}$ \\
32 & 13.66063 & -37.6904 & 68.8 $\pm12.3$ & 111.6 $\pm21.3$ & 121.8 $\pm19.8$ & 93.7 $\pm16.4$ & 83.6 $\pm19.5$ & 21.6 & 20.1 & 64.1 & 1.99 & 200 & 18.6 & 4.8 $\pm1.5$ & 6.3 $\pm1.0$ & 3.8 $\pm0.3$ & 41$^{ D}$ \\
33 & 13.66166 & -37.71191 & 108.9 $\pm14.0$ & 232.2 $\pm24.5$ & 83.5 $\pm14.5$ & 60.1 $\pm13.9$ & $<$ 52.1 & 23.7 & 17.6 & 85.1 & 2.61 & 196 & 18.3 & 3.4 $\pm1.1$ & 3.6 $\pm0.6$ & 15.9 $\pm1.3$ & 40$^{}$ \\
34 & 13.66296 & -37.64646 & 67.9 $\pm11.3$ & 125.1 $\pm11.8$ & 153.4 $\pm9.6$ & 99.1 $\pm9.0$ & $<$ 48.8 & 25.5 & 18.9 & 85.7 & 2.19 & 211 & 16.9 & 8.2 $\pm2.8$ & 2.5 $\pm0.4$ & 0.8 $\pm0.1$ & -- \\
35 & 13.66828 & -37.66738 & 178.0 $\pm14.9$ & 249.8 $\pm25.8$ & 161.3 $\pm19.2$ & 73.7 $\pm14.2$ & $<$ 22.7 & 27.2 & 17.6 & 82.0 & 1.61 & 210 & 19.4 & 5.5 $\pm1.5$ & 4.4 $\pm0.7$ & 1.0 $\pm0.3$ & -- \\
36 & 13.66902 & -37.68159 & 283.7 $\pm15.5$ & 279.0 $\pm26.0$ & 191.9 $\pm19.3$ & 99.1 $\pm15.7$ & $<$ 27.2 & 17.6 & 17.6 & 44.6 & 1.59 & 169 & 20.2 & 5.8 $\pm1.5$ & 8.9 $\pm1.0$ & 38.5 $\pm1.4$ & 45$^{ N}$ \\
37 & 13.67178 & -37.71849 & 156.0 $\pm12.8$ & 198.9 $\pm28.5$ & 177.7 $\pm18.7$ & 101.5 $\pm16.4$ & $<$ 27.5 & 17.8 & 17.6 & 156.7 & 2.61 & 170 & 19.9 & 5.5 $\pm1.4$ & 5.7 $\pm1.1$ & 14.4 $\pm3.4$ & 46$^{ D}$ \\
38 & 13.67336 & -37.71276 & 144.0 $\pm18.0$ & 282.9 $\pm24.6$ & 155.7 $\pm20.7$ & 106.0 $\pm16.9$ & 111.1 $\pm21.6$ & 23.8 & 17.6 & 96.0 & 2.34 & 196 & 19.2 & 5.4 $\pm1.6$ & 5.6 $\pm1.1$ & 11.2 $\pm2.4$ & -- \\
39 & 13.67353 & -37.67228 & 116.3 $\pm12.4$ & 176.0 $\pm26.2$ & 175.6 $\pm17.4$ & 76.1 $\pm15.1$ & 56.7 $\pm14.2$ & 17.6 & 17.6 & 91.5 & 1.43 & 169 & 20.0 & 5.4 $\pm1.4$ & 12.4 $\pm1.2$ & $<$ 1.4 & 51$^{}$ \\
40 & 13.67601 & -37.64914 & 63.5 $\pm8.7$ & 168.0 $\pm17.2$ & 172.2 $\pm11.7$ & 108.7 $\pm10.2$ & $<$ 50.5 & 21.9 & 17.6 & 108.1 & 1.90 & 188 & 18.8 & 6.5 $\pm1.8$ & 6.6 $\pm1.0$ & 16.0 $\pm1.7$ & 49$^{ D}$, 50$^{}$ \\
41 & 13.67867 & -37.66828 & 106.2 $\pm14.5$ & 185.2 $\pm24.6$ & 201.1 $\pm17.1$ & 117.0 $\pm15.1$ & $<$ 47.7 & 18.1 & 17.6 & 171.8 & 1.34 & 171 & 21.1 & 5.3 $\pm1.2$ & 8.6 $\pm1.1$ & 9.7 $\pm1.6$ & 51$^{}$, 52$^{ D}$ \\
42\tablefootmark{e} & 13.6793 & -37.71962 & 1142.0 $\pm23.9$ & 1276.0 $\pm39.1$ & 531.9 $\pm24.7$ & 239.9 $\pm17.2$ & 134.2 $\pm22.5$ & 17.6 & 17.6 & 31.3 & 2.48 & 169 & 21.3 & 13.5 $\pm2.9$ & 52.5 $\pm4.5$ & 93.5 $\pm5.2$ & 53A$^{}$/B$^{}$/C$^{ N}$ \\
43 & 13.68396 & -37.67418 & 224.0 $\pm15.6$ & 327.1 $\pm22.8$ & 193.3 $\pm18.8$ & 81.1 $\pm15.1$ & $<$ 38.9 & 17.6 & 17.6 & 64.7 & 1.14 & 169 & 20.4 & 5.6 $\pm1.4$ & 10.2 $\pm1.0$ & 14.3 $\pm0.3$ & 56$^{ M}$ \\
44 & 13.68525 & -37.667 & $<$ 32.7 & 121.7 $\pm21.5$ & 114.3 $\pm16.8$ & 78.0 $\pm15.6$ & $<$ 43.3 & 17.6 & 17.6 & 108.4 & 1.21 & 169 & 20.4 & 3.3 $\pm0.9$ & 4.9 $\pm0.7$ & 7.4 $\pm0.3$ & 58$^{}$, 59$^{}$ \\
45 & 13.68527 & -37.60789 & 321.9 $\pm22.5$ & 186.3 $\pm16.0$ & 174.9 $\pm9.1$ & 92.0 $\pm9.3$ & 36.0 $\pm9.8$ & 35.8 & 17.6 & 99.9 & 3.62 & 241 & 16.1 & 11.1 $\pm4.2$ & 2.9 $\pm0.6$ & 2.1 $\pm0.2$ & -- \\
46\tablefootmark{f} & 13.68682 & -37.68647 & 72.0 $\pm10.5$ & 84.0 $\pm22.7$ & 100.8 $\pm18.8$ & $<$ 36.1 & $<$ 28.7 & 24.2 & 17.6 & 108.1 & 1.13 & 198 & 18.0 & 4.3 $\pm1.5$ & 2.9 $\pm0.4$ & 3.7 $\pm0.2$ & 60$^{}$ \\
47 & 13.68824 & -37.69757 & 84.7 $\pm9.9$ & 143.6 $\pm18.5$ & 103.8 $\pm20.5$ & $<$ 41.8 & $<$ 31.1 & 17.6 & 17.6 & 84.6 & 1.39 & 169 & 17.8 & 4.6 $\pm1.7$ & 3.5 $\pm0.4$ & 9.8 $\pm0.5$ & -- \\
48 & 13.68853 & -37.64615 & 213.9 $\pm9.5$ & 223.1 $\pm14.6$ & 136.0 $\pm12.1$ & 73.0 $\pm9.5$ & $<$ 28.2 & 17.6 & 17.6 & 85.7 & 1.87 & 169 & 19.4 & 4.6 $\pm1.2$ & 7.2 $\pm0.7$ & 31.0 $\pm0.5$ & 61$^{ D}$, 62$^{}$ \\
49 & 13.68957 & -37.6324 & 80.4 $\pm9.5$ & 143.5 $\pm11.7$ & 57.1 $\pm9.5$ & 25.2 $\pm7.2$ & $<$ 28.8 & 17.6 & 17.6 & 112.8 & 2.48 & 169 & 18.1 & 2.4 $\pm0.8$ & 4.0 $\pm0.5$ & 15.0 $\pm0.1$ & 63$^{ D}$ \\
50 & 13.6899 & -37.75424 & $<$ 44.6 & 71.9 $\pm14.5$ & 50.0 $\pm8.7$ & $<$ 26.4 & $<$ 20.1 & 29.3 & 17.6 & 149.3 & 3.88 & 218 & 14.6 & 4.7 $\pm2.3$ & 1.7 $\pm0.3$ & 1.0 $\pm0.2$ & -- \\
51 & 13.6941 & -37.67355 & 114.2 $\pm13.9$ & 171.2 $\pm18.9$ & 78.5 $\pm18.5$ & $<$ 44.5 & $<$ 28.5 & 17.6 & 17.6 & 34.4 & 0.88 & 169 & 19.8 & 2.5 $\pm0.8$ & 3.8 $\pm0.5$ & 6.2 $\pm0.4$ & 64$^{ N}$ \\
52 & 13.69464 & -37.63263 & 96.9 $\pm12.0$ & 114.9 $\pm13.0$ & 40.4 $\pm10.2$ & 58.7 $\pm7.3$ & $<$ 29.2 & 17.6 & 17.6 & 121.1 & 2.46 & 169 & 18.7 & 1.5 $\pm0.6$ & 4.7 $\pm0.7$ & 13.5 $\pm0.7$ & 66$^{ D}$ \\
53\tablefootmark{f} & 13.69557 & -37.72143 & 84.9 $\pm14.5$ & 174.9 $\pm16.9$ & 180.7 $\pm15.9$ & 88.3 $\pm16.4$ & $<$ 66.2 & 25.7 & 17.8 & 114.7 & 2.23 & 205 & 17.6 & 8.3 $\pm2.6$ & 4.3 $\pm0.9$ & 0.7 $\pm0.2$ & -- \\
54 & 13.69738 & -37.68962 & 355.8 $\pm30.3$ & 222.7 $\pm34.1$ & 103.8 $\pm23.4$ & $<$ 48.1 & $<$ 29.0 & 28.7 & 17.7 & 1.7 & 0.89 & 216 & 19.2 & 3.6 $\pm1.2$ & 3.3 $\pm0.7$ & 1.4 $\pm0.2$ & -- \\
55 & 13.69845 & -37.63464 & $<$ 39.7 & 151.2 $\pm16.7$ & 110.4 $\pm11.1$ & $<$ 22.3 & $<$ 29.1 & 23.0 & 17.6 & 146.0 & 2.36 & 193 & 19.1 & 3.9 $\pm1.1$ & 4.9 $\pm0.9$ & 13.8 $\pm2.2$ & 66$^{ D}$, 68$^{ D}$ \\
56 & 13.69853 & -37.71241 & 50.5 $\pm10.4$ & 75.3 $\pm17.3$ & 73.1 $\pm16.3$ & 63.3 $\pm16.7$ & $<$ 31.9 & 22.4 & 18.1 & 162.9 & 1.76 & 193 & 17.4 & 3.5 $\pm1.4$ & 3.2 $\pm0.5$ & 9.9 $\pm0.4$ & 67$^{}$ \\
57\tablefootmark{f} & 13.69963 & -37.70912 & $<$ 42.2 & 157.0 $\pm20.6$ & 79.1 $\pm19.7$ & 53.6 $\pm17.1$ & 54.1 $\pm16.9$ & 39.5 & 23.2 & 100.9 & 1.58 & 291 & 17.4 & 3.8 $\pm1.5$ & 6.4 $\pm1.1$ & 11.6 $\pm1.2$ & 67$^{}$ \\
58 & 13.69998 & -37.77926 & 149.9 $\pm19.6$ & 46.1 $\pm8.0$ & 56.5 $\pm8.1$ & 45.0 $\pm7.3$ & 45.8 $\pm9.7$ & 25.0 & 17.6 & 161.8 & 4.96 & 201 & -- & -- & 0.5 $\pm0.1$ & 0.9 $\pm0.2$ & -- \\
59 & 13.70007 & -37.726 & 51.0 $\pm9.1$ & 117.0 $\pm17.2$ & 76.6 $\pm9.5$ & $<$ 44.5 & $<$ 62.9 & 17.6 & 17.6 & 110.1 & 2.36 & 169 & 17.4 & 3.7 $\pm1.2$ & 3.4 $\pm0.4$ & 2.2 $\pm0.1$ & 69$^{ D}$ \\
60 & 13.70137 & -37.64064 & 121.5 $\pm14.7$ & 171.9 $\pm14.9$ & 38.1 $\pm10.5$ & $<$ 22.0 & $<$ 29.4 & 20.0 & 17.6 & 62.8 & 2.08 & 180 & 19.9 & 1.2 $\pm0.4$ & 5.0 $\pm0.8$ & 4.1 $\pm0.9$ & -- \\
61 & 13.7045 & -37.59367 & $<$ 2.1 & 200.0 $\pm17.5$ & 83.7 $\pm12.0$ & 67.4 $\pm8.1$ & 35.4 $\pm10.8$ & 29.1 & 17.6 & 143.9 & 4.37 & 217 & -- & -- & 0.6 $\pm0.1$ & 1.3 $\pm0.3$ & -- \\
62\tablefootmark{a,f} & 13.70522 & -37.69339 & 405.2 $\pm32.8$ & 286.3 $\pm34.4$ & 67.7 $\pm21.1$ & 71.4 $\pm14.8$ & $<$ 28.3 & 32.2 & 20.5 & 7.3 & 0.79 & 247 & 19.5 & 2.2 $\pm0.9$ & 4.3 $\pm0.9$ & 2.3 $\pm0.6$ & -- \\
63 & 13.70544 & -37.65772 & 99.7 $\pm11.2$ & 139.0 $\pm15.6$ & 133.4 $\pm13.7$ & 57.7 $\pm14.5$ & $<$ 31.6 & 20.4 & 17.6 & 122.3 & 1.27 & 182 & 19.4 & 4.5 $\pm1.2$ & 3.8 $\pm0.6$ & 9.0 $\pm0.7$ & 71$^{}$, 75$^{}$ \\
64 & 13.7069 & -37.66612 & 82.3 $\pm10.5$ & 105.1 $\pm19.8$ & 84.7 $\pm16.7$ & $<$ 34.3 & $<$ 29.2 & 17.9 & 17.6 & 137.1 & 0.90 & 170 & 20.8 & 2.3 $\pm0.7$ & 3.4 $\pm0.5$ & 1.9 $\pm0.3$ & -- \\
65\tablefootmark{e} & 13.70871 & -37.63924 & 695.0 $\pm19.0$ & 760.6 $\pm20.3$ & 410.6 $\pm12.5$ & 256.9 $\pm10.4$ & 135.1 $\pm19.2$ & 17.6 & 17.6 & 115.1 & 2.16 & 169 & 23.2 & 8.3 $\pm1.5$ & 34.7 $\pm3.5$ & 47.3 $\pm3.7$ & 77$^{}$, 79$^{ D}$ \\
66 & 13.70922 & -37.67468 & 459.4 $\pm17.1$ & 759.8 $\pm33.3$ & 412.8 $\pm17.5$ & 169.3 $\pm15.7$ & 65.9 $\pm14.3$ & 18.7 & 17.6 & 71.5 & 0.54 & 174 & 22.1 & 9.5 $\pm1.9$ & 15.3 $\pm2.4$ & 33.6 $\pm4.3$ & 76A$^{}$/B$^{}$/C$^{ D}$ \\
67 & 13.71079 & -37.68357 & 117.9 $\pm14.9$ & 119.4 $\pm31.2$ & 87.0 $\pm17.5$ & $<$ 35.9 & $<$ 34.2 & 23.5 & 17.6 & 152.8 & 0.36 & 195 & 20.3 & 2.6 $\pm0.8$ & 4.1 $\pm0.6$ & 5.4 $\pm0.8$ & 74$^{}$, 80$^{ M}$ \\
68\tablefootmark{e} & 13.71241 & -37.63999 & 451.3 $\pm20.4$ & 496.2 $\pm19.1$ & 211.5 $\pm12.5$ & 94.5 $\pm10.9$ & $<$ 30.4 & 17.6 & 17.6 & 100.1 & 2.13 & 169 & 22.6 & 4.5 $\pm0.9$ & 26.2 $\pm4.5$ & 63.3 $\pm8.0$ & 77$^{}$, 79$^{ D}$ \\
69\tablefootmark{e,f} & 13.715 & -37.74574 & 66.8 $\pm6.2$ & 159.3 $\pm7.3$ & 82.2 $\pm10.0$ & 47.0 $\pm8.9$ & $<$ 27.3 & 38.9 & 19.3 & 151.9 & 3.12 & 263 & 14.8 & 7.3 $\pm3.4$ & 1.1 $\pm0.2$ & 1.3 $\pm0.2$ & -- \\
70 & 13.71525 & -37.60965 & 122.3 $\pm14.1$ & 77.6 $\pm15.5$ & 95.8 $\pm13.3$ & 63.4 $\pm7.2$ & $<$ 33.7 & 25.1 & 17.6 & 75.0 & 3.65 & 202 & 16.4 & 5.8 $\pm2.2$ & 1.0 $\pm0.2$ & 1.0 $\pm0.2$ & -- \\
71 & 13.71538 & -37.66106 & 469.3 $\pm10.3$ & 429.1 $\pm16.0$ & 234.0 $\pm14.9$ & 94.1 $\pm14.6$ & $<$ 32.5 & 17.6 & 17.6 & 132.6 & 1.12 & 169 & 21.5 & 5.8 $\pm1.3$ & 26.6 $\pm2.0$ & 40.0 $\pm0.4$ & 84$^{}$ \\
72 & 13.7158 & -37.7751 & $<$ 38.9 & 42.8 $\pm14.2$ & 41.1 $\pm8.8$ & 26.8 $\pm6.4$ & $<$ 24.7 & 22.5 & 17.6 & 136.5 & 4.57 & 191 & 15.3 & 3.2 $\pm1.5$ & 1.7 $\pm0.1$ & 1.6 $\pm0.4$ & 82$^{}$ \\
73 & 13.71586 & -37.64561 & 121.0 $\pm10.9$ & 160.7 $\pm16.1$ & 123.0 $\pm12.9$ & 65.4 $\pm10.9$ & $<$ 31.9 & 17.6 & 17.6 & 9.5 & 1.88 & 169 & 21.4 & 3.1 $\pm0.7$ & 3.8 $\pm0.7$ & 7.7 $\pm0.9$ & 83$^{}$ \\
74 & 13.71658 & -37.69378 & 182.1 $\pm15.1$ & 223.4 $\pm27.2$ & 80.9 $\pm17.4$ & 44.8 $\pm13.1$ & $<$ 26.4 & 17.6 & 17.6 & 67.7 & 0.55 & 169 & 20.0 & 2.5 $\pm0.8$ & 3.6 $\pm0.6$ & 6.6 $\pm0.5$ & 81$^{ D}$, 85$^{ M}$ \\
75 & 13.71823 & -37.67739 & 82.6 $\pm13.2$ & 165.0 $\pm17.3$ & 116.9 $\pm17.3$ & $<$ 45.7 & $<$ 31.5 & 18.7 & 17.6 & 169.8 & 0.34 & 174 & 21.2 & 3.0 $\pm0.8$ & 4.2 $\pm0.7$ & 3.3 $\pm0.2$ & 86$^{ D}$ \\
76 & 13.72187 & -37.72963 & 311.0 $\pm11.8$ & 402.7 $\pm16.5$ & 277.9 $\pm13.7$ & 153.7 $\pm11.2$ & 69.3 $\pm21.0$ & 19.7 & 17.6 & 92.5 & 2.25 & 179 & 19.1 & 9.8 $\pm2.6$ & 13.6 $\pm1.3$ & 29.3 $\pm2.1$ & 87$^{}$, 88$^{ D}$, 90$^{}$ \\
77\tablefootmark{c} & 13.72858 & -37.67794 & 95.4 $\pm25.8$ & 251.9 $\pm45.1$ & 131.2 $\pm33.1$ & $<$ 59.3 & $<$ 54.1 & 23.6 & 17.6 & 79.3 & 0.41 & 196 & 19.9 & 4.1 $\pm1.4$ & 2.9 $\pm0.6$ & 1.3 $\pm0.3$ & -- \\
78\tablefootmark{f} & 13.73056 & -37.61221 & 93.2 $\pm13.1$ & 114.0 $\pm16.0$ & 70.4 $\pm7.7$ & 34.2 $\pm6.9$ & $<$ 25.3 & 26.9 & 17.6 & 114.6 & 3.66 & 209 & 16.9 & 3.7 $\pm1.3$ & 1.8 $\pm0.2$ & 7.3 $\pm1.1$ & 92$^{}$, 94$^{}$ \\
79 & 13.73111 & -37.69486 & 72.3 $\pm13.5$ & 83.9 $\pm5.7$ & 60.5 $\pm15.7$ & $<$ 34.8 & $<$ 25.9 & 22.9 & 17.7 & 19.9 & 0.50 & 193 & 19.9 & 1.9 $\pm0.7$ & 3.4 $\pm0.6$ & $<$ 0.3 & -- \\
80 & 13.73168 & -37.63264 & 94.0 $\pm12.7$ & 85.0 $\pm16.2$ & 65.2 $\pm12.8$ & $<$ 24.3 & $<$ 31.2 & 23.5 & 17.6 & 158.3 & 2.66 & 195 & 17.3 & 3.2 $\pm1.2$ & 1.4 $\pm0.2$ & 0.4 $\pm0.1$ & -- \\
81 & 13.73247 & -37.64725 & 117.1 $\pm18.1$ & 109.6 $\pm25.9$ & 100.1 $\pm14.9$ & 64.3 $\pm12.7$ & $<$ 32.1 & 28.3 & 17.6 & 111.4 & 1.95 & 214 & 17.8 & 4.5 $\pm1.5$ & 2.7 $\pm0.5$ & 0.7 $\pm0.2$ & -- \\
82 & 13.73717 & -37.68686 & 199.4 $\pm8.7$ & 229.9 $\pm16.0$ & 104.5 $\pm16.0$ & $<$ 42.3 & $<$ 29.4 & 18.1 & 17.6 & 43.9 & 0.41 & 171 & 21.7 & 2.5 $\pm0.7$ & 11.9 $\pm1.1$ & 12.0 $\pm0.8$ & 100$^{ D}$ \\
83\tablefootmark{f} & 13.73955 & -37.73539 & $<$ 45.2 & 142.6 $\pm19.3$ & 118.1 $\pm15.8$ & 66.4 $\pm10.4$ & 55.7 $\pm15.3$ & 25.0 & 17.6 & 99.2 & 2.42 & 201 & 17.7 & 5.3 $\pm1.8$ & 3.0 $\pm0.5$ & 2.8 $\pm0.2$ & 102$^{}$ \\
84 & 13.73966 & -37.70701 & $<$ 45.1 & 101.1 $\pm21.2$ & 90.9 $\pm17.0$ & 63.6 $\pm13.7$ & 55.5 $\pm13.7$ & 19.6 & 17.6 & 55.2 & 1.07 & 178 & 19.1 & 3.2 $\pm1.0$ & 4.2 $\pm0.6$ & 8.8 $\pm0.9$ & 103$^{ D}$, 104$^{}$ \\
85\tablefootmark{a} & 13.73999 & -37.67157 & 121.4 $\pm13.5$ & 83.6 $\pm13.2$ & 44.1 $\pm13.7$ & $<$ 25.4 & $<$ 32.1 & 25.6 & 17.7 & 87.0 & 0.93 & 204 & 19.2 & 1.5 $\pm0.6$ & 2.4 $\pm0.5$ & 3.5 $\pm0.7$ & -- \\
86 & 13.74367 & -37.76147 & $<$ 35.1 & 90.4 $\pm17.4$ & 64.8 $\pm8.1$ & 29.7 $\pm6.8$ & $<$ 30.0 & 22.5 & 17.6 & 161.2 & 3.68 & 191 & 14.9 & 5.5 $\pm2.5$ & 1.0 $\pm0.1$ & 2.2 $\pm0.3$ & -- \\
87 & 13.74515 & -37.65286 & 31.6 $\pm9.3$ & 102.2 $\pm16.0$ & 117.9 $\pm11.7$ & 68.3 $\pm12.1$ & $<$ 50.2 & 17.6 & 17.6 & 162.1 & 1.88 & 169 & 17.6 & 5.4 $\pm1.8$ & 2.4 $\pm0.5$ & 1.4 $\pm0.3$ & 106$^{}$ \\
88 & 13.7472 & -37.63815 & 67.7 $\pm11.2$ & 118.7 $\pm14.5$ & 53.8 $\pm9.5$ & 34.1 $\pm8.1$ & $<$ 30.8 & 22.7 & 17.6 & 85.3 & 2.62 & 192 & 17.3 & 2.7 $\pm1.0$ & 1.1 $\pm0.2$ & 1.0 $\pm0.2$ & -- \\
89 & 13.74939 & -37.65717 & 68.4 $\pm11.0$ & 139.3 $\pm14.7$ & 113.7 $\pm11.6$ & 63.9 $\pm12.4$ & $<$ 51.0 & 17.6 & 17.6 & 100.5 & 1.77 & 169 & 17.9 & 5.0 $\pm1.6$ & 6.1 $\pm0.7$ & 2.2 $\pm0.3$ & -- \\
90 & 13.74954 & -37.68211 & 59.8 $\pm11.8$ & 152.8 $\pm21.7$ & 76.0 $\pm15.4$ & $<$ 27.5 & $<$ 30.3 & 18.4 & 17.6 & 50.9 & 0.84 & 173 & 20.0 & 2.4 $\pm0.7$ & 3.3 $\pm0.6$ & 3.3 $\pm0.5$ & -- \\
91 & 13.74967 & -37.7648 & $<$ 47.5 & 91.7 $\pm15.4$ & 47.2 $\pm7.4$ & 38.2 $\pm6.7$ & 39.6 $\pm9.4$ & 17.9 & 17.6 & 82.8 & 3.82 & 170 & 14.4 & 4.6 $\pm2.3$ & 0.8 $\pm0.1$ & $<$ 1.4 & 110$^{}$ \\
92 & 13.74994 & -37.7382 & 131.8 $\pm9.1$ & 124.5 $\pm20.2$ & 128.3 $\pm14.9$ & 81.0 $\pm10.1$ & $<$ 45.1 & 20.1 & 17.6 & 131.8 & 2.54 & 181 & 18.7 & 4.9 $\pm1.4$ & 5.5 $\pm0.6$ & 11.4 $\pm0.6$ & 108$^{}$ \\
93 & 13.75217 & -37.67653 & 124.8 $\pm12.0$ & 135.9 $\pm19.3$ & 122.9 $\pm16.0$ & 42.3 $\pm13.6$ & $<$ 30.6 & 17.6 & 17.6 & 34.0 & 1.07 & 169 & 19.8 & 3.9 $\pm1.1$ & 4.6 $\pm0.6$ & 20.8 $\pm0.6$ & 109$^{ D}$ \\
94 & 13.75535 & -37.65889 & $<$ 39.8 & 134.4 $\pm20.1$ & 82.7 $\pm11.7$ & 40.4 $\pm11.8$ & $<$ 30.6 & 17.6 & 17.6 & 137.0 & 1.82 & 169 & 18.1 & 3.5 $\pm1.1$ & 5.7 $\pm0.7$ & 2.6 $\pm0.4$ & -- \\
95 & 13.7576 & -37.73879 & 89.2 $\pm8.1$ & 147.4 $\pm15.5$ & 95.0 $\pm14.1$ & 70.8 $\pm10.1$ & $<$ 46.5 & 17.6 & 17.6 & 114.8 & 2.58 & 169 & 18.8 & 3.6 $\pm1.1$ & 4.2 $\pm0.6$ & 13.0 $\pm1.1$ & 112$^{ M}$, 113$^{}$, 116$^{}$ \\
96 & 13.75849 & -37.61225 & 31.1 $\pm9.5$ & 45.3 $\pm12.0$ & 76.3 $\pm6.5$ & 55.3 $\pm5.8$ & $<$ 40.0 & 24.3 & 17.6 & 157.5 & 4.04 & 199 & 14.6 & 7.2 $\pm3.4$ & 3.6 $\pm0.3$ & 4.8 $\pm0.8$ & -- \\
97 & 13.75961 & -37.69747 & 204.9 $\pm19.8$ & 204.7 $\pm30.4$ & 119.0 $\pm18.0$ & 58.2 $\pm14.4$ & $<$ 22.9 & 22.6 & 17.6 & 77.9 & 1.09 & 191 & 19.9 & 3.7 $\pm1.1$ & 4.2 $\pm0.7$ & 1.9 $\pm0.4$ & -- \\
98 & 13.75982 & -37.66504 & 99.4 $\pm9.7$ & 175.9 $\pm20.4$ & 133.1 $\pm14.4$ & 84.3 $\pm12.5$ & $<$ 47.1 & 19.7 & 17.6 & 157.0 & 1.68 & 179 & 18.0 & 5.7 $\pm1.8$ & 3.6 $\pm0.6$ & $<$ 3.3 & 114$^{ D}$ \\
99 & 13.76095 & -37.74045 & 91.6 $\pm10.1$ & 118.5 $\pm12.6$ & 134.7 $\pm13.8$ & 62.1 $\pm10.4$ & $<$ 44.1 & 18.9 & 17.6 & 75.2 & 2.66 & 175 & 18.4 & 5.4 $\pm1.6$ & 3.5 $\pm0.6$ & 6.3 $\pm1.7$ & 113$^{}$, 116$^{}$ \\
100 & 13.76397 & -37.64142 & $<$ 23.4 & 197.9 $\pm20.6$ & 78.7 $\pm8.0$ & 52.9 $\pm6.0$ & $<$ 29.1 & 28.7 & 17.6 & 11.3 & 2.77 & 216 & 17.3 & 3.9 $\pm1.3$ & 2.5 $\pm0.4$ & 11.5 $\pm1.9$ & 115$^{ N}$ \\
101 & 13.76469 & -37.71347 & 294.2 $\pm19.4$ & 297.5 $\pm30.2$ & 260.9 $\pm25.8$ & 107.1 $\pm13.0$ & 57.1 $\pm17.1$ & 24.4 & 17.6 & 174.5 & 1.60 & 199 & 21.4 & 6.6 $\pm1.5$ & 21.1 $\pm2.1$ & 58.5 $\pm3.1$ & 118A$^{}$ \\
102 & 13.76554 & -37.72287 & 568.5 $\pm26.0$ & 623.9 $\pm21.9$ & 314.7 $\pm16.7$ & 158.6 $\pm11.9$ & 78.8 $\pm18.9$ & 18.9 & 17.6 & 101.4 & 1.95 & 175 & 22.9 & 6.6 $\pm1.3$ & 25.5 $\pm2.4$ & 84.2 $\pm6.2$ & 119A$^{}$/B$^{}$/C$^{ D}$ \\
103 & 13.76655 & -37.65499 & 87.3 $\pm9.7$ & 117.0 $\pm13.5$ & 64.2 $\pm10.5$ & 37.4 $\pm8.5$ & $<$ 29.1 & 17.6 & 17.6 & 30.8 & 2.24 & 169 & 18.0 & 2.8 $\pm0.9$ & 4.0 $\pm0.5$ & 3.8 $\pm0.2$ & 120$^{ N}$ \\
104 & 13.76762 & -37.68285 & 261.3 $\pm9.2$ & 296.0 $\pm22.7$ & 152.0 $\pm16.7$ & 60.7 $\pm14.7$ & $<$ 24.0 & 17.6 & 17.6 & 129.2 & 1.37 & 169 & 20.2 & 4.5 $\pm1.2$ & 9.6 $\pm1.1$ & 17.3 $\pm1.4$ & 122$^{ D}$ \\
105 & 13.76886 & -37.69168 & 92.5 $\pm10.7$ & 139.6 $\pm25.2$ & 99.5 $\pm18.0$ & 51.4 $\pm14.4$ & $<$ 20.7 & 17.6 & 17.6 & 54.3 & 1.32 & 169 & 20.1 & 3.0 $\pm0.9$ & 4.1 $\pm0.6$ & 5.1 $\pm0.7$ & 123$^{}$, 124$^{ N}$ \\
106 & 13.77213 & -37.71663 & 97.2 $\pm15.9$ & 73.1 $\pm21.6$ & $<$ 45.0 & $<$ 32.2 & $<$ 19.9 & 17.6 & 17.6 & 101.7 & 1.82 & 169 & 20.8 & -- & 4.8 $\pm0.8$ & 7.3 $\pm1.5$ & 118B$^{ D}$ \\
107 & 13.77388 & -37.70336 & 88.9 $\pm9.0$ & 75.2 $\pm24.6$ & $<$ 50.7 & $<$ 31.1 & 25.6 $\pm6.1$ & 17.6 & 17.6 & 6.1 & 1.53 & 169 & 20.8 & -- & 16.1 $\pm1.3$ & 1.0 $\pm0.1$ & 125$^{}$ \\
108 & 13.77401 & -37.71065 & 103.5 $\pm20.2$ & $<$ 66.1 & $<$ 51.1 & 43.9 $\pm13.3$ & $<$ 48.9 & 17.6 & 17.6 & 145.5 & 1.68 & 169 & 20.1 & -- & 4.2 $\pm0.8$ & 4.5 $\pm0.9$ & -- \\
109 & 13.77646 & -37.63977 & $<$ 38.3 & 68.8 $\pm11.5$ & 65.6 $\pm6.4$ & 62.7 $\pm4.3$ & $<$ 36.6 & 17.6 & 17.6 & 161.3 & 3.11 & 169 & 14.3 & 6.6 $\pm3.2$ & 1.1 $\pm0.1$ & 0.7 $\pm0.1$ & -- \\
110\tablefootmark{f} & 13.77852 & -37.57327 & 164.6 $\pm24.4$ & 199.5 $\pm2.8$ & 88.4 $\pm11.8$ & 92.7 $\pm8.9$ & 59.1 $\pm7.4$ & 41.6 & 21.5 & 134.4 & 6.23 & 287 & -- & -- & 1.3 $\pm0.2$ & 2.8 $\pm0.6$ & -- \\
111 & 13.77887 & -37.65766 & 115.2 $\pm24.1$ & $<$ 65.6 & 65.5 $\pm14.2$ & 21.1 $\pm4.7$ & $<$ 26.0 & 18.5 & 17.6 & 68.8 & 2.44 & 173 & 18.1 & 2.8 $\pm1.0$ & 1.7 $\pm0.2$ & 3.8 $\pm0.2$ & -- \\
112\tablefootmark{f} & 13.78084 & -37.74006 & 85.4 $\pm16.2$ & 112.3 $\pm14.4$ & 69.2 $\pm12.9$ & 33.6 $\pm9.5$ & $<$ 20.3 & 25.1 & 17.6 & 88.2 & 2.78 & 202 & 15.4 & 5.2 $\pm2.4$ & 1.2 $\pm0.1$ & $<$ 0.9 & -- \\
113 & 13.78087 & -37.69666 & $<$ 36.6 & 157.5 $\pm26.1$ & 105.2 $\pm18.6$ & 83.4 $\pm12.3$ & $<$ 55.0 & 19.0 & 17.6 & 160.5 & 1.66 & 176 & 19.6 & 3.5 $\pm1.1$ & 5.7 $\pm0.8$ & 12.8 $\pm0.5$ & 127$^{ D}$ \\
114 & 13.78164 & -37.68459 & 148.6 $\pm13.5$ & 232.3 $\pm27.9$ & 110.1 $\pm16.1$ & 99.2 $\pm12.3$ & $<$ 57.3 & 17.6 & 17.6 & 51.1 & 1.76 & 169 & 20.0 & 3.4 $\pm0.9$ & 4.9 $\pm0.7$ & 9.0 $\pm0.3$ & 126$^{ D}$ \\
115 & 13.78551 & -37.69492 & 87.8 $\pm11.4$ & 99.6 $\pm27.0$ & 90.9 $\pm16.9$ & 67.3 $\pm12.4$ & $<$ 57.1 & 20.8 & 17.6 & 61.5 & 1.79 & 184 & 19.5 & 3.0 $\pm0.9$ & 4.9 $\pm0.8$ & 4.8 $\pm0.5$ & 128$^{}$ \\
116 & 13.78642 & -37.65862 & 136.4 $\pm8.5$ & 154.6 $\pm12.2$ & 123.4 $\pm10.3$ & 56.1 $\pm9.9$ & $<$ 25.6 & 17.6 & 17.6 & 120.5 & 2.61 & 169 & 18.2 & 5.1 $\pm1.5$ & 4.2 $\pm0.3$ & 12.9 $\pm0.2$ & 129$^{ M}$ \\
117 & 13.78843 & -37.72107 & 61.2 $\pm9.7$ & 69.3 $\pm13.5$ & 74.4 $\pm12.9$ & 42.0 $\pm9.1$ & $<$ 35.7 & 21.0 & 17.6 & 42.0 & 2.24 & 185 & 17.2 & 3.7 $\pm1.4$ & 2.1 $\pm0.3$ & 1.8 $\pm0.3$ & -- \\
118 & 13.78884 & -37.67058 & 32.4 $\pm10.3$ & 79.4 $\pm17.5$ & 37.7 $\pm12.2$ & $<$ 19.3 & $<$ 23.9 & 19.5 & 17.6 & 169.3 & 2.29 & 178 & 18.2 & 1.6 $\pm0.7$ & 2.0 $\pm0.3$ & 1.0 $\pm0.2$ & -- \\
119 & 13.79118 & -37.79867 & 75.4 $\pm6.6$ & 116.8 $\pm11.8$ & 90.9 $\pm6.8$ & 61.1 $\pm7.3$ & 53.1 $\pm9.9$ & 21.6 & 17.6 & 105.9 & 5.41 & 187 & 14.5 & 8.6 $\pm4.0$ & 6.1 $\pm0.5$ & 9.2 $\pm0.8$ & 131$^{}$, 133$^{ N}$ \\
120 & 13.79592 & -37.74146 & 70.8 $\pm15.8$ & 191.5 $\pm16.5$ & 89.3 $\pm11.3$ & 62.8 $\pm8.6$ & $<$ 32.2 & 21.5 & 17.6 & 43.8 & 3.01 & 187 & 15.9 & 6.0 $\pm2.4$ & 2.1 $\pm0.3$ & 13.9 $\pm1.5$ & 134$^{}$, 135$^{}$ \\
121 & 13.79713 & -37.68832 & 190.8 $\pm17.6$ & 193.5 $\pm28.4$ & 206.6 $\pm18.0$ & 124.8 $\pm13.7$ & 69.8 $\pm17.4$ & 23.5 & 17.6 & 60.5 & 2.18 & 195 & 20.6 & 5.8 $\pm1.4$ & 8.8 $\pm1.2$ & 26.3 $\pm2.7$ & 137B$^{ D}$/D$^{}$ \\
122 & 13.79992 & -37.60993 & $<$ 55.5 & $<$ 42.0 & 36.8 $\pm6.8$ & 38.8 $\pm6.0$ & 33.9 $\pm8.5$ & 17.6 & 17.6 & 100.5 & 4.93 & 169 & -- & -- & 0.4 $\pm0.1$ & 0.7 $\pm0.1$ & -- \\
123\tablefootmark{e} & 13.80164 & -37.69454 & 317.2 $\pm18.8$ & 309.7 $\pm24.0$ & 470.7 $\pm18.6$ & 230.6 $\pm13.0$ & 49.8 $\pm16.5$ & 32.7 & 19.8 & 97.6 & 2.27 & 244 & 23.0 & 9.6 $\pm1.8$ & 35.8 $\pm3.9$ & 116.7 $\pm11.7$ & 137A$^{ D}$/B$^{ D}$/C$^{ D}$ \\
124\tablefootmark{e} & 13.80328 & -37.69018 & 127.9 $\pm16.0$ & 173.9 $\pm21.3$ & 127.0 $\pm13.6$ & 82.5 $\pm11.6$ & 51.5 $\pm15.7$ & 23.8 & 17.6 & 73.6 & 2.35 & 196 & 21.6 & 3.1 $\pm0.7$ & 13.7 $\pm3.0$ & 62.8 $\pm10.3$ & 137A$^{ D}$/B$^{ D}$/D$^{}$ \\
125 & 13.80763 & -37.73738 & 97.6 $\pm11.0$ & 185.7 $\pm16.2$ & 119.3 $\pm8.7$ & 71.6 $\pm5.2$ & 37.6 $\pm10.9$ & 24.0 & 17.6 & 98.0 & 3.05 & 197 & 16.8 & 6.6 $\pm2.3$ & 2.9 $\pm0.4$ & 6.2 $\pm1.0$ & 139$^{ D}$ \\
126\tablefootmark{e} & 13.80799 & -37.69424 & 109.4 $\pm15.7$ & 164.4 $\pm23.5$ & 63.5 $\pm12.8$ & 61.6 $\pm10.5$ & 59.2 $\pm14.5$ & 27.1 & 17.6 & 1.5 & 2.46 & 210 & 22.2 & 1.4 $\pm0.4$ & 27.6 $\pm4.8$ & 58.5 $\pm14.4$ & 137A$^{ D}$/B$^{ D}$/C$^{ D}$ \\
127\tablefootmark{a} & 13.812 & -37.71453 & $<$ 68.3 & 154.1 $\pm21.5$ & 81.6 $\pm13.3$ & 41.2 $\pm7.3$ & $<$ 18.5 & 30.6 & 20.3 & 79.6 & 2.64 & 239 & 16.7 & 4.5 $\pm1.7$ & 1.4 $\pm0.2$ & 1.5 $\pm0.3$ & -- \\
128 & 13.81317 & -37.73794 & 81.3 $\pm7.8$ & 85.4 $\pm16.2$ & 56.6 $\pm8.0$ & $<$ 16.5 & $<$ 21.4 & 19.9 & 17.6 & 97.0 & 3.17 & 180 & 16.9 & 3.0 $\pm1.1$ & 6.7 $\pm0.7$ & 5.9 $\pm1.2$ & 140$^{ D}$, 142$^{}$ \\
129 & 13.81882 & -37.74327 & 149.2 $\pm17.0$ & 240.3 $\pm23.2$ & 86.1 $\pm6.7$ & 51.4 $\pm5.1$ & 39.3 $\pm10.5$ & 26.1 & 17.6 & 112.1 & 3.43 & 206 & 16.4 & 5.1 $\pm1.9$ & 1.6 $\pm0.3$ & 3.3 $\pm0.8$ & 141$^{}$ \\
130 & 13.82135 & -37.72594 & 84.5 $\pm8.3$ & 135.4 $\pm11.9$ & 72.5 $\pm9.4$ & 22.2 $\pm7.2$ & $<$ 19.6 & 17.6 & 17.6 & 35.4 & 3.05 & 169 & 16.8 & 3.9 $\pm1.4$ & 2.8 $\pm0.3$ & 10.9 $\pm0.4$ & 143$^{}$ \\
131\tablefootmark{e,f} & 13.827 & -37.60063 & 145.2 $\pm15.6$ & 196.4 $\pm1.0$ & 80.9 $\pm6.8$ & 90.4 $\pm6.8$ & 40.3 $\pm8.3$ & 37.3 & 17.6 & 119.0 & 5.92 & 246 & 13.7 & 9.8 $\pm5.2$ & 1.0 $\pm0.1$ & 1.4 $\pm0.3$ & -- \\
132\tablefootmark{c} & 13.83214 & -37.77987 & 91.9 $\pm20.4$ & 226.1 $\pm30.4$ & 98.9 $\pm10.4$ & 49.2 $\pm8.4$ & $<$ 46.4 & 28.5 & 20.0 & 155.0 & 4.88 & 229 & 15.5 & 7.2 $\pm3.0$ & 1.5 $\pm0.3$ & 3.7 $\pm1.0$ & 144$^{ N}$ \\
133\tablefootmark{a,b} & 13.83245 & -37.69946 & -- & -- & 62.8 $\pm11.8$ & 36.6 $\pm7.3$ & $<$ 20.8 & 32.8 & 20.7 & 34.7 & 3.15 & 250 & 16.3 & 3.8 $\pm1.6$ & 1.2 $\pm0.2$ & 1.0 $\pm0.2$ & -- \\
134 & 13.83273 & -37.65971 & 20.8 $\pm4.9$ & $<$ 27.6 & 35.6 $\pm7.7$ & $<$ 18.1 & $<$ 13.1 & 17.6 & 17.6 & 72.0 & 3.86 & 169 & -- & -- & 0.3 $\pm0.1$ & 0.6 $\pm0.1$ & -- \\
135\tablefootmark{f} & 13.83837 & -37.73078 & 161.1 $\pm21.8$ & 125.2 $\pm11.2$ & 190.2 $\pm10.6$ & 68.7 $\pm6.7$ & $<$ 22.8 & 28.2 & 22.2 & 80.0 & 3.54 & 240 & 16.2 & 11.8 $\pm4.4$ & 3.0 $\pm0.5$ & 11.3 $\pm2.7$ & 145$^{ N}$, 146$^{ N}$ \\
136\tablefootmark{a} & 13.84535 & -37.65303 & $<$ 63.7 & 103.2 $\pm17.4$ & 47.4 $\pm10.2$ & 31.7 $\pm5.4$ & $<$ 17.5 & 23.5 & 19.6 & 89.9 & 4.42 & 206 & 14.0 & 5.2 $\pm2.9$ & 1.0 $\pm0.1$ & 3.0 $\pm0.5$ & -- \\
137 & 13.84906 & -37.66142 & $<$ 23.4 & 60.7 $\pm12.6$ & 43.1 $\pm8.7$ & $<$ 17.6 & $<$ 20.4 & 17.6 & 17.6 & 76.8 & 4.29 & 169 & 14.5 & 4.2 $\pm2.1$ & 0.9 $\pm0.1$ & 6.1 $\pm1.0$ & 147$^{ N}$ \\
138 & 13.85394 & -37.72831 & $<$ 5.4 & 74.7 $\pm12.3$ & 45.3 $\pm8.1$ & 39.6 $\pm4.7$ & $<$ 28.7 & 27.4 & 17.6 & 37.5 & 3.89 & 211 & 13.1 & 6.7 $\pm4.1$ & 0.5 $\pm0.1$ & 1.0 $\pm0.2$ & -- \\
139\tablefootmark{f} & 13.86793 & -37.67635 & 33.6 $\pm9.9$ & $<$ 30.8 & 85.5 $\pm10.1$ & 55.4 $\pm7.2$ & $<$ 22.5 & 36.1 & 19.2 & 51.1 & 4.51 & 253 & 13.6 & 10.7 $\pm5.9$ & 1.2 $\pm0.2$ & 6.4 $\pm1.5$ & 151$^{ N}$, 152$^{}$ \\
140\tablefootmark{a} & 13.87281 & -37.65585 & $<$ 45.8 & 89.7 $\pm14.1$ & 45.4 $\pm9.9$ & 30.5 $\pm5.2$ & $<$ 26.8 & 23.5 & 17.6 & 154.8 & 5.13 & 195 & 15.2 & 3.6 $\pm1.7$ & 1.6 $\pm0.2$ & 9.6 $\pm1.1$ & 153$^{}$ \\
141 & 13.88103 & -37.66699 & 28.7 $\pm8.4$ & 63.7 $\pm11.5$ & 48.4 $\pm10.6$ & 42.0 $\pm7.0$ & $<$ 24.6 & 17.6 & 17.6 & 71.5 & 5.10 & 169 & 13.5 & 6.2 $\pm3.6$ & 0.4 $\pm0.1$ & 1.5 $\pm0.3$ & -- \\
142\tablefootmark{a} & 13.88628 & -37.65171 & $<$ 58.3 & 205.4 $\pm18.6$ & 46.6 $\pm7.4$ & 35.2 $\pm5.9$ & $<$ 26.3 & 28.0 & 17.6 & 176.5 & 5.63 & 213 & 13.9 & 5.3 $\pm2.9$ & 1.4 $\pm0.1$ & 7.3 $\pm0.6$ & 157$^{}$ \\
143 & 13.88951 & -37.69609 & 154.2 $\pm17.6$ & 65.0 $\pm16.0$ & 49.1 $\pm6.4$ & $<$ 20.4 & $<$ 22.6 & 21.4 & 18.3 & 138.3 & 4.88 & 190 & 13.7 & 5.9 $\pm3.2$ & 2.2 $\pm0.2$ & 8.1 $\pm1.0$ & 158$^{}$ \\
144 & 13.89147 & -37.72044 & 76.1 $\pm7.1$ & 92.1 $\pm10.8$ & 70.9 $\pm5.9$ & 48.5 $\pm5.2$ & $<$ 22.9 & 21.0 & 17.6 & 41.8 & 4.85 & 185 & 16.2 & 4.4 $\pm1.7$ & 3.3 $\pm0.3$ & 17.5 $\pm0.7$ & 159$^{}$ \\
145 & 13.89599 & -37.6646 & $<$ 5.2 & $<$ 23.9 & 27.0 $\pm7.5$ & 25.6 $\pm6.0$ & $<$ 29.6 & 20.9 & 17.6 & 115.0 & 5.59 & 184 & -- & -- & 0.5 $\pm0.1$ & 1.8 $\pm0.4$ & -- \\
146\tablefootmark{a,b} & 13.90158 & -37.69759 & -- & -- & 47.0 $\pm8.6$ & 32.2 $\pm6.4$ & $<$ 25.9 & 23.4 & 17.6 & 79.7 & 5.22 & 195 & 13.9 & 5.3 $\pm2.8$ & 3.9 $\pm0.3$ & 6.1 $\pm0.5$ & 161$^{}$ \\
\end{longtable}
\tablefoot{
\footnotesize
\\
\tablefoottext{a}{Sources labelled as tentative by \textit{getsources}. See $\S$~\ref{sec:selection} for more details.}\\
\tablefoottext{b}{Sources added from source extraction for SPIRE-images only. See $\S$~\ref{subsec:settings} and $\S$~\ref{sec:selection} for more details.}\\
\tablefoottext{c}{Strongly deblended with another discarded source candidate. Flux of discarded source candidate was added to this source.}\\
\tablefoottext{d}{Position incorrectly shifted due to background galaxy}\\
\tablefoottext{e}{Uncertain temperature estimate.}\\
\tablefoottext{f}{Size estimate likely too high.}
}
\end{landscape}

\twocolumn

\section{Hubble images}
\label{app:hubble}

\begin{table}
\caption{Information on Hubble images.}
\label{tbl:hubble}
\centering
\small
\renewcommand{\arraystretch}{1.2}
\resizebox{\columnwidth}{!}{%
\begin{tabular}{ccccccc}
        \hline\hline
        Region name & R & G & B & PropID & VisitNum \\
        \hline
        ACS 1 & F814W & F555W & F435W & 9492 & 01 \\ 
        ACS 2/3\tablefootmark{a} & F814W & F555W & F435W & 9492 & 02/03 \\
        ACS 4 & F814W & F555W & F435W & 9492 & 04 \\
        ACS 5 & F814W & F555W & F435W & 9492 & 05 \\
        ACS 6 & F814W & F555W & F435W & 9492 & 06 \\
        ACS WIDE\tablefootmark{a} & F814W & F606W & F475W & 10915 & 09/10/11 \\
        ACS 7 & F814W & R+B\tablefootmark{c} & F606W & 13515 & 01 \\
        ACS 8 & F814W & R+B\tablefootmark{c} & F606W & 13515 & 02 \\
        ACS 9 & F814W & R+B\tablefootmark{c} & F606W & 13515 & 03 \\
        ACS 10 & F814W & R+B\tablefootmark{c} & F606W & 13515 & 04 \\
        WFPC2 02 & F814W & R+B\tablefootmark{c} & F555W & 8584 & 02 \\
        WFPC2 09-12\tablefootmark{a} & F814W & R+B\tablefootmark{c} & F606W & 10915 & 09/10/11 \\
         & F814W & R+B\tablefootmark{c} & F606W & 9162 & 12 \\
        WFC3 02 & F814W & F555W & F438W & 13743 & 02 \\
        WFC3 04 & F814W & F555W & F438W & 13743 & 04 \\
        WFC3\tablefootmark{b} & - & - & - & 13515 & 02/03/04 \\
        \hline
\end{tabular}
}
\tablefoot{ 
\footnotesize 
\tablefoottext{a}{Mosaic}\\ 
\tablefoottext{b}{Mosaic of monochromatic exposures with the F336W filter}\\
\tablefoottext{c}{Coaddition of the R and B images with Montage}
}
\end{table}

\begin{figure}
\centering
\includegraphics[width=\columnwidth]{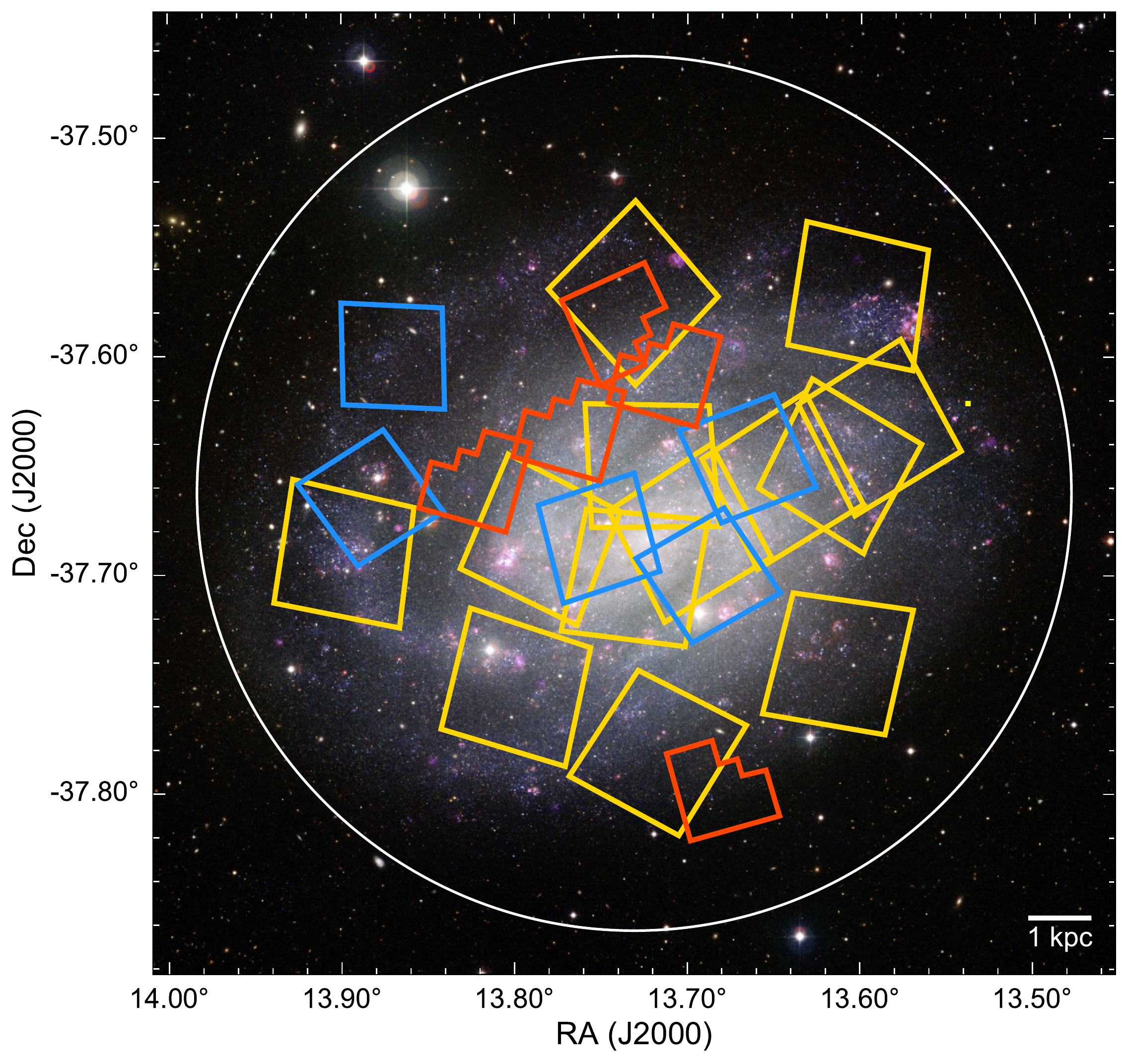}
\caption{Optical image of NGC~300 (Blue: 475 nm and OIII; Green: 605 nm; Orange: SII; Red: H$\alpha$) overplotted with selected regions of archival Hubble data used in this work. 
        The colours of the regions indicate the Hubble instruments (Yellow: ACS/WFI--Advanced Camera for Surveys/Wide Field Imager; Red: WFPC2--Wide Field and Planetary Camera 2; Blue: WFC3/UVIS--Wide Field Camera 3). 
The white circle indicates our chosen ROI, inside which we searched for sources.
Credit background image: ESO.}
\label{fig:hubble}
\end{figure}

In Table \ref{tbl:hubble} we list all Hubble images we used for the visual comparison and in Figure \ref{fig:hubble} we show their outlines overplotted on the ESO/WFI image of NGC~300.
The ACS/WFC images have a resolution of 0.05" per pixel and an effective field of view of 202"$\times$202".
The WFPC2 observations have half the resolution of the ACS/WFC images (0.1" per pixel) with an L-shaped field of view approximated by a 150"$\times$150" square missing one quadrant.
The WFC3/UVIS images have the best resolution with 0.04" per pixel while covering a field of view of 164"$\times$164".

\end{appendix} 

\end{document}